\documentclass[apj,numberedappendix]{emulateapj}
\pdfoutput=1
\usepackage{mathptmx}
\usepackage{ulem}

\def\gtorder{\mathrel{\raise.3ex\hbox{$>$}\mkern-14mu
             \lower0.6ex\hbox{$\sim$}}}
\def\ltorder{\mathrel{\raise.3ex\hbox{$<$}\mkern-14mu
             \lower0.6ex\hbox{$\sim$}}}

\slugcomment{Accepted to ApJ}
\shorttitle{Giant Extra-galactic Sparks?}
\shortauthors{Kulkarni et al.}

\begin{document}

\title{Giant Sparks at Cosmological Distances?}
\author{
S. R. Kulkarni,\altaffilmark{1} 
E. O. Ofek,\altaffilmark{2}
J. D. Neill,\altaffilmark{3}
Z. Zheng\altaffilmark{4} \&\
M. Juric\altaffilmark{5,6} 
}
\altaffiltext{1}{Caltech Optical Observatories 249-17,
California Institute of Technology, Pasadena, CA~91125}
\altaffiltext{2}{Department of Particle Physics \&\ Astrophysics, 
Weizmann Institute of Science, Rehovot 76100, Israel}
\altaffiltext{3}{Space Radiation Laboratory 290-17,
California Institute of Technology, Pasadena, CA~91125}
\altaffiltext{4}{Department of Physics \&\ Astronomy, University of Utah, 
115 South 1400 East \#201, Salt Lake City, UT 84112}
\altaffiltext{5}{Steward Observatory, 933 N. Cherry Avenue, Tucson, AZ 85721}
\altaffiltext{6}{Large Survey Synoptic Survey Telescope, 933 N. Cherry Avenue, 
Tucson, AZ 85721}

\begin{abstract}

Millisecond duration bright radio pulses at 1.4-GHz with high
dispersion measures (DM) were reported by Lorimer et al.,
Keane et al., and Thornton et al.  Their all-sky
rate is $\approx 10^4$/day above $\sim$1 Jy.  Related events are
``Perytons'' -- similar pulsed, dispersed sources, but most certainly
local.  Suggested models of fast radio bursts (FRBs) can originate in
the Earth's atmosphere, in stellar coronae, in other galaxies, and even at
cosmological distances.  Using physically motivated assumptions
combined with observed properties, we
explore these models.  In our analysis, we focus on the 
Lorimer event: a 30 Jy, 5-ms duration burst with
DM$=$ 375 cm$^{-3}$ pc, exhibiting a steep frequency-dependent pulse
width (the {\it Sparker}). To be complete, we drop the
assumption that high DMs are produced by plasma propagation and
assume that the source produces pulses with
frequency-dependent arrival time (``chirped signals''). Within this
framework we explore a scenario in which Perytons, the {\it Sparker}, and
the FRBs are all atmospheric phenomenon occurring at different heights.
This model is {\it ad hoc} in that we cannot explain why Perytons at
higher altitudes show greater DMs or exhibit narrower pulses.  Nonetheless,
we argue the {\it Sparker} may be a Peryton.  We end with two remarks.
First, the detection of a single FRB by an interferometer with a kilometer
(or longer) baseline will prove that FRBs are of
extra-terrestrial origin.  Second, we urge astronomers to pursue
observations and understanding of Perytons since they form (at least)
a formidable foreground for the FRBs.

\end{abstract}

\section{Introduction}
\label{sec:Introduction}

The subject of radio transients seems to have finally come of age.
The Galactic list starts with GCRT~J1745$-$3009, an erratic source
in the meter-wave band \citep{hlk+05} and is followed by neutron
stars which produce strong pulses occasionally, the so-called
Rotating Radio Transients (RRATs; \citealt{mll+06}). Radio afterglows
appear to routinely follow giant flares from soft-gamma repeaters
\citep{fkb99}.  Recently, an entirely new type of radio source
\citep{zbs+11} was unexpectedly discovered first in the hard X-ray
band, Swift\,J164449.3+573451 \citep{bkg+11}. To within the exquisite
astrometric precision afforded by radio VLBI the radio counterpart
coincides with the nucleus of a small star-forming galaxy at $z=0.35$
\citep{ltc+11}.  The Lorentz factor of this relativistic explosion,
$\sim10$, is smaller by an order of magnitude to those inferred in
gamma-ray bursts.  A plausible model for the source is blazar
activity initiated by feeding a tidally disrupted star to a nuclear
black hole \citep{bgm+11}.  A recent summary of the rates of
extra-galactic radio transients can be found in \citet{fko+12}.

\citet{lbm+07}  reported the discovery of an intense (30\,Jy) and
short duration (5-ms) burst in the decimeter band (1.4\,GHz). This
transient was found as a result of the archival analysis of the
Parkes Multi-Beam pulsar data obtained towards the  Magellanic
Clouds.  The dispersion measure (DM)
of this burst, 375\,cm$^{-3}$\,pc, considerably exceeded  the sum
of the estimated dispersion measure contributed by the interstellar
medium of our own Galaxy and that contributed by the Magellanic
Clouds. Follow-up observations did not show any repeating burst.

Lorimer and colleagues proposed that most of the dispersion measure
arose from electrons in the inter-galactic medium (IGM). Using
currently accepted values for the density of the IGM they estimated
the redshift to this event -- which, here after, we call the {\it
Sparker} -- to be about 0.12 ($\sim 500$\,Mpc).

The {\it Sparker} would be the first impulsive radio transient event
seen from outside the Local Group. If so, this discovery assumes a
seminal stature.  Specifically, the sharp pulse will enable astronomers
to probe the column density, magnetic field and turbulence of the
IGM \citep{mk13,Cordes2013}.  The discovery appeared timely given
that several countries have undertaken massive investments in radio
astronomy in the meter and decimetric bands --  the Low Frequency
Array (LOFAR; \citealt{fvd+06}), the  Murchison Wide Field Array
(MWA; \citealt{bwk+07}) and the Australian Square-Kilometre-Array
Pathfinder (ASKAP; \citealt{jtb+08}).  In short, in the parlance
so popular with funding agencies, the discovery reported by
\cite{lbm+07} could be a transformational finding.

The discovery of the {\it Sparker} motivated further archival
searches.  Additional transients were found with some features
similar to that of the {\it Sparker} \citep{bbe+11}. Some of these
bursts exhibited a trajectory in a plane of  arrival-time ($t$) and
frequency ($\nu$) as follows: $t(\nu)\propto \nu^{-n}$ but with
$n\approx 2$. We remind the reader that for a pulse traveling through
cold plasma, $n$ is exactly 2 \citep{rl79}.  Furthermore, some of the bursts
showed ``lumpy'' emission (that is, the broad band spectrum could
not be described by a simple power law).  Most troubling was that
these events were detected in many beams.  These sources were dubbed
{\it Perytons} by the the discoverers.  Burke-Spolaor and colleagues
argue that Perytons are atmospheric phenomenon and explain the
detection in all (most) beams to pickup by distant side lobes.
\cite{kbb+12} found additional Perytons in a second re-analysis and
noted that a cluster of Perytons were separated by about 22\,s and
suggested that Perytons are artificial signals.  Regardless, it is
now accepted that Perytons are terrestrial in origin. The DM of the
{\it Sparker} was noted to be similar to that inferred for Perytons,
and as a result some doubt was cast on the extra-galactic nature
of the {\it Sparker}.

Another archival search of the Parkes Galactic Plane Survey data
found a transient event in a single beam and with a dispersion
measure of 746\,cm$^{-3}$\,pc \citep{ksk+12}.  Earlier this year,
\cite{tsb+13} reported the finding of four short duration bursts.
One of these events showed a frequency dependent arrival time with
$n=2$ to within the precision offered by the measurements.  These
bursts with peak fluxes of about 1\,Jy also showed dispersion
measures (ranging from 553\,cm$^{-3}$\,pc to 1104\,cm$^{-3}$\,pc)
in considerable excess of that expected from the Galactic interstellar
medium.  Unlike the Perytons these four events were found in only
one beam.  This archival analysis drew data from the ``High Time
Resolution Universe'' (HTRU) survey which in turn used a digital
filter bank \citep{kjv+10}, whereas the older Parkes data were
obtained with an analog filter bank.  Thornton and colleagues, like
\cite{lbm+07} before, argue that the excess dispersion arose primarily
in the IGM and infer redshifts ranging from $z=0.45$ to $0.96$.
These authors quote an all-sky rate of $\dot{\mathcal{N}}\approx
10^4\,{\rm events\ day}^{-1}$.  This is a remarkably high rate for
an extra-galactic population (assuming no repetitions).

%Keane et al.  FRB
%Detected in one beam
%l=25.4,b=-4.0
%8ms width (but entirely consistent with expected instrumental
%DM broadening. Delta t (intrinsic) < 3 ms.
%Observed peak flux density 0.4 Jy
%Consistent with nu^-2
%WHY NO GALACTIC SCATTERING? (see Bannister paper)

Curiously, the brightest burst in \citet{tsb+13} exhibits an
asymmetric pulse shape, with a rise time smaller than the decay
time.  Furthermore, for this event and the {\it Sparker}, the observed
pulse width is frequency dependent with the pulse width,
$\Delta\tau$ $\propto \nu^{-m}$
and $m\approx 4$.  Such a characteristic pulse frequency-dependent
broadening is also seen in pulsars with large DMs and attributed
to multi-path scattering as the radio pulse traverses through
inhomogeneous structures in the interstellar medium. In contrast,
the Perytons show symmetrical pulse profiles.

To summarize, analysis of the Parkes Multi-Beam data with two
different pulsar backends (one analog and the other digital) taken
during the course of pulsar searches in the 1.4\,GHz band at the
Parkes Observatory have shown three  types of impulsive radio bursts:
Perytons \citep{bbe+11}, the {\it Sparker} \citep{lbm+07} and Fast
Radio Bursts \citep[FRBs,][]{ksk+12,tsb+13}.  There is agreement that Perytons
are of terrestrial origin. In contrast, the {\it Sparker} and FRBs
have been argued to arise from extra-galactic sources.

%S------------------------------------------------------------------------------
\section{The Rationale and Layout of the paper}
\label{sec:RationaleLayout}
%-------------------------------------------------------------------------------

The inference that the {\it Sparker} and FRBs are of extra-galactic
origin is not unreasonable. However, great claims need great proofs.
It is important to explore if there are plausible explanations of
the excess electron column density arising either in our own Galaxy
or in its extended environs. It is this exploration of alternative
frameworks that is the primary purpose of this paper.

We focus on three observational clues for FRBs: 
\begin{enumerate}
\item[I.] The arrival time of the pulses vary as $\nu^{-2}$ where
$\nu$ is the sky frequency of the pulse.  
\item[II.] For two events the width of the pulse scales as $\nu^{-4}$.
\item[III.] The all-sky rate of the FRBs is $\dot\mathcal{N}\approx
10^4$ events per day.  
\end{enumerate} 
Other clues include the DM, the peak flux, the pulse duration, and limits on the repetition rate.  We have a preliminary measure of how the source count scales with flux from \citet{tsb+13}.  However, we have little information about their angular distribution (isotropic versus Galactic).

The first version of this paper was completed and submitted (to the
Astrophysical Journal) a few months after the \cite{lbm+07} paper
was published.   The primary result of that manuscript was that,
if the frequency dependent arrival time was due to propagation,
then the {\it Sparker} had to be located beyond the Local Group.
However, after inspection of the raw data of the {\it Sparker}
(kindly provided by D.  Lorimer) we developed some doubts about the
celestial nature of the event and so we withdrew the manuscript.
Subsequently, the emergence of Perytons further questioned the
celestial origin of the {\it Sparker}.  The publication by \cite{tsb+13}
showed that the {\it Sparker} was not unique.  Furthermore, the
rash of papers attempting to explain the origin of FRBs shows the
general interest in exploring the extra-galactic nature of FRBs.  Our
interest was revived -- whence this paper.

The paper is quite long and so a summary of the goals is likely to
help prepare the reader as she/he gets ready to read the rest of
the paper.  The goals of this paper are three fold:
\begin{enumerate}
\item[A.] Accepting that the $\nu^{-2}$ arrival time pulse sweep arises from
	propagation in cold plasma, we attempt to constrain the size ($L$)
	and distance ($d$) to the nebula which contains this cold plasma.
	Clearly $d$ is smaller than the distance to the source, $D$.
\item[B.] The events, by virtue of being impulsive, must arise in compact
	regions.  We investigate whether the proposed models would allow
	for decimetric radio pulses to propagate freely from the explosion
	site.
\item[C.] Given the difficulty of an extra-galactic origin for the {\it
	Sparker} and FRBs, we consider the possibility that the trajectory
	of the pulse in the arrival-time-frequency plane is 
	a property of the source itself\footnote{a ``chirped
	signal'' in the parlance of electrical engineering} and that
	Perytons, the {\it Sparker} and FRBs are all local sources.
	We confront  this ``unified'' model with the observations.
\end{enumerate}

The outline is as follows.  The {\it Sparker} by its sheer brilliance,
by having the lowest DM of the proposed Fast Radio Burst family,
and by having a DM similar to Perytons still claims an important
position in this discussion. As such we review this event in
considerable detail.  In \S\ref{sec:TheSpark} we summarize the basic
observations of the {\it Sparker}. In \S\ref{sec:DistanceSize} we
posit an intervening nebula that can account for the excess DM
inferred for the {\it Sparker}.  Using H$\alpha$ surveys, {\it
Galaxy Evolution Explorer} ({\it GALEX}) ultraviolet (UV) data, and
the fact that the decimetric signal from the {\it Sparker} cannot
be heavily absorbed by the ionized nebula, we exclude portions of
the $L$-$d$ phase space.  We conclude that the {\it Sparker} cannot
be located in our Galaxy, in the SMC, nor even in the Local Group.
We investigate potential caveats to this important conclusion: a
porous nebula (\S\ref{sec:PorousNebula}), a nebula ionized by shocks
instead of UV photons (\S\ref{sec:Caveat}) and the possibility that
the hot corona of a star can provide the excess DM
(\S\ref{sec:StellarCorona}).

In \S\ref{sec:ExtraGalacticOrigin} we conclude that the simplest
explanation is that the {\it Sparker}, and by implication the FRBs,
is located well outside the Local Group.  In
\S\ref{sec:Progenitors}--\ref{sec:GiantFlares} we review the proposed
models for FRBs. We check whether the models allow for successful
propagation of decimetric radiation from the site of the explosion,
and separately we check if the large daily rate of FRBs can be
accommodated by the proposed models.  We find several proposed
models fail on the first test and all but one physically motivated
models are severely challenged by the large daily rate of FRBs. We
find that a model in which the radio pulses arise from giant flares
from young extra-galactic magnetars is attractive on both counts
(\S\ref{sec:GiantFlares}).  In \S\ref{sec:ISS} we investigate the
frequency dependent broadening seen in one FRB and the {\it Sparker}
and conclude that this broadening (if due to propagation) is best
explained as due to multi-path propagation in dense interstellar
medium in the vicinity of the progenitor star.

In \S\ref{sec:CaveatNonDispersedSignal} we abandon the assumption
that the $\nu^{-2}$ arrival time pulse sweep is due to propagation,
but instead attribute the frequency sweep as a property of the
source itself.  We investigate  plausible man-made,  solar and
stellar sources.  The Perytons are undeniably local phenomena and
yet share many features with the {\it Sparker} and the FRBs.  In
\S\ref{sec:Perytons} we present a plausible model unifying these
three phenomena with the Perytons taking place close to the Parkes
telescope and the FRBs the furthest away with the {\it Sparker} in
between. We readily admit that our model for ``unifying'' the
Perytons, the {\it Sparker}, and the FRBs is not based on a physically
motivated model.

We summarize in \S\ref{sec:Conclusions}.  In short, there is little
doubt that Perytons are terrestrial signals. We are struck by and
troubled by the DM of the {\it Sparker} being the same as the mode
of the DMs of the Perytons. It is not unreasonable to conclude that
the {\it Sparker} is a Peryton that occurred in the first (or so)
Fresnel zone for the Parkes telescope. It is not a great leap to
conclude that FRBs are simply distant versions of the {\it Sparker}.
Of the extra-galactic models, we favor the model in which FRBs
result from giant flares from young magnetars.  The model can explain
the high daily rate of FRBs.

We end this section by noting that unlike in 2007 we now have the
{\it Sparker} and at least four FRBs. Given this situation, a reader,
at first blush, may wonder why is it important to discuss one
specific case (the {\it Sparker}) in some detail.  In our opinion,
when one is confronted by a new and astonishing phenomenon, it is
almost always useful to approach the observations with elementary
but robust analyses. In some cases it may well be that a simpler
explanation would suffice (e.g., the event of \citealt{ksk+12}, and
other claimed FRBs at low Galactic latitudes are arguably RRATs
hiding behind \ion{H}{2} regions).\footnote{Indeed, as we go to
press, a plausible intervening \ion{H}{2} region which can account
for the large inferred DM has been identified \citep{bm14} for the
FRB reported by \cite{ksk+12}.  In a similar vein, we note that the
same caution would apply to Arecibo events found close to the
Galactic equator.} Second, while we are not able to make concrete
progress (establish or reject an extra-galactic hypothesis) we are
open to the idea that astronomers at Parkes have indeed uncovered
a most fantastic phenomenon -- brilliant sparks at extra-galactic
distances.  Consistent with our (presently) agnostic view, we detail
in \cite{Zheng14} the potential use of FRBs to probe intergalactic
matter.

\begin{figure*}[ht]
\centerline{\includegraphics[width=17cm]{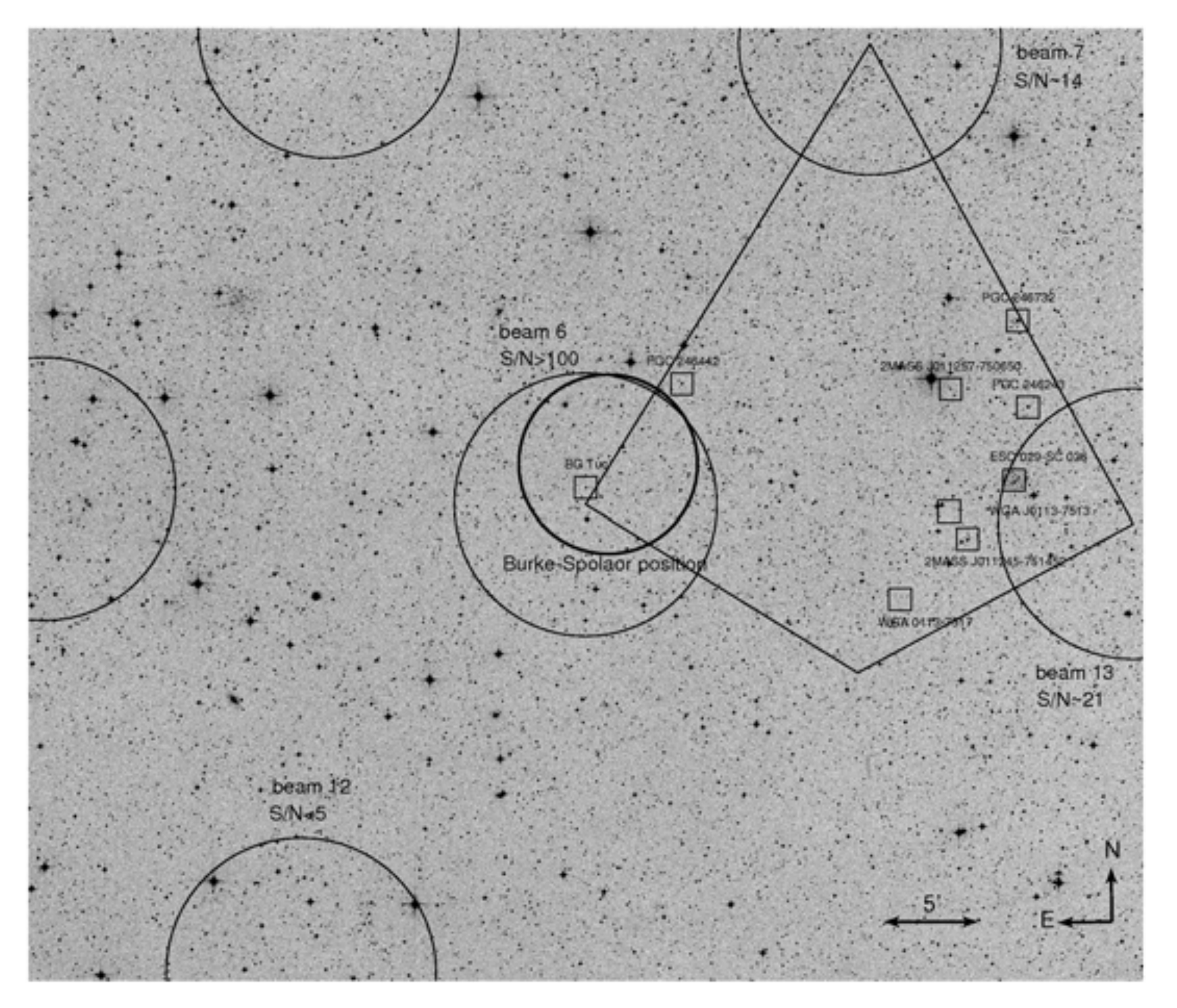}}
\caption{\small A localization of the {\it Sparker}.  The black
circles of radius 7.05 or 7.25\,arc-minute represent the Parkes
beams (see Table~\ref{tab:Beams} for further details).  The polygon
is  a conservative localization region for the {\it Sparker} 
(see \S\ref{sec:BetterLocalization} for details). 
The black smaller circle marked ``Burke-Spolaor position'' is the
best fit position for the {\it Sparker} obtained by \cite{bbe+11}.
 The background is from the Digitized
Sky Survey with the original data coming from the UK Schmidt Second
Epoch Survey (IIIaF+RG610). Objects noted in SIMBAD or NED are
noted.  The object marked as ESO 029-SC 036 is a cluster of galaxies.
\label{fig:SparkerLocation}} 
\end{figure*}

%S----------------------------------------------------------------------
\section{The Spark}
\label{sec:TheSpark}
%-----------------------------------------------------------------------

The event reported by \cite{lbm+07} was found in a re-analysis of
data obtained with the 13-beam system mounted at the  Cassegrain
focus of the Parkes 64-m radio telescope. The data from which this
pulse was discovered was originally obtained to look for pulsars
in the Small Magellanic Cloud (SMC).  For each of the two linear
polarizations, the signal from each of the thirteen beams (sky
frequency 1.28--1.52\,GHz; \citealt{swb+96}) was fed into a filter
bank with 96 channels, each 3\,MHz wide, and followed by square-law
detection. The detected signal from the two polarization signals
was summed, filtered, and digitally sampled at the rate of 1\,kHz
\citep{mlc+01}.  The authors searched the data stream for single
pulses in the range 1\,ms to 1\,s and DMs between 0 and 500\,cm$^{-3}$\,pc.

A single intense burst of short duration ($<5$~ms; epoch, UTC 2001
Aug 24, 19:50:01) with best fit DM of $375\pm1$\,cm$^{-3}$\,pc  was
identified. The burst was so intense that it saturated one of the
beams [signal-to-noise ratio (SNR) reported as $\gg$23 \citep{lbm+07},
for simplicity we adopt SNR$>$100] and was readily detected
in two adjacent beams (see \S\ref{sec:BetterLocalization} for further
analysis).  The peak flux was estimated to be $S\approx30\pm10$~Jy.
The high precision with which the DM was inferred means that the
data are consistent with a cold plasma dispersion model to equally
good precision.

Our attempts to better localize the position of the {\it Sparker}
by accounting for the measured signal levels in different beams
with the (far field) Parkes Multi-Beam response function (updated
by L.  Staveley-Smith following \citealt{swb+96}) did not converge
to a well defined region.  In order to make progress we adopt a
localization that makes use of approximate circular symmetry of
the beams.  This localization region is a polygon (aka ``kite'')
and is displayed graphically in Figure~\ref{fig:SparkerLocation}
and numerically in Table~\ref{tab:Polygon}
(\S\ref{sec:BetterLocalization}). When a single position is needed
(e.g., to compute foreground extinction) we use the position of the
beam in which the signal was saturated: RA=01h18m06.0 Dec=-75:12:19.
(J2000).

\subsection{Energetics}
\label{sec:Energetics}

We adopt the following values for the {\it Sparker} at the fiducial
frequency, $\nu_0=1.4\,$GHz:  peak flux density, $S_0=30\,$Jy,
measured pulse width of $\Delta\tau_0=5$\,ms and intrinsic pulse
width, $\Delta t=1\,$ms.  The broad band spectrum of the {\it
Sparker} appears not to be well determined (D. Lorimer, pers. comm.
and our own analysis).  The usable data for the {\it Sparker} are
from the unsaturated beams and since the response is a function of
frequency one can expect the data to suffer from chromatic effects.

In contrast to the {\it Sparker} the four events reported by
\cite{tsb+13} are found only in one beam. Thus the broad band
spectrum of these events can be expected to be less prone to chromatic
problems discussed above. The four events show reasonably good SNR
across the entire 1.28--1.52\,GHz range. This rough uniformity
suggests that a power law spectrum is adequate to describe the
broad-band spectrum and the power law index, $\alpha$ is not too
far from zero.  We  assume a power law model for the fluence spectrum,
$\mathcal{F}(\nu)\propto\nu^\alpha$. When a specific value is called
for, we use $\alpha = -1$.

The broadband fluence of the {\it Sparker} is
\begin{eqnarray}
	\mathcal{F}&\approx
	&-\frac{S_0\tau_0\nu_0}{\alpha+1}\Big(\frac{\nu_0}{\nu_l}\Big)^{\alpha+1},
	\ \ {\rm for\ }\alpha < -1, \\
	\label{eq:Fluence1}
	&=& S_0\nu_0\tau_0{\rm ln}\Big(\frac{\nu_u}{\nu_l}\Big)\ \ {\rm for\ }\alpha=-1;
	\label{eq:Fluence2}
\end{eqnarray} 
where $\tau_0$ is the optical depth at $\nu_0$, and $\nu_u$ and $\nu_l$ are the upper and lower cutoff frequencies
of the broad band spectrum and we assume that $\nu_0 \ll \nu_u$.
A conservative estimate of the bolometric fluence is obtained with
$\alpha=-1$ and setting the log factor to 10. With these two
assumptions we find $\mathcal{F}\approx 2.1\times 10^{-14}\,{\rm
erg\,cm^{-2}}$. Thus the isotropic radiated energy in the radio
band alone is
\begin{equation} 
	\mathcal{E}_R \approx 2.5\times 10^{30}D_{\rm kpc}^2 \,{\rm erg} 
	\label{eq:ER}
\end{equation}
where $D_{\rm kpc}$ is the distance to the
source.\footnotemark\footnotetext{We use the convention that a
particular quantity is normalized with the appropriate physical
unit displayed as a subscript in Roman font. Thus $D_{\rm kpc}$ is
the distance to the source in units of kpc.}

\subsection{Constraints from Brightness Temperature}
\label{sec:BrightnessTemperature}

The brightness temperature can be computed from the Rayleigh-Jeans
formula, $S_0 = 2kT_B(\nu_0^2/c^2)\pi(R/D)^2$, where $R$ is the
radius of the source.  Provided that there are no relativistic flows
$R<c\Delta t$, and the minimum brightness temperature is
\begin{equation}
	T_B(\nu_0) \lesssim 1.6\times 10^{24}D_{\rm kpc}^2\,
	\Delta t_{\rm ms}^{-2}\,{\rm K}.
	\label{eq:TB}
\end{equation}
The emission mechanism is either incoherent or coherent. We will
consider the first option. It is well known that brightness
temperatures in excess of $10^{12}\,$K lead to severe Compton losses
\citep{kp69}.  If, however, there is a relativistic outflow (with
a bulk Lorentz factor, $\Gamma$), then the observed duration is
compressed by the forward motion (towards us) and also the flux
enhanced (for the same reason).  As a result the inferred brightness
temperature (Equation~\ref{eq:TB}) is $\approx \Gamma^3$ larger
than that in the rest frame of emission (\citealt{PadmanabhanIII},
pp.  490).  Limiting the brightness temperature in the source frame
to $10^{12}$\,K then leads us to
\begin{equation}
	D_{\rm kpc}^2\lesssim \frac{\Gamma^3}{10^{12}}.  
	\label{eq:DkpcGamma}
\end{equation}	
Gamma-ray bursts (GRBs), cosmic explosions with the most relativistic
bulk outflows, have inferred values of $\Gamma$ as high as $10^3$.
In this scenario, $D \leq 100$\,pc.  Thus, the observed excess
dispersion measure must clearly arise from the source (or its
circumstellar medium).  We investigate this possibility in
\S\ref{sec:StellarCorona}.  Conversely, should we find incontrovertible
evidence that the {\it Sparker} is located at great distances (say,
even 1\,kpc), then the emission process must be coherent. In
principle, coherent processes can produce (almost) arbitrarily high
brightness temperature radiation -- provided that the emitting
region consists of highly relativistic matter.

%S----------------------------------------------------------------------
\section{The Distance to and the Size of the DM Nebula}
\label{sec:DistanceSize}
%-----------------------------------------------------------------------

The excess DM inferred towards the {\it Sparker} requires that there be an
intervening ionized nebula. Our goal in this section is to derive some
constraints on the size (or diameter), $L$, and the distance, $d$, to the
nebula.  We will assume the following values for Galactic ionized gas: DM
and emission measure (EM; see below) towards the Galactic pole of
25\,cm$^{-3}$\,pc and 2\,cm$^{-6}$\,pc (\citealt{Cox2000}; \S21.1).

For simplicity, we assume the nebula is a sphere of hydrogen with uniform
electron density, $n_e$. Such a nebula will manifest itself via H$\alpha$
recombination radiation and free-free absorption.  Separately, provided the
nebula is photo-ionized, the UV continuum from the ionizing source may be
detectable.  Our approach is to compute the expected signatures
(recombination radiation, UV continuum, free-free absorption) and use
existing observations to constrain the phase space of $d$ and $L$.  We will
accept the IGM solution only if our exploration rules out Galactic or
near-Galactic possibilities.

The primary parameter which determines the strength of the H$\alpha$
emission and free-free absorption is the emission measure.  The dispersion
measure of the intervening  nebula is ${\rm DM}=L_{\rm pc}n_e$ cm$^{-3}$ pc
and the  corresponding EM (in the usual units) is
\begin{equation}
	{\rm EM}={\rm DM}^2L^{-1}_{\rm pc}{\rm cm^{-6}\,pc},
	\label{eq:EMDM}
\end{equation}
where $L_{\rm pc}=L/(1\,{\rm pc})$.

\subsection{Free-free absorption}
\label{sec:EM}

The free-free optical depth is given by 
\begin{equation}
	\tau_{\rm ff}(\nu) = 4.4\times 10^{-7} {\rm EM}\,
	\bigg(\frac{T_e}{8,000\,{\rm K}}\bigg)^{-1.35}
		\bigg(\frac{\nu}{1\,{\rm GHz}}\bigg)^{-2.1} ,
	\label{eq:tauff}
\end{equation}
where $T_e$ is the electron temperature and $\nu$ is the frequency in GHz
(\citealt{L74}; p. 47).  The temperature normalization is appropriate for a
photo-ionized nebula.  Combining Equations~\ref{eq:EMDM} and \ref{eq:tauff}
we find
\begin{eqnarray}
	\tau_{\rm ff}(\nu) &=& 2.7
	\Big(\frac{\nu}{\nu_0}\Big)^{-2.1}
	\Big(\frac{T_e}{8,000\,K}\Big)^{-1.35}\Big(\frac{L}{0.01\,{\rm pc}}\Big)^{-1}
	\label{eq:Lpc}
\end{eqnarray}
where we have set DM=350\,cm$^{-3}$\,pc (accounting for a Galactic 
contribution of 25\,cm$^{-3}$\,pc).  From
this equation we immediately see that invoking very compact nebulae,
$L < 0.01$ pc, is problematic due to producing high optical depths.  The observed fluence spectrum,
$\mathcal{F}(\nu)$,  is related to the true fluence spectrum,
$F(\nu)$, as follows:
\begin{equation}
	\mathcal{F}(\nu) \propto F(\nu)
	\exp\Bigg[-\tau_0 \bigg(\frac{\nu}{\nu_0}\bigg)^{-2.1}\Bigg].
\end{equation}
As discussed earlier (\S\ref{sec:Energetics}) the fluence spectrum
is not well measured.  In the vicinity of frequency $\nu$, we can
make an expansion
\begin{eqnarray}
	\alpha^\prime &\equiv& \frac{d{\rm ln}\mathcal{F}(\nu)}{d{\rm ln}\nu}\Big|_{\nu} \cr
	&=&  \beta +2.1\tau_0\Big(\frac{\nu}{\nu_0}\Big)^{-2.1},
	\label{eq:beta}
\end{eqnarray}
where the intrinsic spectrum in the vicinity of $\nu_0$ is assumed
to be a power law, $F(\nu)\propto \nu^\beta$ with $\beta = -1$. Thus even a modest
amount of optical depth ($2 < \tau_0 < 4$) can result in extra-ordinarily
steep spectrum ($\alpha^\prime = 3.2 - 7.4 (\nu/\nu_0)^{-2.1}$) for the
underlying spectrum. There are two consequences.

First, an intrinsic spectrum as steep as the values discussed above
would be remarkable.  Millisecond pulsars possess the reputation
for having the ``steepest'' spectra.  Examples include PSR\,1937+21
($\alpha=-2.7$; \citealt{bkh+82,E83}) and PSR\,1957+20 ($\alpha=-3$;
\citealt{fbb+90}).  A perusal of the literature shows two sources
which are even steeper: the Sun \citep{gcs+06} and GCRT\,J1745$-$3009
\citep{hrp+07}.  For the latter source, a weak burst was found to
have a spectral index of $\alpha=-13.5\pm 3$. This was measured
over a limited frequency range from 310 to 338~MHz.  The Sun is
much better studied.  For some spiky bursts from the Sun the spectrum
is exponential, consistent with the spectrum emitted by a mono-energetic
electron gyrating in an uniform field (see \S\ref{sec:MonoEnergetic}).
An exponential spectrum can give an arbitrarily high power law index
for frequency (see \S\ref{sec:MonoEnergetic}).

A steep intrinsic spectral index would thus favor (based on the
fact that all the steepest sources are fit with exponentials)
an exponential distribution for $F(\nu) \propto
\exp(-\nu/\nu_c)$, where $\nu_c$ is the characteristic frequency.  In 
this case, Equation~\ref{eq:beta} becomes
	\begin{equation}
	\alpha^\prime = -\frac{\nu}{\nu_c} + 2.1\tau_0\Big(\frac{\nu}{\nu_0}\Big)^{-2.1}.
	\label{eq:nuctau0}
	\end{equation}
It is clear from this equation that a large value of $\tau_0$ ($\gg 1$)
would result in $\alpha^\prime$ varying rapidly even over the 1.2--1.5\,GHz
band of the Parkes pulsar spectrometer.  We see no evidence for such strong
spectral curvature for either the {\it Sparker} or the
FRBs.\footnote{Parenthetically we note that should a pulse be found with
positive and large observed spectral index (that is $\alpha \gg 1$; but not
so high that the pulse is entirely absorbed) then a plausible explanation
is that $\tau_0\gtrsim 1$.}

Next, the bolometric fluence in the exponential model is  
\begin{equation}
	\mathcal{F}=S_0\tau_0\nu_c \exp(\nu_0/\nu_c+\tau_0).
	\label{eq:Bolometric}
\end{equation}
Let us consider the implication of invoking significant free-free
absorption.  For instance, setting  $\tau_0=5$ in
Equation~\ref{eq:nuctau0}, we find $\nu_c\approx \nu_0/12$.  Propagating
this choice of $\nu_c$ to Equation~\ref{eq:Bolometric} results in an
isotropic emission of nearly $5\times 10^{35}\,D_{\rm kpc}^2\,$ erg. Even
at 100\,pc the inferred energy loss in the radio band would severely
challenge what is observed from the most active stars, whose radio
emission is typically measured in hundreds of milli-Janskys \citep{gu02}.

%NOTE TO SRK: 
%Give an example of an active star. How much radio energy is typically
%%given out etc.

Continuing with this theme and setting $\tau_0<5$ we find, from
Equation~\ref{eq:Lpc}, 
\begin{equation}
	L >L_{\rm ff}= 5.6\times 10^{-3}\,{\rm pc}.
	\label{eq:Lmin}
\end{equation}
Note that this severe constraint on $L$ has no dependence on $d$.
It also has no dependence on the angular size of the nebula since
by assumption the angular size of the {\it Sparker} is assumed to
be smaller than that of the putative nebula.  The nebula size,
$L_{\rm ff}$ (Equation~\ref{eq:Lmin}), corresponds to $2\times
10^5\,R_\odot$. 

From Equation~\ref{eq:Lmin} we conclude that  {\it Sparker} cannot
arise from a terrestrial phenomenon. The reader may find it instructive
to read \S\ref{sec:StarsSupernovae} to appreciate typical DM and
EM in any reasonable  stellar context (including compact binaries
with a main sequence companion and so on).  \cite{Luan14} arrived
at the same conclusion independently.  The only way to avoid a
stellar model for FRBs is to invoke high temperatures  -- a possibility
that we treat in depth in \S\ref{sec:StellarCorona}.

\subsection{Dispersion Measure: Galactic Contribution}
\label{sec:GalacticDM}

\begin{figure}
\centerline{\includegraphics[width=8.5cm]{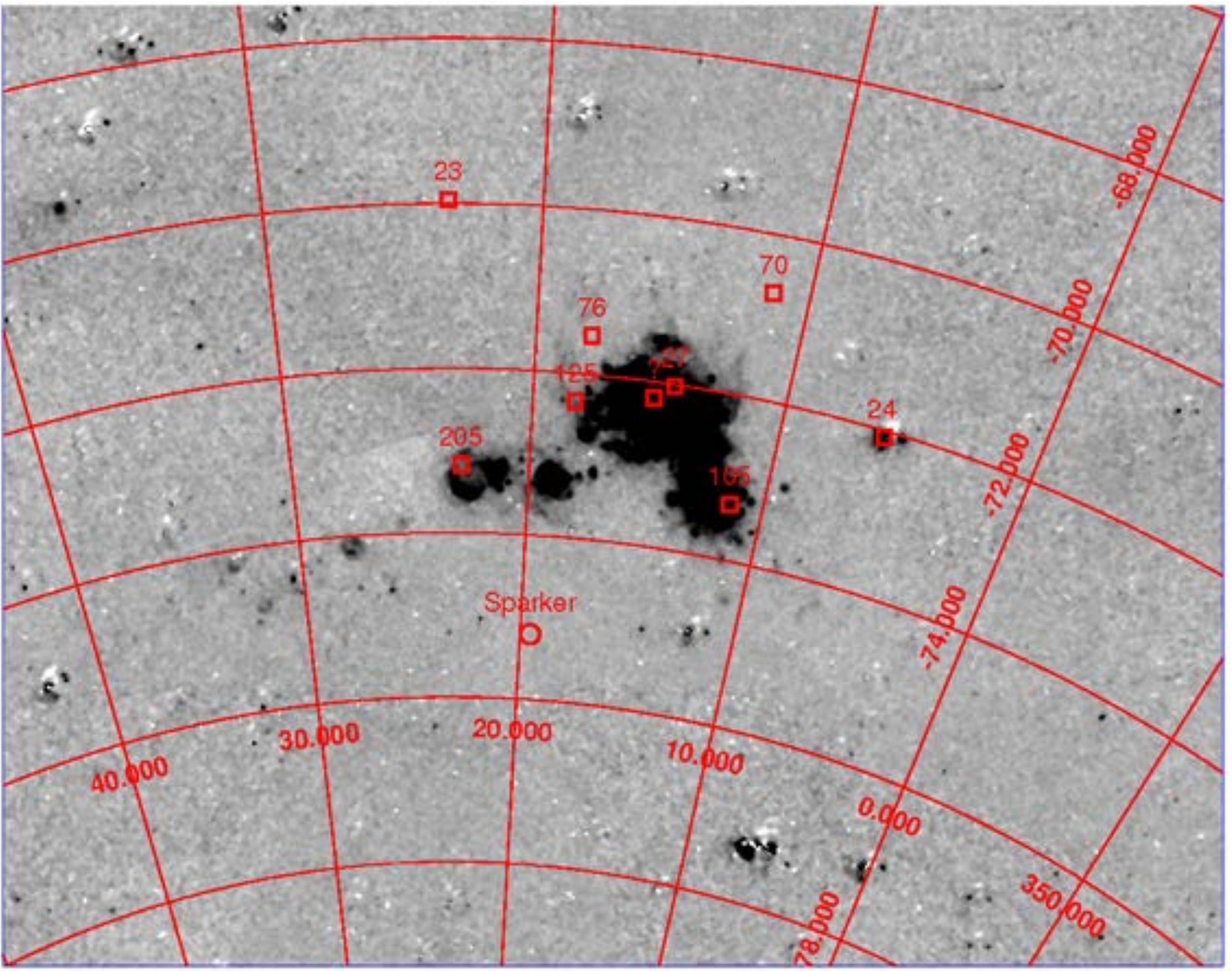}}
\caption{\small
Tangential projection of the known pulsars (marked by squares; the
number next to each square is the dispersion measure of the pulsar)
in the vicinity of the SMC with North up and East to the left.  The
{\it Sparker} is marked by a circle. The background is the diffuse
H$\alpha$ emission obtained from the Southern H$\alpha$ Sky Survey
\citep{gmr+01}.  The grid marks the right ascension (RA) and the
declination (Dec), both in units of degrees. The South Celestial
Pole is towards the bottom of the Figure.  The pulsar-rich globular
cluster, 47~Tucanae, is located at RA $\approx6^\circ$ and Dec
$\approx-72^\circ$ (square box; the mean dispersion measure of the
pulsars in this cluster is 24\,cm$^{-3}$\,pc.).  The Large Magellanic
Cloud (not marked) is located to the North and East of the SMC and
lies outside this map.  
\label{fig:SHASSLargeFOV}} 
\end{figure}

\begin{figure}
\centerline{\includegraphics[width=8.5cm]{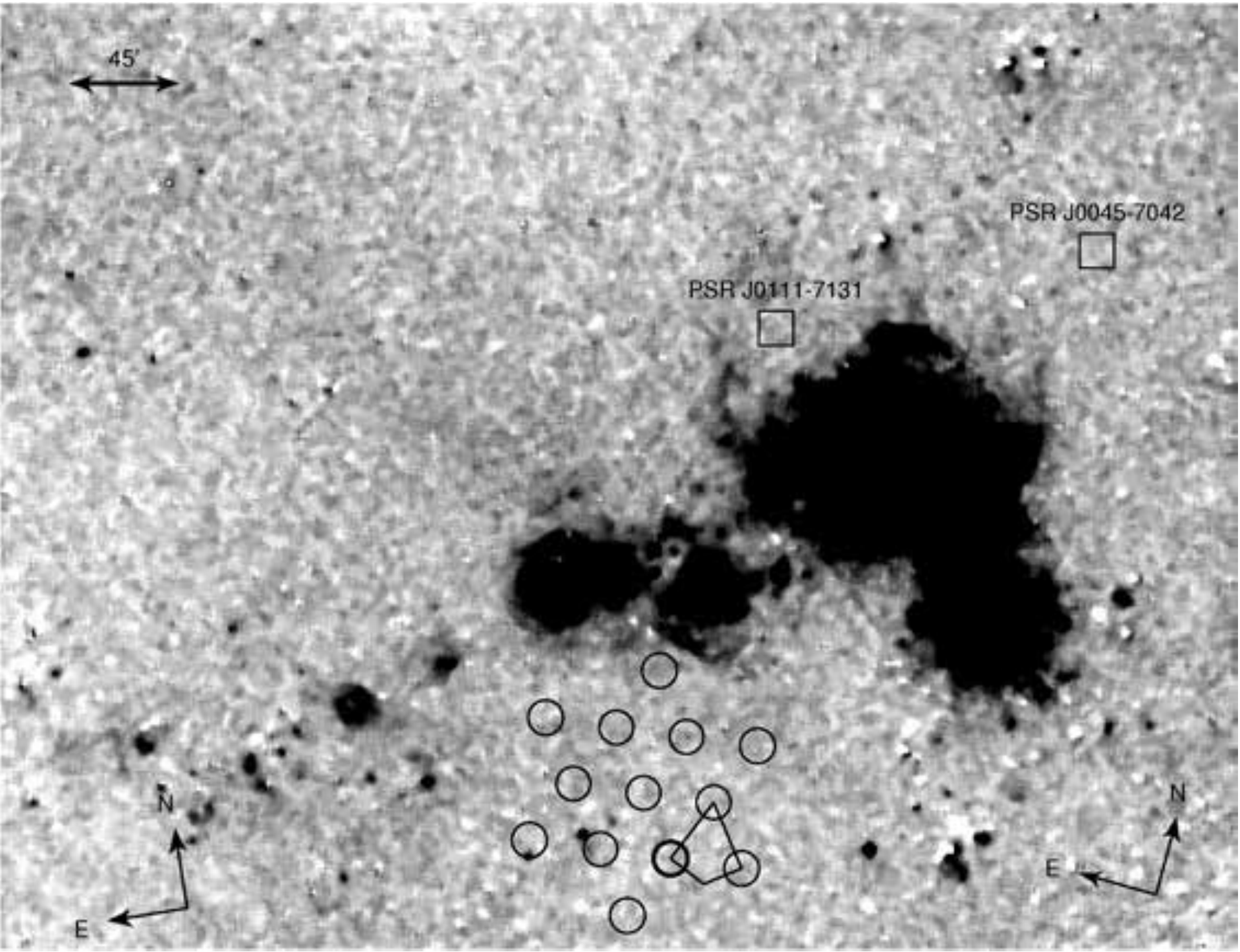}}
\caption{\small 
A zoom-in of the Southern Sky H$\alpha$ Survey containing
the localization of the {\it Sparker}. The beams listed in
Table~\ref{tab:Beams} are shown as circles with radius of
about 7\,arc\,minutes (the main beam size for each of the
beams). The beam in which the {\it Sparker} saturated is shown
with a thicker line. The polygon described in
Table~\ref{tab:Polygon} is also displayed.  Also marked are
two SMC pulsars (open squares and marked by their
names).  Notice the absence of detectable H$\alpha$ emission
towards the {\it Sparker} and the two SMC pulsars. The faint
emission towards the North-East is the Magellanic Stream.}
\label{fig:SHASS}
\end{figure}

In Figure~\ref{fig:SHASSLargeFOV} we present a wide-field overview
of the region of the {\it Sparker}.  The {\it Sparker} is about
three degrees South of the center of the SMC (see
Figure~\ref{fig:SHASSLargeFOV}).  The projected transverse distance
is 3.1\,kpc, assuming a distance of 60\,kpc to the SMC \citep{scg+04}.
In Figure~\ref{fig:SHASS} we present a zoom-in of the field centered
around the SMC.  As noted by \cite{lbm+07} the {\it Sparker} lies
outside the bright \ion{H}{1} and the bright H$\alpha$ boundaries
of the SMC.

%NOTE TO SRK: Check if any new pulsars have been found in this region

In the region of the sky containing the {\it Sparker} and the SMC
there are six
pulsars\footnotemark\footnotetext{\texttt{http://www.atnf.csiro.au/research/pulsar/psrcat}}
\citep{mht+05} and one magnetar \citep{mgr+05}. One pulsar
(PSR\,J0057$-$7201) has a DM of 27\,cm$^{-3}$\,pc \citep{ckm+01}
-- consistent with being a Galactic pulsar located above the Warm
Ionized Medium (WIM) layer.  The DMs of the remaining five pulsars
range from 70 to 205\,cm$^{-3}$\,pc \citep{mmh+91,ckm+01,mfl+06}.
These five pulsars are generally thought to be associated with the
SMC.  As a matter of reference, the pulsars in the Large Magellanic
Cloud (LMC) have an excess (over the Galactic value) of about
100\,cm$^{-3}$\,pc.  Five degrees away and located at a distance
of 4\,kpc (in the inner halo of our Galaxy), the globular cluster
47~Tucanae hosts a hive of pulsars with typical DMs of about
24\,cm$^{-3}$\,pc.

Thus, the first conclusion is that the Galactic contribution to the
observed DM of the {\it Sparker} (assuming, say, a halo location)
is no more than 25\,cm$^{-3}$\,pc.  As can be gathered from discussion
at the beginning of this section, the Galactic contribution to the
EM is negligible.

Next, the angular size of the DM-inducing nebula for the {\it
Sparker} cannot be larger than, say, $\theta_{\rm DM} \sim 5^\circ$.
Otherwise, we would expect a larger DM for the pulsars in the
neighborhood.  This conclusion is true whether the {\it Sparker}
is located in the halo or the SMC.  Thus we obtain our first
constraint:
\begin{equation}
	L < d\theta_{\rm DM}, \label{eq:thetaDM}
\end{equation}
provided that $D<60\,$kpc (the distance to the SMC).

\bigskip

%PSR J0057-7201 00:57:44.0 -72:01:19 DM=27 RM=* (P=0.93 s)
%PSR J0111-7131 01:11:28.8 -71:31:47 DM=76 RM=*
%PSR J0045-7042 00:45:25.7 -70:42:07 DM=70 RM=*
%Magnetar       01:00:43.0 -72:11:34 DM=*  RM=*

\subsection{Dispersion Measure: SMC Contribution}
\label{sec:SMCDM}

\begin{figure}
\centerline{\includegraphics[width=8.5cm]{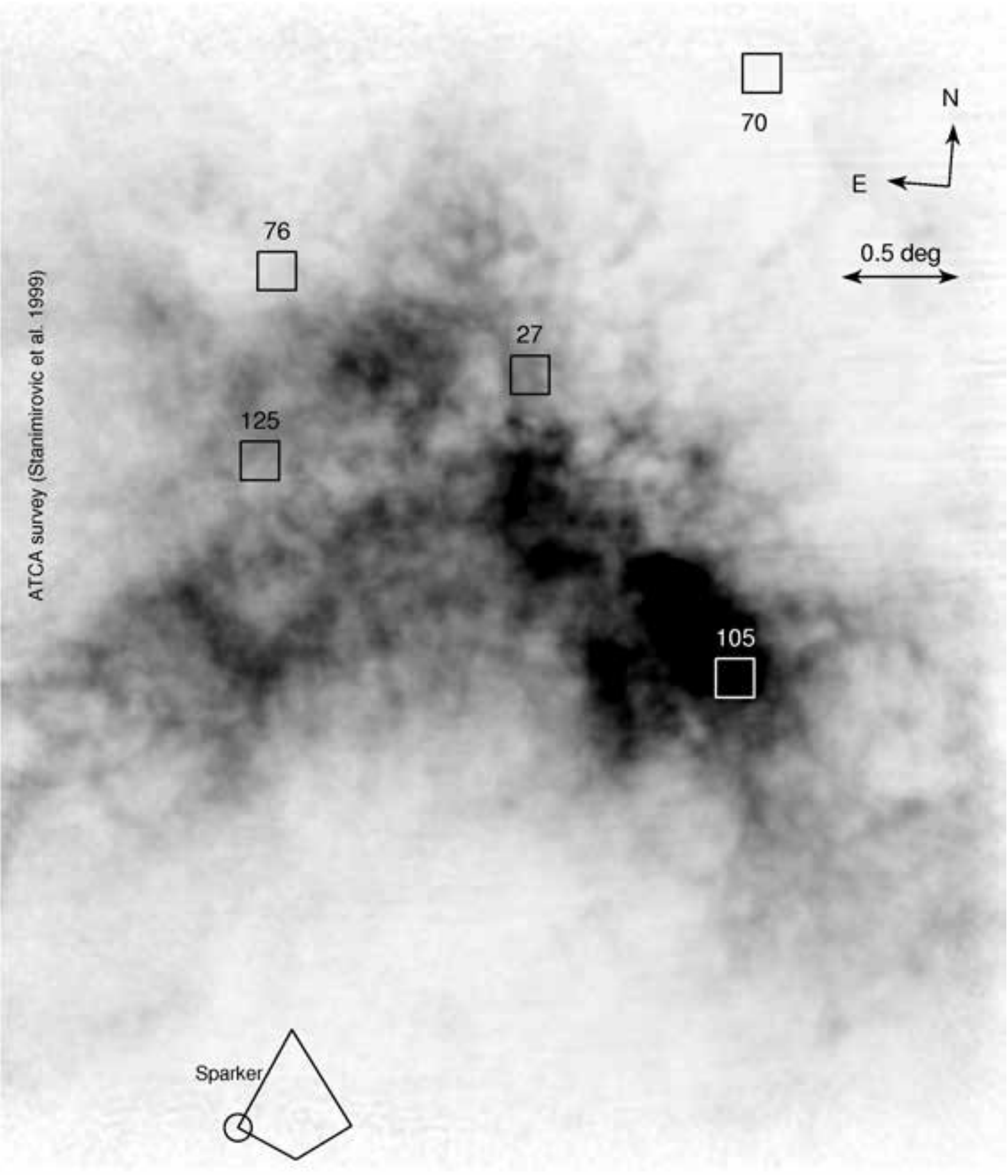}}
\caption{\small
The neutral hydrogen (\ion{H}{1}) column density integrated over
the heliocentric  velocity range 88.5--215.5~km\,s$^{-1}$ in the
direction towards the {\it Sparker}.  The data were obtained with
the ATCA (beam FWHM of 98 arcsec) supplemented by the Parkes 64-m
single dish images.  The polygon at the bottom of the image represents
the {\it Sparker} localization (\S\ref{sec:BetterLocalization}),
while the circle shows the position and FWHM size of the beam in
which the {\it Sparker} saturated the detector (Table~\ref{tab:Beams}).
The gray-scale intensity range is $-3\times 10^{19}$ to $8.4\times
10^{21}\,$ atom~cm$^{-2}$.  Boxes (white and black) show the positions
of known SMC and Galactic pulsars while the number accompanying
each box shows the pulsar measured DM.  At the position of the {\it
Sparker}, the \ion{H}{1} column density is $3.5\times 10^{20}\,$cm$^{-2}$.
From \citet{ssd99}.} 
\label{fig:HI} 
\end{figure}

With respect to Figure~\ref{fig:SHASS} and noting the DMs of
PSR\,J0045$-$7042 (70\,cm$^{-3}$\,pc) and PSR\,J0111$-$7131
(76\,cm$^{-3}$\,pc), we suggest that the SMC has an extended (diameter
of 4 degrees) ionized halo which contributes about 50\,cm$^{-3}$\,pc
(which when added to the Galactic DM yields a total of 75\,cm$^{-3}$\,pc).
Assuming a spherical geometry and a diameter of 4\,kpc, this extended
diffuse halo of the SMC has a mean electron density of 0.0125\,cm$^{-3}$
and the corresponding EM contribution is 0.63\,cm$^{-6}$\,pc.
Incidentally, we note that the \ion{H}{1} column density towards
PSR\,J0045$-$7042 (DM = 70\,cm$^{-3}$\,pc) is $2.1\times
10^{20}\,$cm$^{-2}$ (Figure~\ref{fig:HI}) and is comparable to the
column density arising from the ionized SMC halo.

We conclude that the Galactic+SMC DM contribution is about
75\,cm$^{-3}$\,pc.  Thus, were the {\it Sparker} to be located in
or behind the SMC, the excess DM is 300\,cm$^{-3}$\,pc and
correspondingly the emission measure is\footnote{We ignore the
contributions to the EM from our own Galaxy and the SMC.},
\begin{equation}
	{\rm EM}^S = 9\times 10^4 L_{\rm pc}^{-1} {\rm\ cm^{-6}\,pc}.
	\label{eq:EMSMC}
\end{equation}
The superscript ``$S$'' (``$G$") is used to indicate the expected
EM assuming an origin for the {\it Sparker} at the distance of the
SMC or beyond (or in the Galaxy). For the Galactic case, the excess
DM corresponds to 350--375\,cm$^{-3}$\,pc.

\subsection{Recombination Radiation: H$\alpha$}
\label{sec:Halpha}

An ionized nebula emits recombination radiation (e.g., the Balmer
series).  The H$\alpha$ brightness is determined by the recombination
rate and the fraction of recombinations that result in H$\alpha$
emission (see \citealt{Reynolds84}).  For $T\sim 8,000\,$K and
assuming case~B\footnote{ See Equation~9 of \citealt{vg98}. We adopt
the following recombination coefficient: $\alpha_B =
8.7\times10^{-14}T_4^{-0.89}$\,cm$^{3}$\,s$^{-1}$, where $T_{4}$
is the temperature in units of $10^4$\,K \citep{of06}.\label{fn:CaseB}}
we find $I=1.09\times 10^{-7}\times {\rm EM} {\rm \ erg\ cm^{-2}\
s^{-1}\ sr^{-1}}$,  whence the usual statement that
1\,Rayleigh\footnotemark\footnotetext{A unit of surface brightness
commonly used in aeronomy. One Rayleigh is $10^6/(4\pi)$ photons
per square centimeter per steradian per second.  For the H$\alpha$
line, the intensity in cgs units is $2.41\times 10^{-7}\,{\rm erg\
cm^{-2}\ s^{-1}\ sr^{-1}}$.} of photon intensity corresponds to an
EM of 2.2~cm$^{-6}$~pc.

The Southern H$\alpha$ Sky Survey (SHASS; see Figure~\ref{fig:SHASS})
imaged the entire Southern Sky in a narrow band ($\Delta\lambda_{\rm
SHASS}=16\,$\AA; this corresponds to a velocity width of $\pm
365\,$km\,s$^{-1}$ ) centered on the rest wavelength of H$\alpha$
(6563\,\AA)  and at an angular resolution  of $0.8^{\prime}$
\citep{gmr+01}.  The rms per pixel is about 2\,Rayleigh.  When
dealing with surface brightness it helps to divide the discussion
into two parts: objects with a size bigger (``resolved'') or smaller
(``compact'') than the angular size of the beam(s) of the survey(s).

We first consider the resolved case.  We determined that the SHASS
5-$\sigma$ detection limit for a 1-degree diameter nebula is about
0.5\,Rayleigh.  The upper limit at a scale of one arc\,minute is
naturally larger and was found to be 6\,Rayleigh. We thus find
\begin{eqnarray}
	\label{eq:LSurfaceBrightness}
	\frac{\rm EM}{2.2\ {\rm cm^{-6}\,pc}} &=& 0.45{\rm DM}^2 L_{\rm pc}^{-1}
	\ltorder R_{\rm SHASS},
\end{eqnarray}
where $R_{\rm SHASS}$ is the surface brightness (on the relevant angular
scale). Using Equation~\ref{eq:EMSMC} we  obtain our second constraint:
\begin{eqnarray}
	L &\gtorder& 82\ {\rm kpc}\   {\rm for\ } \theta\sim 1^\circ,\cr
	L &\gtorder& 6.8\ {\rm kpc}\  {\rm for\ }\theta\sim 1^\prime.
	\label{eq:LSurface}
\end{eqnarray}
Here $\theta=L/d$ is the angular diameter of the nebula.  Note that the
size constraint is independent of $d$, provided that the nebula has an
angular size $\gtorder 1^\circ$ or $\gtorder 1^\prime$, respectively.

Next consider the case of a nebula whose angular size is smaller than that
of the resolution of the SHASS.  The expected H$\alpha$ flux is
\begin{equation}
	F_{\rm H\alpha} = h\nu_\alpha R_{\rm SHASS}
				\frac{\pi\theta^2}{4},
\end{equation}
where $h\nu_\alpha$ is the energy of an H$\alpha$ photon.  Combining this
equation with Equation~\ref{eq:EMSMC} (and likewise for a Galactic
location), we find
\begin{eqnarray}
	\label{eq:FHalphaG}
	F^G_{\rm H\alpha} &=& 
	10.5\times 10^{-9}L_{\rm pc}d_{\rm kpc}^{-2}\, {\rm erg\,cm^{-2}\,s^{-1}},\\
	F^S_{\rm H\alpha} &=& 
	7.8\times 10^{-9}L_{\rm pc}d_{\rm kpc}^{-2}\, {\rm erg\,cm^{-2}\,s^{-1}}.
	\label{eq:FHalphaSMC}
\end{eqnarray}

The point source limit for a single pixel of SHASS is $R_{\rm
SHASS}\Delta\Omega$ where $\Delta\Omega \sim 5.4\times 10^{-8}\,$steradian
(corresponding to one SHASS pixel). Given that $R_{\rm SHASS}=6\,$Rayleigh
(see above) we find
\begin{equation}
	F_{\rm H\alpha} \ltorder  7.8\times 10^{-14}
		\,{\rm erg\,\,cm}^{-2}{\rm s}^{-1}.
	\label{eq:SHASSLimit}
\end{equation}
Combining the inequality (Equation~\ref{eq:SHASSLimit}) with
Equations~\ref{eq:FHalphaG} and \ref{eq:FHalphaSMC} we derive the following
constraints:
\begin{eqnarray}
	\label{eq:LSHASSG}
        L_{\rm pc}d_{\rm kpc}^{-2} &\ltorder& 0.75\times 10^{-5} \qquad {\rm (Galactic)} \\
        L_{\rm pc}d_{\rm kpc}^{-2} &\ltorder& 1\times 10^{-5} \qquad {\rm (SMC\ and\ Beyond)}
                \label{eq:LSHASSSMC}
\end{eqnarray}
Combining Equation~\ref{eq:LSHASSG} with Equation~\ref{eq:Lmin} we
find the nebula cannot be located any closer than
\begin{equation}
	d_{\rm min}(H\alpha) \sim 27\,{\rm kpc}.
	\label{eq:dmin}
\end{equation}
Parenthetically, we note that, {\it in principle}, a similar exercise can
be carried out for the two-photon emission, traced by UV observations.

\begin{figure*}
\centerline{\includegraphics[width=17cm]{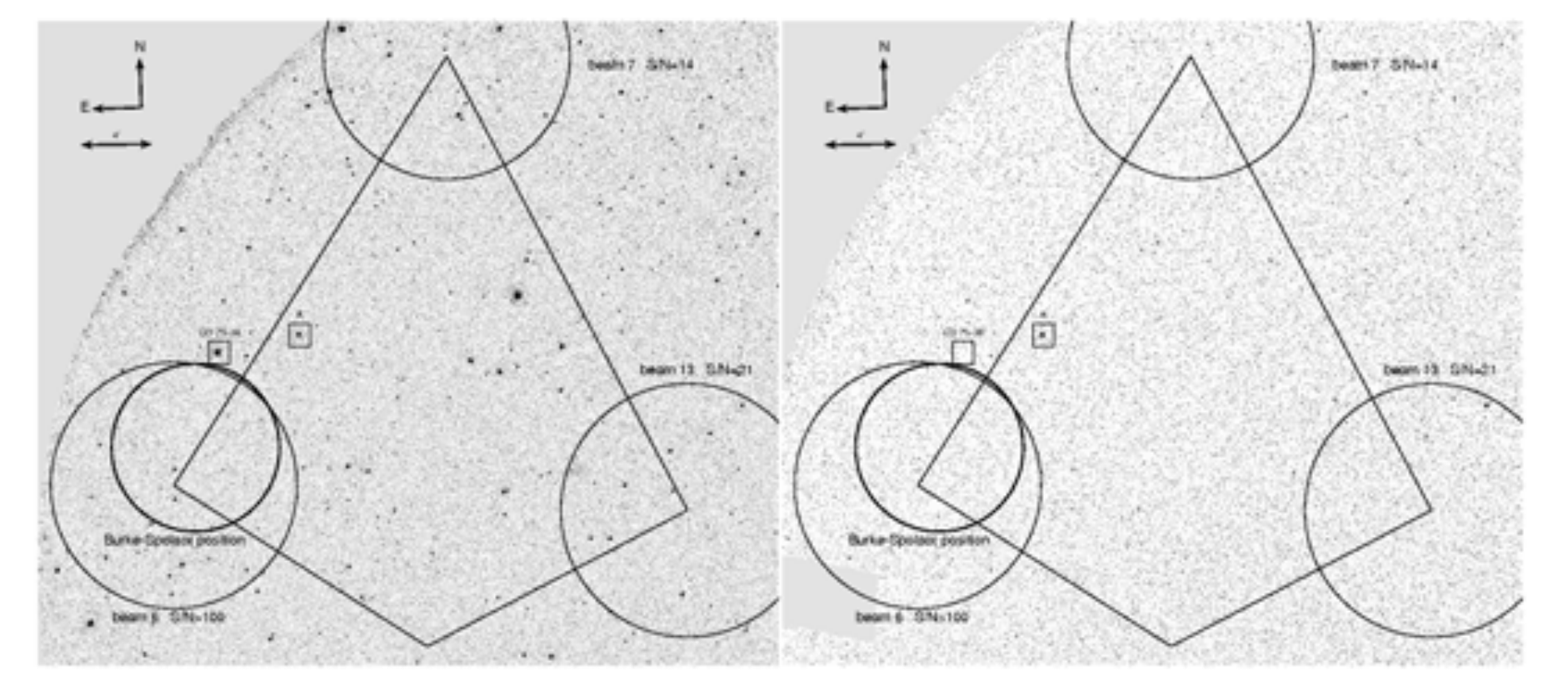}}
\caption{\small
{\it GALEX} NUV (left) and FUV (right) images of the field of the
{\it Sparker}.  The signal-to-noise ratio $=3$ limit corresponds
to $FUV=20.5$\,mag and $NUV=21.4\,$mag.  Here, following standard
convention, the {\it GALEX} magnitudes are defined in the AB system
\citep{Oke74}.  CD-75~38 is a useful comparison star with $V=10.35$,
$B=10.98$, $NUV=15.310\pm 0.001$ and $FUV=22.23\pm 0.08$.  Star A
is the hottest star within the polygon.  See \S\ref{sec:UVContinuum}
for more discussion.  
\label{fig:GALEXpolygon}} 
\end{figure*}

\begin{table}[bth]
\caption[]{\small
The physical parameters of the allowed ionized nebula}
\begin{center}
\begin{tabular}{lllll}
\hline
\hline
$d$     & $L$                              & $\theta$ & $\langle{n_e}\rangle$ & $\dot{N}_R$ \cr
(kpc)	  & (pc)	                          & (arcsec) & cm$^{-3}$            & s$^{-1}$ \cr
\hline
27      &   $5.6\times 10^{-3}$    & 0.04       & $5.4\times 10^4$   & $2.0\times 10^{45}$ \cr
60      &   $2.7\times 10^{-2}$    & 0.09       & $1.1\times 10^4$   & $9.8\times 10^{45}$ \cr
303    &   0.69                            & 0.47       & $4.35\times 10^2$ & $2.5\times 10^{47}$\cr
1000  &   7.5                              & 1.55       & 40                          & $2.7\times 10^{48}$ \cr
\hline
\end{tabular}
\label{tab:Nebulae}
\end{center}
Notes:  $d$ is the distance to the ionized spherical nebula, $L$
is the maximum permitted diameter at that distance, $\theta=L/d$
is the corresponding angular diameter, $\langle{n_e}\rangle=DM/L$
is the corresponding mean electron density in the nebula, and
$\dot{N}_R$ is corresponding inferred rate of recombinations (see
\S\ref{sec:Halpha}).  For the first entry we used the DM appropriate
for a Galactic location, DM = 350\,cm$^{-3}$\,pc. For the remaining
we used DM = 300\,cm$^{-3}$\,pc.  For all distances, $d>27\,$kpc,
the size of nebula is constrained to lie between $L_{ff}$
(Equation~\ref{eq:Lmin}) and the values indicated in the second
column.  
\end{table}

\subsection{UV Continuum}
\label{sec:UVContinuum}

A nebula ionized by one or more hot stars would be accompanied by a strong
underlying stellar continuum.  Here we explore archival data to see if
suitable ionizing stars exist within the {\it Sparker} region.  We then
match the rate of recombination for the nebula (which is a function of $L$
and thence of $d$), $\dot N_R$, to the rate of ionization by possible
ionizing sources, $\dot N_I$. The strongest plausible ionizing source or
the most distant ionizing source then provides the largest $d$.  It is
important to understand that this exercise will only constrain ionizing
sources within (at best) the Local Group. The exercise does not constrain
very distant ionizing sources (e.g., the IGM)

The rate of recombinations is $\dot{N}_R= \pi/6 n_e^2\alpha_BL^3$ where
$\alpha_B$ is the recombination coefficient (see footnote~\ref{fn:CaseB}).
Consistent with the spirit of this section (namely, a photo-ionization
model), we assume $T=8,000\,$K.  For a given distance, $d$, we derive a
maximum nebular diameter, $L$, using the constraints provided in the
previous section and thence $\dot N_R$.  The calculations are summarized in
Table~\ref{tab:Nebulae}.  We find that $\dot{N}_R$ is as small as $2\times
10^{45}\,{\rm s}^{-1}$ and as large as $2.7\times 10^{48}\,{\rm s}^{-1}$
(at 1\,Mpc).  The corresponding minimum luminosity in ionizing photons
(assuming photons with energy just above the Lyman continuum) is
$\dot{N}_Rh\nu_1$ where $h\nu_1$ is the energy of a photon at
the Lyman edge. This luminosity ranges between $4.4\times
10^{34}\,$erg\,s$^{-1}$ and $5.9\times 10^{37}\,$erg\,s$^{-1}$.  The
inferred ionizing rates should be compared with that expected from O and B
stars \citep{sk97}: $\log(\dot{N}_I)=49.85$ for an O3V star [$T_{\rm
eff}=51230\,$K and $\log(L_{\rm bol}/L_\odot)=6.0$] and
$\log(\dot{N}_I)=47.77$ for a B0.5V star [$T_{\rm eff}=32060\,$K and
$\log(L_{\rm bol}/L_\odot)=4.7$].

The {\it Galaxy Evolution Explorer} ({\it GALEX}) UV space telescope
\citep{mfs+05} is well suited to search for and characterize potential hot
(and thus ionizing) sources.  The {\it GALEX} images are shown in
Fig.~\ref{fig:GALEXpolygon}, and have exposure times of 135\,s in both the
far-UV (FUV) channel (center wavelength, 1538\,\AA; FWHM = 226\,\AA) and
the near-UV (NUV) channel (2289\,\AA; FWHM = 794\,\AA).

\begin{figure}
\centerline{\includegraphics[angle=90.,scale=0.35]{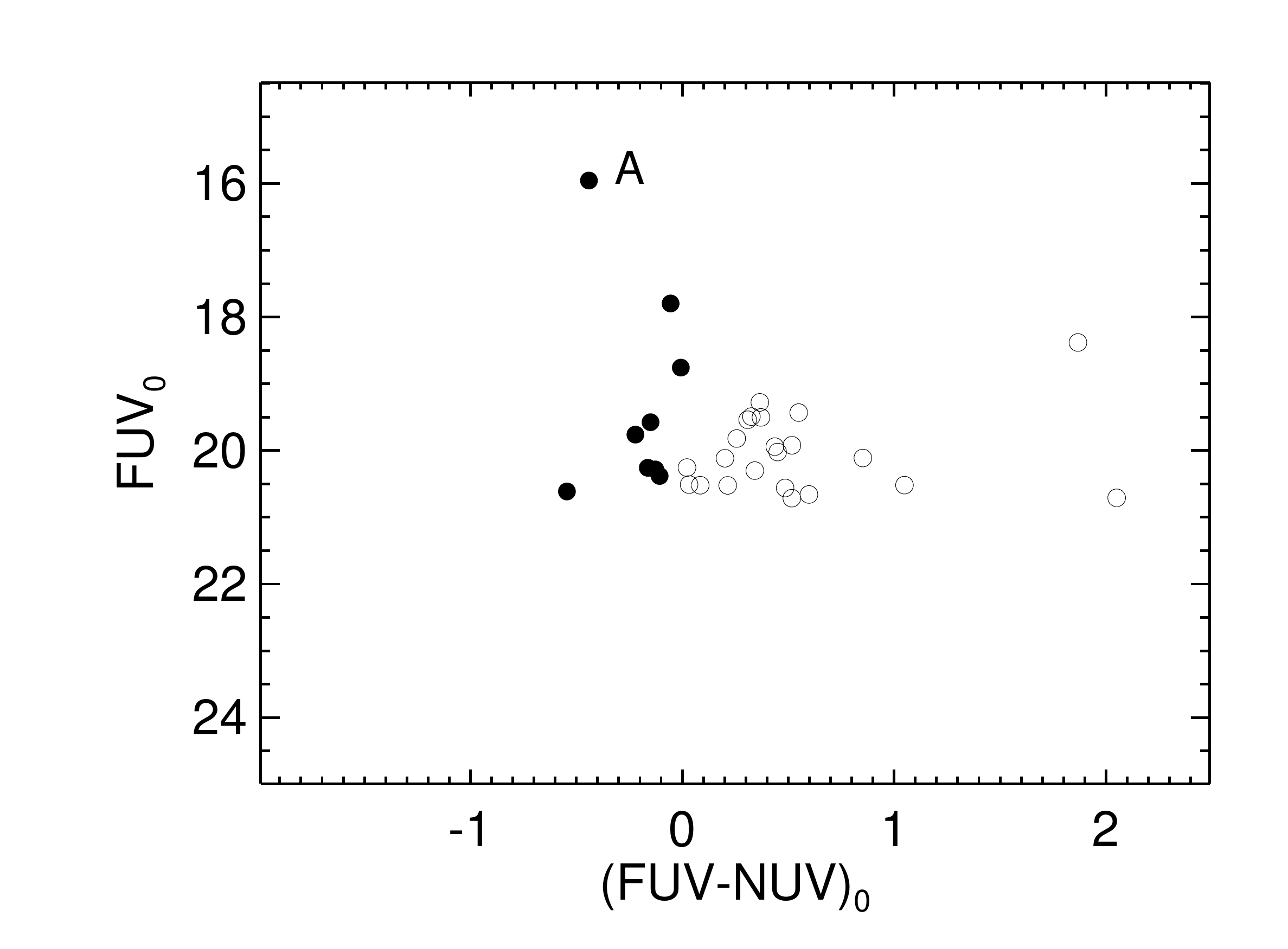}}
\caption{\small
{\it GALEX} Color-Magnitude diagram (CMD) around the {\it Sparker}
region.  The colors and magnitudes were derived from the {\it GALEX}
All-sky Imaging Survey image of this field.  The magnitudes are
extinction corrected [$E(B-V) = 0.054$ giving $A_{FUV} = 0.447$ mag
and $A_{NUV} = 0.445$ mag].  The CMD is restricted to well detected
stars (SNR $>$3, in both bands) within the polygon (see
Figure~\ref{fig:SparkerLocation}).  Stars with $\Delta{\rm UV}<0$
are marked by filled circles.  The {\it GALEX} data were taken from
the 7th data release.  
\label{fig:GalexCMD}} 
\end{figure}

Only hot stars with $T_{\rm eff} \gtorder 2\times 10^4\,$K are capable of
ionizing hydrogen atoms.  The extinction-corrected {\it GALEX}
Color-Magnitude diagram (CMD; assuming all sources are outside the Galaxy)
for detected sources in the {\it Sparker} region is shown in
Figure~\ref{fig:GalexCMD}.  The extinction was corrected using the Galactic
color excess in the direction of the {\it Sparker} \citep{sfd98}, a
total-to-selective extinction ratio derived from the \citet{ccm89}
extinction law (see \S2.3 of \citealt{wms+07}), and assuming $R_{V}=3.08$.
As can be seen from \citet[Figure~7]{brv+07}, the {\it GALEX} color,
$\Delta_{UV}\equiv FUV-NUV$, of stars with $T_{\rm eff} \gtorder 2\times
10^4\,$K is $\Delta_{UV} < 0$.

Although some objects in the polygon region (Figure~\ref{fig:GalexCMD})
have $\Delta_{UV}<0$, we argue that these stars are too faint to be the
ionizing sources responsible for the DM nebula.  The brightest object
(labeled star A in Figures~\ref{fig:GALEXpolygon} and \ref{fig:GalexCMD})
with $\Delta_{UV}<0$ within the polygon region has FUV AB magnitude of 16.
From $\Delta_{UV}$ we compute an effective black-body temperature of
44,000\,K.  There are two possibilities: star A is either a foreground
white dwarf or a  main sequence B  star located at 530\,kpc.  We note that
the number of ionizing photons emitted by a 44,000\,K WD is smaller by a
factor of about $10^6$, relative to a main sequence star with the same
effective temperature. Thus if star A is a foreground white dwarf then its
ionizing capacity is negligible.  Should star A be a young star in the
outskirts of our Local Group (530\,kpc) then it has the ability to power a
nebula with $L=0.69\,$pc (see Table~\ref{tab:Nebulae}) and this nebula
could account for the excess DM. However, the {\it Sparker} will either
have to arise in this nebula or, if behind, be closely aligned with this
star (recall that the angular size of the nebula is 0.4$^{\prime\prime}$;
see Table~\ref{tab:Nebulae}).

Using the same arguments as above, we conclude that all the other
blue objects with fainter FUV magnitudes are not consistent with
being hot main sequence stars in the SMC or closer according to the
models in \citet{brv+07}.  These objects are more likely to be
foreground white dwarfs or background unresolved star-forming
galaxies.

\begin{figure*}
\centerline{\includegraphics[width=7in]{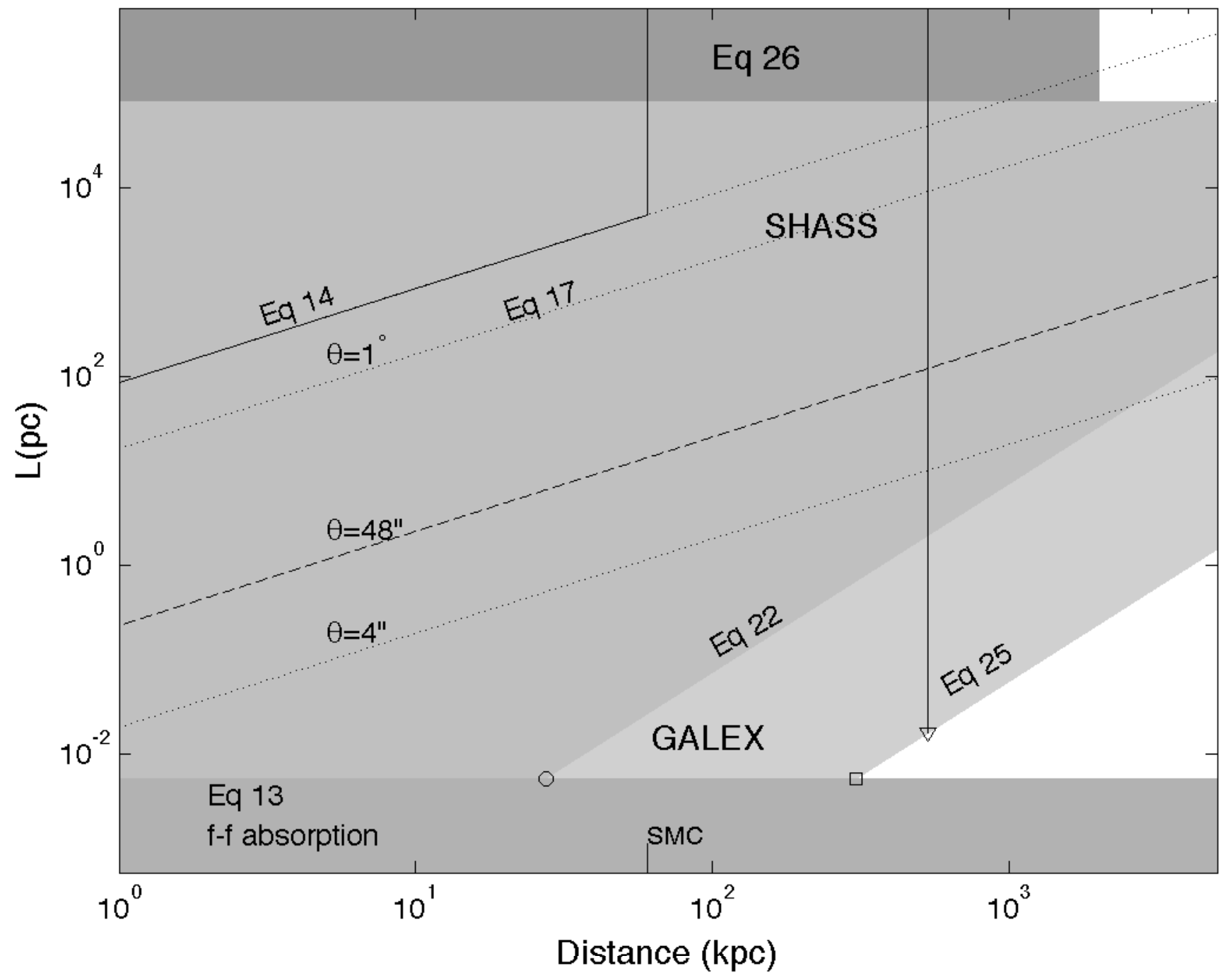}}
\caption{\small
Parameter space of the size of the nebula ($L$ in pc) and
the distance to the nebula ($d$ in kpc) based on the DMs
of pulsars in the vicinity of the {\it Sparker}
(Equation~\ref{eq:thetaDM}), the lack of radio free-free
absorption of the burst itself (Equation~\ref{eq:Lmin}),
the surface brightness limit from SHASS
(Equation~\ref{eq:LSurface}), and the point source limit
obtained from SHASS (Equation~\ref{eq:LSHASSG}).  The dashed
line marks the lower edge of the phase space excluded by
the lack of suitably powerful ionizing source (\S\ref{eq:LGalex}).
The constraint on the mass of the interstellar halo leads
to the top rectangle (marked as
Equation~\ref{eq:MassInterStellarHalo}).  The circle
[$d=d_{\rm min}(H\alpha)\sim 27\,$kpc] and square [$d=303\,$kpc]
mark the minimum allowed distance based on the absence of
H$\alpha$ and Lyman continuum ({\it GALEX}) data.  The SHASS
constraint is limited to a distance of several Mpc due to
the small width of the SHASS H$\alpha$ filter ($\pm 8\,$\AA).
For this reason the plot cuts off at 5 Mpc.
\label{fig:Constraints}} 
\end{figure*}

To summarize: we did not find any suitable ionizing source capable
of maintaining a DM = 350\,cm$^{-3}$\,pc (Galactic origin) or DM =
300\,cm$^{-3}$\,pc (SMC or beyond location) nebula.  We translate
this constraint as follows.  We equate the Lyman continuum flux of
the hottest and brightest star in the localization region to the
recombination rate of a photo-ionized nebula of diameter $L$.  Since
this is the maximum possible luminosity we derive our fifth constraint.
\begin{equation} 
  	L_{\rm pc}d_{\rm kpc}^{-2} \ltorder 6\times 10^{-8}.  
  	\label{eq:LGalex} 
\end{equation}
When Equation~\ref{eq:LGalex} is combined with the free-free
constraint (Equation~\ref{eq:Lmin}) we find $d\gtorder 303\,$kpc.
This demand is marked by an open square in Figure~\ref{fig:Constraints}.
The minimum distance estimate can be further improved by determining
the luminosity class of star~A (via spectroscopic observations).
In any case, even if star A is indeed a young  main sequence B star
in the outskirts of the Local Group, then the {\it Sparker} is
located at a distance of 530\,kpc. This is shown by the inverted
triangle in Figure~\ref{fig:Constraints}.  In this framework star~A
cannot supply any more ionizing photons than that required for the
minimum size nebula (at this distance; see Table~\ref{tab:Nebulae})
we can exclude any nebula with a size larger than that of the minimum
nebula at 530\,kpc or larger distance -- whence the vertical line
in Figure~\ref{fig:Constraints}.

\subsection{A Large Ionized Galactic Corona?}
\label{sec:LargeIonizedCorona}

At this point one can imagine a location for the {\it Sparker} at
the edge of our Local Group (though the progenitor population would
have to be non-stellar and exotic). Is there any constraint on such
a hypothesis?

To start with, we note that the DM contribution from the diffuse
interstellar medium (ISM) in the Galactic halo and the Local Group
is less than 5\,cm$^{-3}$\,pc \citep{b07,w06}.   Nonetheless, let
us be bold and postulate that our Local Group is adorned by a large
ionized corona of radius $R_{\rm C,kpc}$ and attribute all of DM$^S$
to this corona.

With the DM$^S$ fixed, assuming that the corona is composed of only
Hydrogen, the mass of ionized gas in this corona is
\begin{equation} 
	M_C  = 2.8\times 10^{7} R_{\rm C,kpc}^2\,M_\odot.
	\label{eq:MassInterStellarHalo} 
\end{equation}
A corona with size of 100\,kpc and 1\,Mpc would have a mass of
$2.8\times 10^{11}\,M_\odot$ and $2.8\times 10^{13}\,M_\odot$.  To
our knowledge there is no indication of such a massive interstellar
halo in our Local Group (\citealt{b07}; also, M. Shull, pers. comm.).
As a result, we derive our last constraint, namely, we  exclude any
large structure on the scale of the Local Group, giving a constraint
on the minimum distance of about 1\,Mpc.

% should we explicitly quantify this constraint to connect it with Fig 7?

\subsection{Allowed Phase Space}
\label{sec:AllowedPhaseSpace}

The following discussion is aided by inspecting
Figure~~\ref{fig:Constraints}.  There are two regions of phase space
that are not excluded.  First we discuss the small triangle in the
lower right corner.  Second we discuss the small rectangle in the
upper-right corner.  It is important to note that the phase space
diagram is -- consistent with the SHASS bandwidth -- limited to
distances less than 5\,Mpc (assuming Hubble expansion).

Let us consider the lower triangle region.  Any allowed nebula
cannot be located any closer than $d_{\rm min}(H\alpha)=27\,$kpc
(using only the H$\alpha$ data and Equation~\ref{eq:Lmin}; marked
by an open circle).  If we assume photo-ionization, then the
intervening nebula is beyond 303\,kpc (marked by an open square).
However, there is no reason to believe that the outer reaches of
our Galaxy is peppered with any dense interstellar clouds ($n_e\sim
435\,$cm$^{-3}$; see Table~\ref{tab:Nebulae}), nor stars capable
of ionizing such compact nebulae.  Given the paucity of stars at
such distances, postulating such a nebulae routinely (not only for
the {\it Sparker} but for each FRB) is most artificial.  On the
other hand, a host galaxy located well outside the Local Group would
be entirely allowed by the observations.

Next let us consider the upper right region in
Figure~\ref{fig:Constraints}. Here we are allowed to have large
nebulae ($L\gtrsim 20$\,kpc) but at great distances ($d>2\,$Mpc).
This requirement is easily met by the IGM.

We conclude that the circumstantial evidence is not consistent with
a Galactic or SMC origin nor even a Local Group origin.  The 
following possibilities are allowed: the excess DM arises in the IGM or in a galaxy
well outside the Local Group or both.  In \S\ref{sec:PorousNebula}
and \S\ref{sec:Caveat} we discuss possible loopholes in reaching
this conclusion.

\section{A Porous Nebula?}
\label{sec:PorousNebula}

The discussion thus far in this section is based on the assumption
of a homogeneous intervening nebula.  We now consider the implications
of a porous nebula.  Specifically, we assume that the nebula is
composed of $N$ ionized clumps of size $l$.  For mathematical
simplicity, we assume that both the nebula and the clumps are cubes.
We define the volume filling factor as $\phi_V=Nl^3/L^3$. Let
$n_c=N/L^3$ be the number density of clumps. Since the cross-section
of the clumps is $l^2$, on average we encounter $n_cl^2L$ clumps
along a given line-of-sight. The average dispersion  measure is
then
\begin{equation}
	{\rm DM}=n_cl^2L \times n_el=\phi_Vn_eL,
	\label{eq:DMphi}
\end{equation}
and the average emission measure is
\begin{equation}
	{\rm EM}=n_cl^2L \times n_e^2l=\phi_Vn_e^2L,
	\label{eq:EMphi}
\end{equation}
where $n_e$ is the electron number density in the clumps.
Therefore,
\begin{equation}
	\frac{\rm DM^2}{\rm EM} = \phi_V L.
        \label{eq:phiL}
\end{equation}
That is, the size of the nebula inferred from the {\rm DM} and {\rm EM} 
is the filling factor times the physical size of the region.

In the previous section, a unity volume filling factor is assumed
(i.e., no clumps). All constraints related to the nebula size
inferred from the DM and EM are affected if the filling factor is
not unity, with some strengthened and some weakened. Mathematically,
the effect is to replace $L$ in the relevant constraint equations
with $\phi_V L$.

Apparently, the constraint from the spatial range of the nebula
limited by the DM of pulsars (Equation~\ref{eq:thetaDM}) is not
affected.  The constraint from free-free absorption
(Equation~\ref{eq:Lmin}) becomes $\phi_VL>L_{\rm ff}$ and the
corresponding line in Figure~\ref{fig:Constraints} moves up by a
factor of $1/\phi_V$.  Similarly, $L$ in Equation~\ref{eq:LSurface}
is also replaced by $\phi_V L$ and the corresponding line in
Figure~\ref{fig:Constraints} moves up.  Constraints from
Equations~\ref{eq:thetaDM} and \ref{eq:LSurface} are thus strengthened
by requiring a larger size of the nebula region and shrinking the
allowed region (that above the line) in the parameter space.
Replacing $L$ by $\phi_V L$ in Equations~\ref{eq:LSHASSG} and the
UV constraints (\S\ref{sec:UVContinuum})  also moves up the
corresponding lines in Figure~\ref{fig:Constraints}, but this change
expands the allowed region (region below each line) and thus weakens
the constraints.  We note, however, that because the relevant lines
move up by the same factor, the minimum distance set by combining
equation \ref{eq:Lmin} and equation \ref{eq:LSHASSG} (or that by
equation \ref{eq:Lmin} and equation \ref{eq:LSHASSSMC}) remains
unchanged.

In summary, a porous nebula does not change the minimum distance
to the nebula. The free-free constraint {\it increases} the minimum
size of the nebula. Thus, on both grounds Galactic models are
excluded even more strongly.

\section{Caveat: A Nebula Not Powered by Photo-Ionization}
\label{sec:Caveat}

In the previous section the strongest constraint on the minimum
distance to the {\it Sparker} came from examining the {\it GALEX}
UV data. This constraint is meaningful only if the DM nebula is
photo ionized.  However, one could think of free electrons being
produced by other mechanisms. Three mechanisms come to mind: cosmic
ray ionization, radiative shocks and a flash ionized nebula.  The
{\it GALEX} limits would be rendered useless should we be able to
develop a plausible case for any of these mechanisms.  Separately,
dust extinction would attenuate UV photons and also potentially
dilute the {\it GALEX} constraints.

Historically, cosmic rays were the first to be suggested as ionizing
sources for the diffuse ISM (see \citealt{Spitzer78}).  The ionization
cross-section is dominated by low energy (non-relativistic) protons
and ions (see \citealt{Webber98}).  The estimated cosmic ray
ionization rate lies in the range (3--300)$\times 10^{-17}\,$s$^{-1}$
per H atom \citep{wmh+03,lrh04}. We can easily show that matching
the recombination time to any of  the nebular parameters listed in
Table~\ref{tab:Nebulae} to the ionization timescale would require
a cosmic ray flux $10^7$ times larger than the above value.

\subsection{Ionization by Shocks}
\label{sec:RadiativeShocks}

Shocks, usually the product of supernova blast waves or stellar
winds, provide an alternative ionization mechanism. The amount of
DM generated depends on the properties of the medium (most notably,
the ambient particle density $n_0$), the energy carried by the shock
and the fraction transferred to the ISM. Here and below, unless stated otherwise,
when computing the H$\alpha$ flux a nominal temperature of $10^4\,$K is
assumed.

On one side of the energy spectrum are strong, high velocity, shocks
such as those originating in supernova blast waves (e.g.,
\citealt{hm07}).  For example, an $E=10^{51}$\,erg supernova that
expends, say, $f=10$\% of the total energy on ionization of the surrounding
ISM requires an ambient H~I density of:
\begin{equation}
	n_0 = 8.5 f^{-1/2} E_{51}^{-1/2} {\rm DM}_{300}^{3/2}\, {\rm
	cm}^{-3}
\end{equation}
to produce the expected levels of DM. 
The size ($\sim 35$~pc) and the emission measure (${\rm
EM} \sim 2600\,{\rm cm^{-6}\,pc}$) of the resulting nebula would
make it easily detectable by SHASS, even if located within the  SMC
or beyond.  

On the low energy side, we find that typical (e.g., \citealt{rhc+88})
fully developed radiative shocks are incapable of producing the
required levels of DM. As reviewed by \cite{dm93}, the extent of
the radiative zone is solely determined by the column density,
$N_{\rm rad}$, of the shocked material. For typical shock speeds
of $60 < {\rm v}_s < 150$~km\,s$^{-1}$ and the Alfv\'en Mach
number\footnote{$M_A$ is the Alfv\'en Mach number, $M_A = {\rm
v}_s/{\rm v}_A$ and ${\rm v}_A$ is the Alfv\'en velocity.}, $M_A
\gg 1$, the shocked column is $N_{\rm rad} \sim 10^{17.5} {\rm v}_ {s,7}^4\,
{\rm cm}^{-2}$ \citep{mhs+87}. The total column of ionized hydrogen
is larger, and estimated to be on order of $N_{\rm rad} \sim 10^{18.5}
{\rm v}_{s,7}^4\, {\rm cm}^{-2}$ for shocks with $80 < {\rm v}_s < 140$~km\,s$^{-1}$
and $M_A \sim 10$ \citep{rhc+88,dm93}. This corresponds to ${\rm
DM} \sim 1 {\rm v}_{s,7}^4\, {\rm cm^{-3}\,pc}$, and length scales of
order $1 n_0^{-1}\, {\rm pc}$, again insufficient to explain the
{\it Sparker} DM.

We may invert the question and look for the ${\rm v}_s$ required to produce
${\rm DM} \sim 300$. The answer depends on the cooling function and
timescales in the radiative zone, compared to the shock propagation
speed. Essentially, the leading edge of the shock must move forward
fast enough to accumulate the required DM before the ionized medium
at the trailing edge has cooled sufficiently to recombine. Assuming
a strong and steady shock, full ionization and isobaric radiative
cooling, and by approximating the cooling curve of \citet{bh89}
with $\Lambda = 10^{-23} T_7^{-1}\ {\rm erg\, cm^3\, s^{-1}}$,  for
$10^5 < T < 10^7\, {\rm K}$, we find that ${\rm v}_s \sim 840 \rm
(DM_{300})^{1/5}\, km\, s^{-1}$. 
The timescale for the
development of the full radiative shock is $t \sim 3\times 10^4
n_0^{-1}\ {\rm yr}$; again, a lower limit. From the velocity and timescale
it follows that the lower limit on the size of the
radiative region is $L \sim 7.5 n_0^{-1}\ {\rm pc}$, excluded by
existing constraints to beyond the SMC.  The timescale is rather long --
a typical supernova shock would have significantly slowed down by
then. Increasing the ambient density, $n_0$, would shorten the
timescale, but would then run afoul of constraints set by SHASS,
due to the increased emission measure for a smaller, denser, region.

Therefore, this possibility requires  a highly elongated radiative
shock with an almost edge-on viewing geometry. It is within the
realm of possibilities that such a situation may 
take place for an RRAT located in the Galactic
plane. However,  the requirements of a supernova shock and
edge on geometry means that we cannot routinely invoke this
explanation for most FRBs.

\subsection{Flash Ionized Nebula?}
\label{sec:Flash}

In the previous section we assumed that the DM causing nebula was
already present when the {\it Sparker} event took place. We now
consider the possibility that the {\it Sparker} was accompanied by
a soft X-ray flash ({\it Flasher!}\,) and this flash resulted in
the ionization of the nebula.

The soft X-ray flash has to be powerful enough to produce a nebula
with electron column density of DM = 300\,cm$^{-3}$\,pc or $1.16\times
10^{21}\,{\rm cm}^{-2}$ and there has to be enough circumburst gas
to provide the necessary number of electrons.  The number of electrons
within the flash ionized nebula is
\begin{eqnarray}
	N_{\rm e} &=& \frac{4\pi}{3}\Big(\frac{L}{2}\Big)^3 n_e = 4.6\times 10^{53}
		\frac {\rm DM} {300\,{\rm cm^{-3}\,pc}} 
		\bigg(\frac {L_{\rm pc}} 
                                   {10^{-2}\,{\rm pc}}
                        \bigg)^2.
\end{eqnarray}
The timescale for ionization at radius $r$ is
\begin{equation}
	\tau_{\rm ion} = \Bigg[\sigma_1\bigg(\frac{\nu}{\nu_1}\bigg)^{-3}
				\frac{N_e/\Delta t_{\rm X }}{4\pi r^2}
			\Bigg]^{-1},
\end{equation}
where $\Delta t_{\rm X}$ is the duration of the soft X-ray flash,
$\sigma_1(\nu/\nu_1)^{-3}$ is the photoelectric absorption at
frequency $\nu$ and $\sigma_1=6\times 10^{-18}\,$cm$^{2}$ is the
cross-section at the Lyman edge ($h\nu_1=13.6\,$eV).  At the edge
of the nebula ($r=L/2$) the ionization time is determined by the
luminosity and ionization cross-section (which is dominated by the
photo-ionization of Hydrogen) and is
\begin{equation}
	\tau_{\rm ion} = 0.001 \Delta t_{\rm X} (\nu/\nu_1)^3.
\end{equation}
Provided that $\Delta t_{\rm X}\sim \Delta t$ we find $\tau_{\rm ion}$ is
is much smaller than the delay between the propagation in the decimeter
band (say 1.4\,GHz) and that by a photon at high energies. This justifies
assuming an instantaneous creation for the flash ionized nebula.

The energy of the X-ray flash is 
\begin{eqnarray}
	E_{\rm ion} &>& N_e h\nu_1=
		 1\times 10^{43} 
			\bigg(
			\frac {\rm DM} {300\rm\,cm^{-3}\,pc}
			\bigg)
			\Big( 
			\frac{L_{\rm pc}}{10^{-2}\rm pc}
			\Big)^2 \,{\rm erg}.
	\label{eq:Eion}
\end{eqnarray}
The (isotropic) energy budget is quite impressive even for the smallest
allowed value of $L$. In particular, the isotropic bolometric yield of the
rare hyper-flares from soft $\gamma$-ray repeaters (SGRs) can be as high as
$10^{47}\,$erg -- but with most of the release in the hard X-ray band.
Furthermore, the estimate of Equation~\ref{eq:Eion} does not account for
radiation at energies lower or higher than $h\nu_1$.  Should the nebula be
a few parsecs in size, then the {\it Sparker} results from a cataclysmic
event.

The {\it Sparker} took place about thirteen years ago.  Given the recombination
timescale  of
\begin{eqnarray}
	\tau_{\rm R} &=& (n_e\alpha_B)^{-1}\cr
		&=& 4.2 \bigg(\frac{\rm DM}{300\,\,{\rm cm}^{-3}{\,\rm pc}}\bigg)^{-1}
			\bigg(\frac{L}{10^{-2}\,\rm pc}\bigg)\,{\rm yr},
	\label{eq:RecombinationTimeScale}
\end{eqnarray}
there  still exists an opportunity to search for the flash ionized nebula.
The flux level is the same as that estimated for the \ion{H}{2} region
model (\S\ref{sec:Halpha}).

We note, however, that if a comparable amount of the required energy
for the ``flasher'' is emitted as X-rays or $\gamma$-rays, then it
would be readily detectable by the interplanetary
network\footnotemark\footnotetext{http://heasarc.gsfc.nasa.gov/docs/heasarc/missions/ipn.html}
(IPN) up to the SMC distance.  \citet{lbm+07} reported that the
IPN, which has almost full sky coverage, did not detect any GRBs
or SGR hyper flares temporally associated with the {\it Sparker}.

As in the previous section we can probably invoke this framework
for a single source such as the {\it Sparker}. However, it would
be difficult to do so for an entire population with a daily rate
of $10^4$ and not have the expected EUV/X-ray flashes remain
undetected by past and existing missions.

\section{Stellar Coronal Model}
\label{sec:StellarCorona}

Taking a contrarian view
\cite{lsm13} propose a scenario in which a stellar corona provides
the observed DM. This means that the actual electromagnetic pulse
(EMP) takes place somewhere inside the corona and the radio pulse
accumulates the DM as it propagates towards the observer.  In this
section we use EMP to indicate the pre-dispersed pulse as distinct from
the {\it Sparker}, which we use to indicate the observed, dispersed
pulse.

The simplest expectation for this model is that FRBs should be
concentrated towards the Galactic plane.  The reported events are
all at high latitudes which is obviously not a good omen for the
model (unless they are all very nearby).  We eagerly await the
analysis of low latitude fields from
Parkes and Arecibo. Robust detection of FRBs in these data sets
would certainly boost this model.

The emission
mechanism is either via coherent or incoherent processes.  Coherent
emission within a corona (which consists of dense non-relativistic
plasma) may be problematic. On the other hand, it is possible to
imagine a sudden deposition of energy (e.g., magnetic reconnection)
which results in ultra-relativistic shock.  A radio pulse can
plausibly be produced in the post-shock gas via incoherent synchrotron
emission (see \citealt{Blandford77}).  From Equation~\ref{eq:DkpcGamma}
we find $\gamma\sim 10^4$.  The size of the emitting region is
$2\gamma^2c\Delta t$.  For $\Delta t=1\,$ms, the size of the emitting
region is $6\times 10^{15}\,$cm which is much larger than any
plausible corona.  Independent of this concern it would be useful
to investigate possible modifications of the spectrum of the radio
pulse as it propagates through the coronal plasma and coronal photon
field.

Now let us return to some basic considerations of the model
(independent of how the EMP was generated).  We start with a simple
model: a corona with a homogeneous electron density, $n_e$ and
radius $L$.  We assume that the EMP is generated at radius $R_*$
(which is not necessarily the photospheric radius).  Then ${\rm DM}
= n_e(L_{\rm pc}-R_{\rm pc})$ where $R_{pc}=R_*/{(\rm 1\,pc)}$ and
$L_{\rm pc} = L/({\rm 1\,pc})$.

For high temperatures ($T>3\times 10^5\,$K), the free-free absorption
coefficient per unit length (\citealt{L74}, p. 47) is
\begin{equation}
	\alpha(\nu) = 9.79\times 10^{-3} \frac{n_e n_i}{\nu^2T^{3/2}}{\rm ln}\Big[\frac{4.7\times 10^{10}T}{\nu}\Big]
\,{\rm cm}^{-1}.
\end{equation}
Normalizing  $\nu=\nu_0= 1.4\times 10^9$\,Hz and setting $T=10^8T_8$, we find
\begin{eqnarray}
	\tau(\nu) &\approx& 3.4\times 10^{-7}\Big(\frac{\nu}{\nu_0}\Big)^{-2} T_8^{-3/2}
	\Big(\frac{{\rm DM}_3^2}{L_{\rm pc}}\Big),
\end{eqnarray}
where ${\rm DM=10^3DM_3\,cm^{-3}\,pc}$. Let us say $\tau(\nu_0)\lesssim
3$ ({\it cf.} Equation~\ref{eq:beta} and subsequent discussion).  Thus, we
have
\begin{eqnarray}
	L_{\rm pc}-R_{\rm pc} &\gtrsim& 1\times 10^{-7}T_8^{-3/2}{\rm
	DM_3}^2\Big(\frac{\nu}{\nu_0}\Big)^{-2}.
\end{eqnarray}
This length scale corresponds to about $4\,R_\odot$.  Going forward we will
set $\nu=\nu_0$. 

The mean density, the mass  and the thermal content of the corona is 
\begin{eqnarray}
	n_e &=& 1\times 10^{10} T_8^{3/2} {\rm DM_3}^{-1}\ {\rm cm}^{-3},\cr
	M_c &=&  1\times 10^{-12}{\rm DM_3}^5T_8^{-3}\,M_\odot,\cr
	Q_c &=& 5.1\times 10^{37}{\rm DM_3}^5 T_8^{-2}\,{\rm erg}.
\end{eqnarray} 
For the corona to be in approximate hydrostatic equilibrium we must
have the thermal energy (in each of electron and proton)
be less than the gravitational potential energy (per H atom) or
$3k_BT < GMm_H/L$; here, $m_H=m_p+m_e$.  This is clearly violated
and so we must assume that there is outflow. The characteristic
thermal velocity that matters is
\begin{equation}
	{\rm v} = \sqrt{\frac{3k_BT}{m_H}} = 1580\,T_8^{1/2}\,{\rm
	km\,s^{-1}},
\end{equation}
and the mass flux is 
\begin{eqnarray}
	\dot M &=& 4\pi L^2 n_e m_H \sqrt{\frac{3k_BT}{m_H}},\cr
		&\approx & 5\times 10^{-8}T_8^{-2}{\rm DM_3}^3\,M_\odot\,{\rm yr}^{-1}.
\end{eqnarray}

So even though we started with a static model for the corona we
find that the corona is not dynamically stable and has a strong
outflow.  If so, the assumption of a homogeneous density in the
corona is not correct.  Therefore, we need to adopt a wind equation:
$n_e\propto r^{-2}$.  As noted in \S\ref{sec:StarsSupernovae}, as
long as $L$ is even modestly larger than $R_*$, we can approximate
${\rm EM\approx DM^2/R_{\rm pc}}$, which is similar to the homogeneous
case, provided we identify $R_*$ with $L$.

The free-free luminosity per unit volume is (\citealt{L74}, p. 46)
\begin{equation}
	\epsilon_{\rm ff} = 1.4\times 10^{-27}T^{1/2}n_e^2\ {\rm
	erg\,cm^{-3}\,s^{-1}},
\end{equation}
where we have assumed a pure hydrogen plasma ($n_e=n_i$). The luminosity
(assuming that the plasma is optically thin)\footnote{A dispersion
measure of $10^3\,{\rm cm^{-3}\,pc}$ is Compton thin; the plasma
is also thin for free-free absorption for $h\nu$ comparable to
$k_BT$.} and the bolometric flux density are, respectively,
\begin{eqnarray}
	L_{\rm ff} &=& \frac{4\pi}{3}L^3\epsilon_{\rm ff}
		= 1.7\times 10^{32}T_8^{-1}{\rm DM_3}^4\,{\rm erg\,s^{-1}},\cr
	f_{\rm ff} &=& 1.3\times 10^{-12}D_{\rm kpc}^{-2}T_8^{-1}{\rm DM_3}^4\, 
		{\rm erg\,cm^{-2}\,s^{-1}}.
\end{eqnarray}
The cooling  and the hydrodynamical timescales are 
\begin{eqnarray}
	t_{\rm ff} &=& 3k_B T n_e/\epsilon_{\rm ff},\cr
	&=&3.43\,T_8^{-1}{\rm DM_3}   \,{\rm day},
\end{eqnarray}
and 
\begin{eqnarray}
	t_{\rm h} = L/{\rm v}  
	= 0.5\,T_8^{-2}{\rm DM_3}^{2}\,{\rm hr}.
\end{eqnarray}
Since the corona is optically thin and $t_{\rm h}<t_{\rm ff}$ we
expect to see a bright X-ray source with typical photon energy of
$2.7k_BT=23T_8\,{\rm keV}$ lasting for $0.5T_8^{-2}\,$ hours
(after the radio burst). However, we note that X-ray emission will
be seen for at least a similar duration as the corona inflates to
provide the necessary DM. Thus we will have X-ray emission, preceding
and succeeding the EMP, with a fluence of
\begin{equation}
	F_{\rm ff} = 2.3\times 10^{-9}D_{\rm kpc}^{-2}\,T_8^{-3}{\rm DM_3}^6\,{\rm erg\,cm^{-2}}.
\end{equation}

X-ray missions are more sensitive at lower energies and so better
constraints on this model can be obtained by considering missions
which operated primarily in the classical X-ray band or the soft
X-ray band.  In order to compute the X-ray light curve we would
need to know the boundary conditions at the base of the corona.
Since the proposed model is not sufficiently developed, any further
calculation of this sort is premature.  We can reasonably assume
that the duration of the X-ray emission at lower energies (keV
range) is longer than the ${\rm 1T_8^{-2}DM_3^2}$\,hr discussed
above.

In summary, an expectation of the coronal class of models is {\it
pre-cursor} hard X-ray emission followed by an X-ray afterglow that
becomes softer with time. Given a daily FRB rate of $\dot{\mathcal{N}}\approx
10^4\,{\rm day}^{-1}$ the number of X-ray sources we expect to see
is $\dot{\mathcal{N}}\tau_X$ where $\tau_X$ is the duration over
which the X-ray signal is above the detection level. For $T=10^8\,$K
we expect about 400 sources at any given time in the sky.  According
to \cite{kbb+13}, at any given time, there are $4\times 10^{-4}$
X-ray transients per square degrees on the sky with a flux threshold
greater than $3\times 10^{-12}\,{\rm erg\,cm^{-2}\,s^{-1}}$
[0.2--2\,keV band], or about 16 sources over the entire sky.  Most
of these are identified with sources which are expected to be
variable from other considerations (e.g., known flare stars primarily;
see \citealt{Vikhlinin98}).  Clearly, coronal models with $T=10^8\,$K
are not favored on observational grounds.

Let us consider even hotter coronas, say $T=3\times 10^8\,$K.
Relative to $T=10^8\,$K coronal model, the duration of the event
is reduced by a factor of ten (from an hour to 6 minutes) and the
flux  decreased by a factor of three.  With a mean temperature of
70\,keV, this short lived object may even be mistaken for a long
duration GRB! Given $\dot{\mathcal{N}}$ we would expect ten nearby
(100\,pc) events every day each with a fluence of $10^{-7}\,{\rm
erg\,cm^{-2}}$.  The Burst Alert Telescope (BAT) can detect GRBs
with fluence [15--150\,keV] brighter than $10^{-8}\,{\rm erg\,cm^{-2}}$
(though most of the GRBs are considerably brighter). A search through
the BAT catalog \citep{sbb+11} would provide observational feedback
to the coronal model.

\section{An Extragalactic Origin}
\label{sec:ExtraGalacticOrigin}

In \S\ref{sec:DistanceSize} using basic theory and archival H$\alpha$
and {\it GALEX} data, we attempted to constrain the size ($L$) and
the location (distance, $d$) to an intervening ionized nebula that
could account for the excess (over Galactic value, if the {\it
Sparker} was located in our Galaxy or in the vicinity of the
Magellanic Clouds) of the dispersion measure inferred from the
frequency-dependent arrival time of the pulse from the {\it Sparker}.
The allowed phase space for $L$ and $d$ is summarized in
Figure~\ref{fig:Constraints}.  We concluded that the nebula cannot
be located in our own Galaxy or the SMC and is not even allowed to
be on the periphery of our Galaxy.  After investigating possible
caveats (\S\ref{sec:PorousNebula}--\ref{sec:Caveat}) we concluded
that the excess of dispersion measure arises in another galaxy  or
in the intergalactic medium (IGM) or both.  Having reached this
conclusion, the only issue is to apportion the dispersion measure
between the inter-galactic medium (IGM) and ionized gas within the
host galaxy. For the {\it Sparker}, in accord with \cite{lbm+07},
a red-shift range $0.1\lesssim z\lesssim 0.3$ is reasonable.\footnote{
As can be seen from Figure~\ref{fig:SparkerLocation}, there is no
distinctive galaxy within the localization region.  The most notable
galaxy lies outside the polygon beyond the NorthWestern tip.}

A similar analysis can be applied to the four FRBs reported by
\cite{tsb+13}, but that is not educational. What is useful is to
take the best constraints from the whole set of the Parkes events.
In particular, the $L_{\rm ff}$ scales as DM$^{2}$ ({\it cf.}  Equations~\ref{eq:EMDM},
\ref{eq:tauff}).  The larger DMs of the \cite{tsb+13} therefore
provide the strongest constraints on compact intervening nebulae
and for stellar models (\S\ref{sec:StellarCorona}).

We conclude that the {\it Sparker} and the four Parkes events have
to be extra-galactic -- {\it provided that the frequency dependent
arrival time is a result of propagation through cold plasma.}  In
this section we investigate the consequences of the {\it Sparker}
being located in a distant galaxy. Anticipating the later discussion
to include FRBs, we set the nominal distance to 1\,Gpc.  We will
now revisit the issue of energetics and brightness temperature ({\it cf.}
\S\ref{sec:Energetics}-\ref{sec:BrightnessTemperature}).

Switching now to parameters typical of FRBs: peak flux of 1\,Jy at 
1.4\,GHz and $\Delta t=1$\,ms, we find the isotropic energy release
in the radio band is 
\begin{equation}
	\mathcal{E}_S\sim 4\times 10^{39}D_{\rm Gpc}^2\,{\rm erg},
	\label{eq:ES}
\end{equation}
assuming  $\alpha=-1$ with a low frequency
cutoff of $\nu_0/10$ ({\it cf.} \S\ref{sec:Energetics}).  However, if the
intrinsic spectrum is an exponential ({\it cf.} \S\ref{sec:EM}) then the
isotropic energy release is larger by $\approx \exp(x_0)/(x_0{\rm
ln}(x_0))$ where $x_0=\nu_0/\nu_c$.

The brightness temperature at $\nu_0$ is $6\times 10^{34}D_{\rm
Gpc}^2$K and  is larger by the factor $x_0^2\exp(x_0)$ at $\nu=\nu_c$.
We compare the {\it Sparker} to Galactic RRATs and giant pulses
from pulsars.  The brightest  RRAT known to date has a peak flux
of 3\,Jy in the 21-cm band.  For the RRAT sample of \cite{mll+06},
we derive brightness temperatures as high as $10^{23}\,$K.  Next,
the highest brightness temperature event to date is a 15-ns wide
giant pulse from PSR\,1937+214 with $T_B>5\times 10^{39}\,$K (in
the 1.65-GHz band; after correction for interstellar scintillation
\&\ scattering; \citealt{spb+04}).  Thus, apparently, pulsars can
produce the high brightness temperatures that we are inferring for 
the {\it Sparker}.

We draw the reader's attention to the dual-frequency (2.7 \&\ 3.5\,GHz)
studies of PSR\,J1824$-$2452A \citep{kbm+06}. The authors report that
the spectral index of $\sim -5.4$ was observed over the frequency
range 2.7--5.4\,GHz. Furthermore,  it was noted that the giant pulse
phenomena is not necessarily broad-band (i.e. the spectrum could be
quenched at lower frequencies).  Finally, many of the giant pulses
are 100\% elliptically polarized. 

Despite the apparent agreement of brightness temperature and potential
spectral similarity there is one big difference between giant pulses
from pulsars and the {\it Sparker}: the size of the emitting
region. The high brightness temperatures exhibited by pulsars is
on nano-second timescales. This translates into sizes for the emitting
regions from a few meters and up. In contrast, the size of
the {\it Sparker} emitting region is $R=c\Delta t\lesssim 300$\,km.  This
is an upper limit due to possible dispersion and scattering broadening.

Before we discuss the proposed models it is useful to discuss the most
general constraint(s) that can be obtained from the observations.
Clearly, the high brightness temperatures of FRBs stand out. 
As first discussed by \citet{wr78}, the high brightness temperature
inferred in the Crab pulsar requires two conditions: an extremely
clean  region, to prevent severe losses due to induced
Compton scattering, and an ultra-relativistic flow which would then
boost the inferred brightness temperature by $\gamma^3$.  Separately,
matter, if present, would be accelerated by the strong electromagnetic
field and rapidly dissipate energy. Propagation will also be impeded.
Were the {\it Sparker} to be an RRAT or a pulsar, albeit at
cosmological distances, then $\gamma\sim 10^4$ to $10^6$ would be
needed in order to prevent induced Compton scattering from significantly
attenuating the radio emission.  In this spirit we draw the reader's attention
to a recent paper by \cite{katz13} where he argues that $\gamma>10^3$
and notes that  a compact source or an expanding highly relativistic
source are both possible. 

In summary, suitable progenitor models
are those which have an ultra-clean emitting region and, in addition,
a low density circumstellar medium so that external absorption is not significant.  This means, almost always, that the
free-free optical depth should not be large (for usual parameters,
the plasma frequency is usually well below the GHz band).

\section{Progenitors}
\label{sec:Progenitors}

Even more remarkable than their inferred extra-galactic nature is
the all-sky rate of {\it Sparker} and associated Parkes events.
\cite{lbm+07}, noting that the {\it Sparker} would have been detected
to $z\sim 0.3$ ($D\sim 1\,$Gpc), derived a local volumetric rate
of 90\ Gpc$^{-3}$\,d$^{-1}$.  For the four Parkes events, \cite{tsb+13}
quote an all-sky-rate of $1.0^{+0.6}_{-0.5}\times 10^4\,{\rm d}^{-1}$
(for fluence above a few Jy\,ms in the 1.4\,GHz band). The co-moving
distances for these events, if most of the  DM is attributed to the
IGM, is [2.8, 2.2. 3.2, 1.7]\,Gpc.  From this we  derive a volumetric
annual rate of 
\begin{equation}
	\Phi_{\rm FRB} = 2.4\pm0.7\times 10^4\,{\rm Gpc^{-3}\,yr^{-1}}.
\end{equation}

The very large volumetric rate of {\it Sparker}s poses great
difficulty for any extra-galactic model.  It is useful to compare
the volumetric rate of {\it Sparker}s to the rates of well established
cosmic explosions (Table~\ref{tab:Rates}). The most frequent stellar
deaths are core-collapse supernovae. The FRBs would claim 10\% of
the core collapse rate.  In this section we review and critique suggested stellar models for
FRBs. We discuss the model of giant flares from soft gamma-ray 
repeaters as possible progenitors to FRBs in the next section.
We do so because in our opinion this model stands out for having
a sound physical basis.

\subsection{Core Collapse Supernovae}
\label{sec:CoreCollapseSupernovae}

Massive stars lose matter throughout their life. The parameter
$A=\dot M/(4\pi {\rm v_w})$ where $\dot M$ is the mass loss rate
and ${\rm v_w}$ is the velocity of the mass losing wind determines
the run of the circumstellar density (see \S\ref{sec:StarsSupernovae}).
Even for type Ib/Ic supernovae (which have the fastest winds and
the smallest mass loss rate) $A_*$ is in the range of  0.01 to 1;
here, $A_* \equiv A/5\times 10^{11}{\,\rm g\, cm^{-1}}$.  As argued
in \S\ref{sec:StarsSupernovae}, this value is sufficient to cause
free-free absorption (in the decimeter band) at a radius of
$10^{14}\,$cm.  Thus for a successful radio burst, the radio emitting
region must be located beyond this radius. For this reason we reject
all ordinary core-collapse supernovae and their more exotic variants:
long  duration GRBs, low luminosity GRBs, as well as the model of
\cite{ep09}.

\subsection{The Blitzar Model}
\label{sec:Blitzar}

To circumvent the fundamental problem of absorption by either the ejecta or
the circumstellar medium, \citet{fr13} propose a novel scenario: the
desired fraction of core-collapse supernovae explode and leave massive
neutron stars which are rotating sufficiently rapidly that they can exceed
the maximum mass of a stable but static neutron star.  The neutron star
spins down via the pulsar mechanism.  Meanwhile the SN debris and
circumstellar medium is slowly cleared up. At some point the super-massive
neutron star can no longer support itself and collapses to a black hole.
During this transmutation a strong radio pulse is emitted ({\it Blitzar!}).
We agree that the Blitzar model is a clever scenario, but below we argue
that the ramifications of the model are not in accord with what we know
about the demographics of pulsars and the energetics of supernovae and
supernova remnants.

%power from Cobalt decay = 7E9exp(-t/111days)erg/s/gram

We consider a simple and hopefully illustrative example.  
Let us say that at the end stage of a super-massive neutron star's
life, just before it collapses into a black hole, it has a spin period
of $P_1=1.5\,$ms, a value that is typical of the faster spinning
millisecond pulsars.  Now let us make the simplifying assumption that
this pulsar was born with a spin period that is half it's final spin
period, that is with $P_0 = 0.75\,$ms.  With $P_0$
and $P_1$ fixed the only free parameter is the time, $\tau$, it
takes for the super-massive neutron star to spin down to $P_1$.
The magnetic field strength prior to the collapse can be computed
in the vacuum dipole framework and is $B=5.2\times 10^{10}\tau_4^{-1/2}\,$G
where $\tau=10^4\tau_4\,$yr.  The spin-down luminosity of the pulsar,
prior to the transmutation, is extra-ordinary: $\dot E=2\times
10^{40}\tau_4^{-1}\,$erg\,s$^{-1}$.  The spin-down luminosity at
birth is $(P_1/P_0)^4 = 16$ times higher.

Now we work out the ramifications of the Blitzar hypothesis.  First,
({\it i}), in a typical late-type galaxy, given the putative birthrate
of FRBs (1 per $10^3$\,yr), we should expect $10\tau_4$ such bright
{\it young} pulsars with magnetic field strengths significantly
above those of millisecond pulsars ($B\lesssim 10^9\,$G).  Next,
({\it ii}), given the ratio of the FRB rate to that of core-collapse
supernovae, one in ten supernovae should exhibit evidence of an
underlying long-lived powerful source of energy.  Let us consider
a specific case and set $\dot E=10^{42}\,$erg\,s$^{-1}$. Assuming
a mean expansion speed of $5\times 10^8\,$cm\,s$^{-1}$ (at late
times), the radius of a supernova two years after the explosion is
$R_S=3\times 10^{16}\,$cm. It is safe to assume that this power
input is rapidly thermalized.  Equating the blackbody luminosity,
$4\pi R_S^2\sigma T_S^4$ to $\dot E$ yields $T\sim 10^3\,$K.  A
search with WISE and Spitzer missions  for mid-IR emission from
nearby and decade old core collapse supernovae would provide useful
upper limits on the rate of Blitzars ({\it cf.} \citealt{hko+13}).

Decreasing the typical time to collapse from $10^4\,$yr to $10^3\,$yr
would alleviate the issue raised in ({\it i}) but exacerbate that
discussed in ({\it ii}).  Increasing $\tau$ to $10^6\,$yr would
alleviate the concern raised in ({\it ii}) but lead to a large
population ($10^3$) of millisecond young ($10^6\,$yr) pulsars -- a
hypothesis that can be immediately refuted given the known demographics
of Galactic pulsars.  Finally, ({\it iii}), by constructions these
events would release, over a timescale of $\tau$,  an energy of
$\Delta E = 1/2(I_0\omega_0^2-I_1\omega_1^2)$, -- which is comparable
to the typical initial rotation energy of the neutron star or
$10^{52}\,{\rm erg}$; here, $I$ is the moment of inertia,
$\omega=2\pi/P$, and the subscripts are as in the previous paragraph.
There is little evidence that the inferred energy release in any
Galactic supernova remnant, including those associated with magnetars,
exceeds $10^{51}\,$erg \citep{vk06}.

\begin{table} %[tbh]
\caption[]{Volumetric Rates of Selected Cosmic Explosions}
\begin{center}
\begin{tabular}{llll}
\hline
\hline
Class & Type & $\Phi$ & Ref\\
	&	& Gpc$^{-3}$\,yr$^{-1}$ & \\
\hline
LSB (low)     & BC  & 100--1800       & [1,2]\\
LSB ( high)   & Obs & 1               & [1]\\
              & BC  & 100--550        & [1]\\
SHB	      & Obs & $>10$           & [3a] \\
	      & BC  & 500-2000       & [3b]\\
In-spiral     & Th  & $3\times 10^3$  & [4]\\
SGR           & Obs & $<2.5\times 10^4$ & [5]\\
Type Ia      & Obs & $10^5$ & [6]\\
Core Collapse & Obs & $2\times 10^5$  & [7]   \\
FRB      & Obs & $\approx 2\times 10^4$  & [8,9]\\
\hline
\end{tabular}
\label{tab:Rates}
\end{center}
{\small
Notes: 
``Obs'' is the annual rate inferred from observations.  ``BC'' is
the observed rate corrected for beaming.  ``Th'' is the rate deduced
from stellar models.  LSB stands for GRBs of the long duration and
soft spectrum variety.  A gamma-ray luminosity of $10^{49}\,$erg\,s$^{-1}$
divides the ``low'' and ``high'' subclasses (see \citealt{gd07}).
SHB stands for GRBs of the short duration and hard spectrum class.
SGR stands for Soft Gamma-ray Repeaters. Here we only include those
giant flares with isotropic energy release $>4\times10^{46}\,$erg.
Refs: [1] \citealt{gd07}; [2] \citealt{skn+06}; 
[3a] \citealt{ngf06}; 
[3b] \citealt{chp+12};
[4] \citealt{kkl+04}; 
[5] \citealt{Ofek07};
[6] \citealt{sb2005};
[7] \citealt{lcl+11}
[8] \citealt{lbm+07};
[9] \citealt{tsb+13}.
}
\end{table}

\subsection{Short Hard Bursts}
\label{sec:ShortHardBursts}

Short hard bursts are well suited as possible progenitors.  After
all, these systems are clean: no supernovae ejecta, and no rich
circumstellar medium.  However, as has been noted earlier, the rates
of the Parkes events far exceeds that of the short hard bursts (see
Table~\ref{tab:Rates}).  Additionally, we offer the following line
of simple reasoning. The very large rate for the Parkes events
suggests that they are not beamed.  The five Parkes events have
$z<1$.  In contrast, the redshift distribution of short hard bursts
is wider. Bearing this in mind we note that the all-sky rate of
short hard bursts is $\approx 0.5\,{\rm day}^{-1}$ \citep{Nakar2007}.
Thus, concordance between these two estimates would require an
inverse beaming factor in excess of $2\times 10^4$!  There is no
evidence for such a large inverse beaming factor\footnote{The beaming
factor is the fraction of the celestial sphere lit up by sources
with strong conical emission.  If $\theta$ is the half-opening angle
of each of the two jets then the beaming factor is $f_b=1-\cos(\theta)$.
The inverse beaming factor is $f_b^{-1}$.}  \citep{Berger2013}.  In
order to preserve the connection between FRBs and coalescence events,
we have to conclude that only a small fraction of coalescence events
produce short hard bursts.

We now discuss specific models related to short hard bursts.
\citet{Totani2013} revives erstwhile models in which the neutron
stars are reactivated as they approach coalescence. This is an
attractive model from the point of view of radio pulse generation,
as well as the fact that the radio emission takes place {\it prior}
to the coalescence.  However, as noted above, in this scenario
Nature is bountiful with coalescence events.  We should expect to
see an event within 100\,Mpc every 3 days once Advanced LIGO turns
on. We admit that we find this scenario to be positively Panglossian
\citep[see, e.g., ][]{Belczynski:12:358}.

Next, it has been noted in \cite{Zhang2014} and \cite{lhr+13}, that
in some short hard bursts the X-ray light curve shows a plateau.
The authors interpret the cessation of this X-ray plateau as marking
the transmutation of the coalescence product -- a supra-massive
neutron star -- into a black hole.  Inspired by the Blitzar model,
\cite{Zhang2014} suggest that the transmutation results in an intense
radio burst.  On general grounds one expects that the merger will be
followed by the ejection of a relatively small amount
($10^{-4}\,M_\odot$--$10^{-2}\,M_\odot$) of sub-relativistic matter
(see \citet{hkk+13}). In \S\ref{sec:StarsSupernovae}, we construct
a simple toy model with spherical ejection, constant shell thickness,
and a coasting velocity and find that decimetric radiation will be
absorbed, via the free-free process, by the expanding shell.
\cite{Zhang2014} argue that a radio pulse would be seen for those
events whose axis of explosion is pointed towards us. However, if
we are seeking a single explanation for FRBs, this model spectacularly
fails on the grounds of demographics.

\subsection{White Dwarf Magnetar}

An entirely new class of models is speculated by \cite{kim13}. These
authors propose that a fraction of the mergers of two white dwarfs lead to
a highly magnetized white dwarf rotating rapidly and that such an object
may produce a strong radio pulse. These authors make the implicit
assumption that the merger takes place with no ejection of material.
However, the merger is not a clean process (e.g., \citealt{mns04,rkm+13}).
The less massive white dwarf, having the lower density, is disrupted first.
The disrupted material forms an accretion disk which then feeds the more
massive star (primary).  Accretion power heats up the primary star as well
as the disk itself.  As a result, one expects a strong stellar wind to
accompany accretion.  As noted in \S\ref{sec:ExtraGalacticOrigin}, the
production of high temperature beams of radiation require a very clean
environment and the few baryons that are present have to be relativistic.
Leaving this general comment aside, we argue that the resulting wind cannot
be any less strong than that seen for Wolf-Rayet stars and thus $A_* \sim
1$. If so, the radio pulse will be absorbed by the free-free process
(\S\ref{sec:StarsSupernovae}).  Calculation of $A_*$ for merger models is
beyond the scope of this paper but proponents are advised to look into this
issue.

\section{Giant Flares from Soft Gamma-ray Repeaters}
\label{sec:GiantFlares}

We finally come to giant flares from soft gamma-ray repeaters which
have been speculated to be the FRB progenitors by \cite{pp07} and
\cite{tsb+13}.  What makes this suggestion worthwhile is a plausible
physical model \citep{Lyubarsky14}. In this model, following the
giant flare, an electromagnetic pulse (Poynting vector) is formed
and propagates outwards. The pulse eventually shocks the magnetized
plasma which constitutes the plerion (inflated by steady power from
the magnetar during the course of its life). Lyubarsky provides
plausible arguments for strong radio emission either from both the
reverse shock or the forward shock.  Specifically, the model supports
an efficiency of $10^{-5}$ to $10^{-6}$ in converting the energy
released  to bolometric  radio emission. Next, the high brightness
temperature is elegantly accounted for by synchrotron maser emission.

The  most spectacular and energetic Galactic giant flare was observed
on 2004 December 27 from SGR~1806$-$20 \citep{hbs+05,pbg+05}.
We will use this event as the benchmark for giant flares from SGRs
and as such its distance enters into our calculations.  For simplicity, we assume a distance of 15\,kpc \citep{Svirski11} for SGR\,1806$-$20 in all our analyses in this work.  This event, at our assumed distance, had a characteristic energy release of $\mathcal{E}_* \equiv 3.6\times
10^{46}\,{\rm erg}$ in the X-ray band \citep{bzb+07}.  If we assume that the isotropic energy release in $\gamma$-rays, $\mathcal{E}_\gamma$, was approximately equal to this characteristic value, then in Lyubarsky's
model this event could explain FRBs with radio emission of
$\mathcal{E}_R\sim 3.6\times 10^{40}\,{\rm erg}$ to ten times this
value. This energy release is sufficient to account for a typical
FRB at say 1\,Gpc.

We now proceed to compute the volumetric rate of SGR flares.  We
do so in two different ways.  \cite{Ofek07} combined the observations
of Galactic SGR giant flares with the limits on giant flares in
nearby galaxies.  Based on these observations, Ofek finds that the
rate of giant flares with energy above $\mathcal{E}_\gamma \gtrsim
3.6\times10^{46}$\,erg is about $(0.4-5)\times10^{-4}$\,yr$^{-1}$ per
SGR with an upper limit on the volumetric rate\footnote{obtained by assuming
5 active SGRs in the Milky Way and assuming 0.01 Milky Way per Mpc$^3$ \citep{Ofek07}} of
\begin{equation}
	\Phi_{GF}(\mathcal{E}_\gamma\lesssim \mathcal{E}_*) <2.5\times 10^4\,{\rm Gpc^{-3}\,yr^{-1}}
\end{equation} 
(and stated in Table~\ref{tab:Rates}).  This upper limit is compatible
with the inclusion of recent giant flares in nearby galaxies:
GRB\,051103 \citep{okn+06} and GRB\,070201 \citep{omq+08}.  Comparison
of the Galactic rate (discussed below)  with the inferred extragalactic
rate implies a gradual cutoff (or steepening) of the flare energy
distribution at $\mathcal{E}_\gamma \ltorder \mathcal{E}_*$ (95\%
confidence).

Giant flares such as that of 2004 December 27 are detectable by the
BAT instrument aboard the {\it Swift} Gamma-Ray Observatory.
\cite{hbs+05} quote\footnote{The BAT rate is computed for events
similar to the 2004 December 27 event also assumed to have a distance
of 15\,kpc.} a detection of giant flares
by BAT of $19(\tau_{\rm GF}/30\,{\rm yr})^{-1}\,{\rm yr}^{-1}$ where
$\tau_{\rm GF}$ is the mean time between Galactic giant flares as
energetic as 2004 December 27.  During the period 2005--2013 BAT
discovered a total of 70 short duration events. Most of these are
genuine short hard gamma-ray bursts \citep{Berger2013}.  The only
short hard event in this sample which has been claimed to be an
extragalactic giant flare is GRB\,050906 [and associated with the
star-burst galaxy IC~328 (distance of 130\,Mpc; \citealt{ltj+08})].
After discounting securely identified and strong candidate short
hard bursts in the BAT sample we are led to the conclusion that
$\tau_{\rm GF}$ easily exceeds 100\,years. We thus reaffirm the
primary conclusion of the Ofek (2007) analysis: there is a break
in the luminosity function of giant flares and the mean time between
flares as bright as the 2004 December 27 event is in excess of a
century.

A second approach is to use the statistics of Galactic (including
satellite galaxies) giant flares including those fainter than
$\mathcal{E}_*$.  The lifetime of the field of X-ray astronomy is,
say, 40 years. During this period we have observed three giant
flares with energy above $\approx10^{44}$\,erg (1979 March 5, 1998
August 27 and 2004 December 27).  Thus we can plausibly assume that
the mean time between giant flares is $\tau_{\rm GF}\approx 25$\,years.
The g-band luminosity of the Milky Way is $1.8\times 10^{10}\,L_\odot$
\citep{ln13}.  The local density in B-band is $1.8\times
10^8\,L_\odot\,{\rm Mpc}^{-3}$ \citep{cdc+01}. Thus the volumetric
rate of giant flares is
	\begin{equation}
	\Phi_{\rm GF} (\mathcal{E}_\gamma\gtrsim 3\times 10^{44}\,{\rm erg}) \approx 4\times 10^5(\tau_{\rm GF}/25\,{\rm yr})^{-1}
	\,{\rm Gpc}^{-3}\,{\rm yr}^{-1}.
	\label{eq:phiGF}
	\end{equation}
This rate applies to events which are brighter than the event of
1998 August 27 (which was approximately 100 times fainter than the
event of 2004 December 27 event).  This simple determination of the
volumetric rate and  the upper bound of Ofek discussed above (which
we remind the reader applies to bursts with $\mathcal{E}_\gamma\lesssim
\mathcal{E}_*$) are consistent with each other.

\subsection{Dense Interstellar Medium}

We draw the readers attention to an important issue.  We have looked
into the environments of several magnetars in our Galaxy.  Almost
all of them, not surprisingly\footnote{Active SGRs are a youthful
population. For instance the true age of the prototype of the giant
flare, SGR~1806$-$20, is only 650\,yr \citep{tck12}.}, are in
star-forming regions (which are rich in both ionized and neutral
interstellar gas) or embedded in a supernova remnant. We find DMs
ranging from ${\rm 100\,cm^{-3}\,pc}$ to nearly ${\rm 10^3\,cm^{-3}\,pc}$.
Furthermore, the X-ray flash could additionally ionize neutral
matter (see \S\ref{sec:Flash}).  Indeed, this causal association
of young SGRs with dense ISM regions provides the most  reasonable
explanation for scattering tails seen in one FRB and in the {\it
Sparker} (and discussed in \S\ref{sec:ISS}).

Consistent with this giant flare hypothesis, it follows that a
significant contribution to the inferred DM arises from the vicinity
(distance comparable to star-forming regions, say $\lesssim 100\,$pc)
of the young magnetar. We advocate 400\,cm$^{-3}\,$pc as a
representative value. In this case, the effective volume of FRBs
is reduced. However, the substantial Poisson error in
Equation~\ref{eq:phiGF} shows that we can easily tolerate a reduction
in the true volume by a factor of a few. In summary, it is not
unreasonable to claim a good match between the true volumetric rate
of FRBs and that of giant flares from SGRs.

Additionally, it may well be that for some FRBs the local ISM is
dense enough that the decimetric signal is attenuated by free-free
absorption.\footnote{Those sources with a free-free optical depth
of say a few would show up with strongly positive spectral index;
see Equation~\ref{eq:nuctau0} and the discussion that follows.}
These may further increase the volumetric rate of FRBs.  Another
consequence is that low frequency (meter wavelength) searches would
find fewer FRBs compared to L-band searches as pointed out in
\cite{Hassall:13:371}.

%Shri:
%
%There are only three SNRs cleanly associated with magnetars.
%1) 1E 1547.0-5408 has an integrated DM of 830 pc cm^-3. The nebula around it has a size of 4.7 pc, so if the entire DM were associated with the nebula, the corresponding EM would be 146.6E3 pc cm^-6
%2) PSR 1622-4950 has an integrated DM of 820 pc cm^-3, "associated" with SNR G333.9+0.0 but its too far south.
%3) 4U 0142 has a integrated DM of 27, no nebula.
%
%In general, we can convert nH to DM as 10^20 cm^-2 == 0.3 pc cm^-3. So all magnetars in the catalog range from 1-200 ish.
%
%4) SGR 1806-20 has a converted DM of 207 pc cm^-3
%5) SGR 1900+14 has a DM of 63
%6) 1E 2259+586 as a DM of 30 pc cm^-3 associated with CTB 109 with a diameter of 26pc --> EM = 70 pc cm^-6
%7) 1E 1841-045: DM = 66 pc cm^-3 associated with Kes 73, diameter ~ 9 pc, --> EM = 1E3 pc cm^-6.
%
%There are only four-five radio magnetars (for a direct  DM measurement) and only three good associations with SNRs (rest are marginal).
%
%
%-- Shriharsh

\subsection{Efficiency of Radio Emission}
\label{sec:EfficiencyRadioEmission}

An important test for self-consistency of the giant flare model for
FRBs is whether giant flares can support the required energetics.
In order to correctly evaluate the isotropic bolometric energy
release of the FRBs we need to know the radio spectrum of FRBs and
in particular whether there is significant emission in bands outside
the 1.4-GHz band.  At present, we have no constraints on this and
so we will assume that $\mathcal{F}=\ln(10) \nu S_\nu\Delta t$ is
a good measure of the true fluence of the source (see
Equation~\ref{eq:Fluence2}). Here $S_\nu$ is the observed peak flux
density. The bolometric isotropic energy release is then
$\mathcal{F}(1+z)4\pi D^2$ where $D$ is the comoving source distance. For
the four FRBs  we find the radio bolometric energy, $\mathcal{E}_R$,
ranges from $10^{39}$\,erg to $10^{41}$\,erg. After accounting for
the local DM contribution, the distances are smaller and as a result
the isotropic release is smaller by a factor of a few. According
to Lyubarski ({\it ibid}) bolometric radio emission can be produced
with an efficiency of $\eta_R=\mathcal{E}_R/ \mathcal{E}_\gamma=
10^{-6}$ to $10^{-5}$. Thus, working backwards this model would
demand energy releases for the four FRBs to range from $10^{44}\,$erg
to $10^{46}\,$erg (where we have adopted $\eta=10^{-5}$).  This
energy range is well matched to the assumptions made in computing
the volumetric rate  (see comments following Equation~\ref{eq:phiGF}).

%\footnote{This
%hypothesis is tempting given that  during the lifetime of X-ray astronomy one giant
%flare has been seen from the satellite dwarf galaxy LMC  whereas only two
%have been seen from the Milky Way. However, the star-formation rate
%of the LMC is in the range 0.25--0.8$\,M_\odot\,{\rm yr}^{-1}$
%\citep{hz09}
%whereas that of the Galaxy is about 0.7--1.4$\,M_\odot\,{\rm yr}^{-1}$ \citep{rw10}.
%So there is no {\it prima facie} evidence to support this view.}

To conclude, radio  emission arising from giant flares of young
magnetars offer the most plausible physical model that can account
for the high brightness temperature of FRBs (whilst not suffering
from free-free absorption) {\it and} also account for the scattering
tails seen in some FRBs. Furthermore, we find good agreement between
the rates of giant flares and of FRBs.

\section{Frequency Dependent Pulse Width}
\label{sec:ISS}

The {\it Sparker} as well as the brightest FRB in the \cite{tsb+13}
sample show a pulse width that is frequency dependent, $\Delta
t(\nu)\propto \nu^m$ with $m\approx -4$. The simplest explanation
(as has been noted by the discoverers) is that this broadening of
the pulse is due to multi-path propagation (``Interstellar Scintillation
\&\ Scattering'' or ISS). The observations of the {\it Sparker}
with its low DM (relative to the FRBs) is the most difficult to
explain -- whence the focus, in this section, on the {\it Sparker}.
Given our post-mortem of extra-galactic models we focus, in this
section, only on the young magnetar model.

First, we summarize the minimum background to understand the basic
physics of multi-path propagation.  The spectrum of the density
fluctuations is usually modeled as a power law with exponent
$q^{-\beta_K}$, between spatial frequency, $q_1=2\pi/l_1$ and
$q_0=2\pi/l_0$. Here, $l_1$ is the so-called inner scale (at which
energy is dissipated) and $l_0$ is the outer scale (at which energy
is injected).  For the electrons in the diffuse ISM, it appears
that the Kolmogorov spectrum ($\beta_K=11/3$) describes the density
fluctuations quite well.  The normalization of the power law is
described by the ``Scattering Measure'' ($SM$). For a given $SM$
one can derive the spatial coherence scale\footnotemark\footnotetext{The
transverse scale length over which the incident rays will accrue
an rms shift of about 1 radian.  This is similar to the  well known
Fried parameter used by aeronomers and astronomers.}, $r_0$.

\begin{figure}
\centerline{\includegraphics[width=8.5cm]{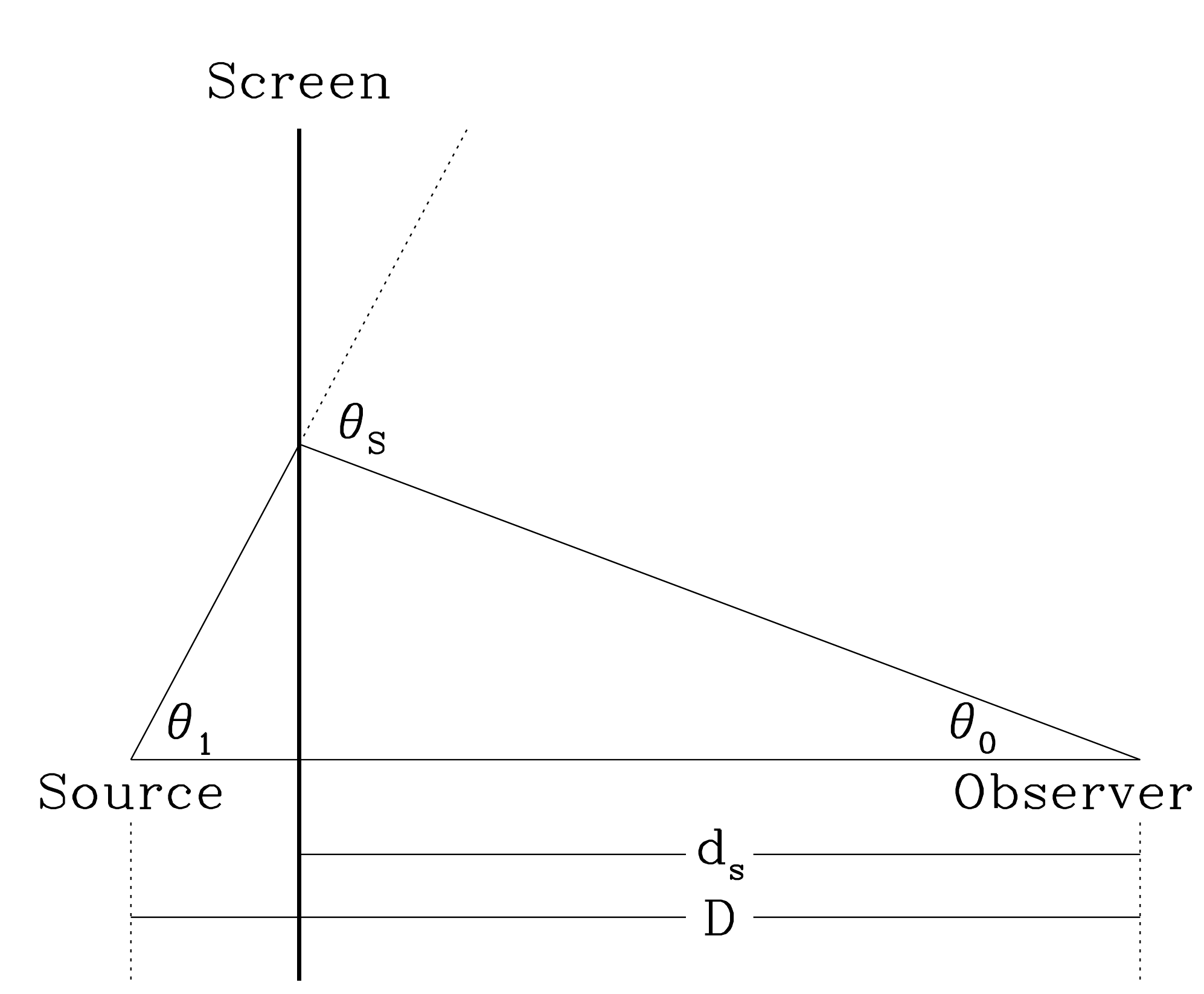}}
\caption{Geometry of the scattering screen.
In the ``thin-screen'' approximation the scattering is confined
to an intervening ``thin'' screen located at $d_s$. 
The screen scatters an incoming ray by the scattering angle,
$\theta_s$ (whose value is directly related to the scattering
strength of the screen). In this example, rays from the source
can reach observer via a direct path and by a scattered path.
The difference between the two arrival paths results in pulse
broadening (amongst other effects). 
\label{fig:ISS}}
\end{figure}

We adopt the ``thin-screen'' approximation (Figure~\ref{fig:ISS})
with the distance to the screen being $d_s$.
With reference to Figure~\ref{fig:ISS}, the rms angle by which a
ray is bent is $\theta_s=1/(kr_0)$; here $k=2\pi/\lambda$.  From
Figure~\ref{fig:ISS} we deduce that $\theta_s=\theta_1+\theta_0$.
In the small angle approximation, $\theta_1 = \theta_0 {d_s}/{(D-d_s)}$
and thus
	\begin{equation}
	\theta_0 = \theta_s\frac{D-d_s}{D}.
	\label{eq:theta_0}
	\end{equation}

A burst of radiation can reach the observer via two extreme paths:
a straight line or via a scattered ray. The time difference between
the two rays gives rise to an exponential scattering tail whose
width is given by
	\begin{equation}
	\Delta\tau \approx \frac{d_s}{2c}\theta_s^2\big(1-\frac{d_s}{D}\big).
	\label{eq:tau2}
	\end{equation}

Equation~\ref{eq:tau2} suggests three locales: ({\it i}) $d_s\ll
D$ (screen close to the observer), ({\it ii}) $d_s\sim D/2$  (screen
midway to the observer and source) and ({\it iii}) $d_s\approx D$
(screen close to the source).  Note that case ({\it i}) and ({\it
ii})  require the same scattering properties but have very different
observational manifestations.\footnote{In particular, for the case
of $d_s\rightarrow D$, the observed angular broadening is severely
suppressed; see Equation~\ref{eq:theta_0}.}

We begin by first estimating the contribution to ISS by the Galactic
ISM. To this end we apply the
NE2001\footnotemark\footnotetext{\texttt{http://rsd-www.nrl.navy.mil/7213/lazio/ne\_model}}
model \citep{cl02} to this line-of-sight and find that the Galactic
ISM contributes a scattering measure ($SM$), in the usual mongrel
and horrific units of $3\times 10^{-4}\,{\rm m}^{-20/3}\,{\rm kpc}$.
The associated Galactic ISS pulse broadening is 0.05\,$\mu$s at 1.4
GHz.  Clearly, the Galactic ISM cannot account for the 5-ms pulse
width of the {\it Sparker}.

\cite{Luan14} provides good arguments why the IGM is unlikely to
have the necessary level of turbulence to result in $\Delta t\approx
5$\,ms (at 1.4\,GHz).  We find the explanation convincing  and so
now focus on the last locale. In this case, we have
	\begin{equation}
	\theta_s^2 \approx 2c\Delta t/l
	\end{equation}
where $l=D-d_s$. 
For $\Delta t=5\,$ms we find $\theta_s=2.04l_{\rm pc}^{-1/2}\,$arcsecond.
The inferred scattering angle, $\theta_s$, can be converted
to {\it SM} using the standard formulation \citep{g97}:
	\begin{equation}
	\theta_s(\nu) = 0.22\,{\rm mas}\big(\frac{\nu}{\rm 1.4\,GHz}\big)^{-11/5}
		\Big(\frac{SM}{10^{-3.5}\,\rm m^{-20/3}\,kpc}\Big)^{3/5}
	\label{eq:thetaSM}
	\end{equation}
where mas stands for milli-arcseconds. From this we deduce 
	\begin{equation}
	{\rm log}({\it SM})=3.1 -\frac{5}{6}{\rm log}(l_{\rm pc}).
	\end{equation}

The most turbulent regions known to date are the following: the
\ion{H}{2} region NGC~6334, $\log(SM)\sim 3.3$ (\citealt{mgr+90});
the Galactic center, $\log(SM)\sim 1.2$ \citep{lca+99}; and the
star-forming Cygnus region, $\log(SM)\sim 1.2$ (\citealt{mmr+95}).
These three regions are rich in gas and stars.  Highly turbulent
screens are usually found at the interfaces of \ion{H}{2} regions,
stellar wind bubbles and the ISM.  {\it This is precisely the sort
of locales where young magnetars are located. }

It is important to check that the scattering medium is not so dense
as to absorb the decimetric pulse.  Turbulence in the nebula results
in variations in density of the electrons, $\langle\delta n_e^2\rangle$.
The EM from the rms variations {\it alone} is \citep{cwf+91}
	\begin{equation}
	{\rm EM}_{\rm SM} = 544\ {\rm cm^{-6}\ pc}
		\Bigg(\frac{SM}{\rm kpc\ m^{-20/3}}\Bigg)
		\Bigg(\frac{l_0}{\rm 1\,pc}\Bigg)^{2/3};
	\end{equation}
here, $l_0$ is the outer scale length of the turbulence spectrum.
This EM should not exceed our previous constraints of $2.7\times
10^7 < {\rm EM} < 6.4\times 10^3\ {\rm cm^{-6}\ pc}$ ({\it cf.}
Table~\ref{tab:Nebulae}).   It is reasonable to assume that the
outer scale length will be a fraction of the size of the nebula
({\it cf.} NGC~6334 and the Galactic center; see \citealt{lca+99}).
Bearing this in mind a scattering measure even as large  as
$\log(SM)\sim 3$ can be accommodated. However, in this extreme case
the scattering screen is not only dense but also very turbulent.
Parenthetically, we wonder whether some FRBs are not detected because
of free-free absorption within the host galaxy (and exacerbating
the all-sky rates of FRBs).

In summary we can explain in the young magnetar model why some FRBs
may exhibit frequency dependent pulses. The radio pulse is broadened
by dense ISM structures that likely form the interface between the
magnetar plerion (or star-forming complex) and molecular clouds
illuminated by young stars.  This hypothesis nicely explains why
scattering tails are not seen in {\it all} FRBs (namely, it is seen
in only those cases where the magnetar is embedded in highly turbulent
structures). In contrast, in the framework where multi-path propagation
takes in place in the IGM one would expect scattering tails to be
seen in all FRBs.

\section{Non-dispersed signal}
\label{sec:CaveatNonDispersedSignal}

The assumption that the frequency-dependent arrival time is due to
propagation through an ionized medium provides the underpinnings
of the discussions in \S\ref{sec:DistanceSize}--\ref{sec:Caveat}.
These considerations led us to reject a stellar, a Galactic, and
even a Local Group origin for the {\it Sparker} and the four Parkes
events. We were led to the conclusion that the {\it Sparker} and
associated events must arise in other galaxies and propose in
\S\ref{sec:GiantFlares} that giant flares from SGRs are the most
plausible progenitor.  The range of models we have considered is
quite comprehensive, yet we must leave no stone unturned.

Motivated thus, in this section we abandon this central assumption.
We will assume that the frequency-dependent arrival time is due to
a property of the source itself.  We start the discussion by noting
that the following three equations denote the same
phenomenon\footnotemark\footnotetext{ In communications a
frequency-dependent arrival time is referred to as a ``chirp''.
Propagation through a cold plasma has a specific chirp signature,
$t\propto \nu^{-2}$.}: $t \propto \nu^{-n}$,  $\dot{\nu} \propto
\nu^{n+1}$ and $\nu \propto t^{-(1/n)}$.

\subsection{Artificial Signals}

The Ultra High Frequency (UHF) band covers the frequency range
0.3--3\,GHz (aka the ``decimetric'' band). Starting from 1.24\,GHz
the frequency allocations are as follows: amateur radio, military,
mobile phone (many
blocks) and cordless phone.  The band 1.4--1.427\,MHz is exclusively
allocated to radio astronomers to undertake passive observations.
Perytons are seen in this band. If Perytons are artificial signals
then the radio astronomy allocation is being (illegally) infringed upon.

It is important to understand that it does not take much for nearby
sources to produce Jy-level signals.  In appropriate units, the
isotropic emitted power of the {\it Sparker}, $1\times 10^6
(D/100\mathrm{km})^2 {\rm\ erg\ s^{-1}}$, is easily emitted by an
orbiting satellite or a terrestrial transmitter\footnote{For
comparison, the power emitted by an active typical cell (mobile)
phone is 0.5 watts or $\sim3\times 10^6$~erg~s$^{-1}$}.  In a similar
vein the signal strength of the GPS signal at a typical location
on the surface of
earth\footnote{\texttt{http://gpsinformation.net/main/gpspower.htm}} is
$-138\ {\rm dBW\,m^{-2}\,MHz^{-1}}$ corresponding to $1.6\times
10^6\,$Jy at the primary carrier frequency (L1) of GPS (1575\,MHz;
2\,MHz wide). Next,  the leisurely drift  (half a second to traverse
300\,MHz of bandwidth) and the quadratic chirp of the Perytons bear
no similarity to artificial signals. Incidentally this  discussion
also shows it will take some effort to post-facto detect (from musty
archives at various radio observatories and monitoring facilities)
radio bursts expected from the past giant flares of SGR\,1900+14
and SGR\,1806$-$20.

\subsection{Solar Flares}
\label{sec:SolarFlares}

A search of the literature revealed Type III solar radio-bursts
\citep{bbg+98} as examples of drifting signals.  Of specific interest
are decimetric Type III bursts (``Type IIIdm''): short pulses of
radiation in the 1--3 GHz range.  The characteristics of typical
Type III burst are: i) a duration of $(\nu/(220\,\mathrm{MHz}))^{-1}$\,s,
ii) a frequency drift of $\dot{\nu}_{\mathrm{GHz/s}} \sim
\nu_{\mathrm{GHz}}^{1.84}$, iii) a strength of
$10-100$\,sfu\footnote{``sfu'' is the solar flux unit, 1~sfu$=
10^4$\,Jy.}, and iv) a brightness temperature in excess of $10^{12}$\,K
indicating that the emission is due to a coherent process. Type
IIIdm bursts usually appear in a series of hundreds to thousands
of bursts, but single bursts have been observed as well (see Figure~7
of \citealt{ib94}).

While their physics is poorly understood, Type IIIdm bursts are
thought to be caused by downward (or upward) directed beams of
non-thermal electrons in the solar corona.  The frequency drift is
believed to be caused by the change in the plasma frequency,
$\omega_p^2 = 4 \pi n_e(r) e^2 / m_e$, a result of the gradient of
the ambient electron density $n_e(r)$ felt by the moving beam.

Except for the weak energetics, the characteristics of the {\it
Sparker} event fit to an order of magnitude the description of a
Type IIIdm burst.  However -- at the time of observation (Aug 24
2001, 19:50:01~UT, or 05:50 local time) -- the Sun was $\sim 7^\circ$
below the horizon at the Parkes radio-telescope site and the angular
distance from the Sun (with respect to the pointing of the telescope)
was $\sim 111^\circ$. This excludes the Sun as the direct origin
of the event.  The hypothesis could still be saved by assuming that
emission from a solar burst was reflected off an orbiting reflector
(e.g., a satellite, or a piece of debris) or the moon\footnote{Such
an event may have been detected during night time at the Bleien
Observatory; see \cite{sbm14}.}.  This would explain the relative
weakness of the event, since, depending on the characteristics of
the reflector and the flare, the signal may be attenuated at will.
However, it would require a series of very fortunate events to have
a very fine-tuned Sun-reflector-Earth configuration occurring at
precisely the right time to reflect a $\nu^{-2}$ Type IIIdm
burst\footnote{This in itself would be unusual given the $t \propto
\nu^{-0.84}$ dependence for typical solar Type IIIdm bursts.} towards
the telescope antenna.  All of the above makes this hypothesis
highly implausible. Additionally, a search of the Virtual Solar
Observatory\footnote{\texttt{http://virtualsolar.org}} revealed no
flares around the time of the {\it Sparker} event.

Other than the Sun, the planet Jupiter is the only significant
source of bursty radio-emission in the Solar system. Jupiter's
emission is dominated by strong ($10^5-10^6$\,Jy) bursts, but
primarily in the decameter band.  Furthermore, at the time of
observation Jupiter was at RA$= 6^{\rm h}37^\prime$,
Dec$= 22^\circ56^\prime$, more than 120$^\circ$ away from the location
of the event.

\subsection{Stellar Flares}
\label{sec:StellarFlares}

A promising source of drifting signals similar to the {\it Sparker}
are the stellar analogs of Type IIIdm bursts.  Flaring at GHz radio
wavelengths has been observed in late-type main sequence stars
\citep{bbd+90} and, as discussed in the previous Section, Type IIIdm
flares are particularly good candidates for a {\it Sparker}-like
signal. For example, a Type III-like burst has recently been observed
in AD Leonis \citep{ob06}, a young, nearby ($D=4.9$\,pc) dM4e star.
Its quiescent $1.5$\,GHz radio luminosity is $5.5 \times 10^{13}\
{\rm erg\, s^{-1}\, Hz^{-1}}$ \citep{jkw89}, equivalent to flux
density levels of $\sim 2\,$mJy, with transient flux density
enhancements of up to $1\,$Jy.

Despite the superficial similarities, the details of stellar flares
and the {\it Sparker} event are in qualitative disagreement.  First,
decimetric bursts observed in flare stars show evidence for
substructures (a series of smaller sub-bursts) not observed in the
{\it Sparker} event [e.g., compare the dynamic spectra in Figures
1 and 5 of \citet{ob06} to Figure~2 in \citealt{lbm+07}].  Second,
the drifts of coronal radio-bursts are typically well fit with a
simple linear dependence or a $t \propto \nu^{-0.84}$ power law in
case of the Sun, significantly different from the observed $t \propto
\nu^{-2}$ drift.  A stellar radio burst compatible with the {\it
Sparker} would need to be one of a kind and unusually fine tuned,
in addition to coming from a yet unknown nearby flare
star.\footnotemark\footnotetext{A \texttt{Simbad} search reveals
no known flare stars in the vicinity of the {\it Sparker}.}  We
therefore consider this explanation unlikely.

We next consider neutron-star analogs of solar Type IIIdm bursts,
recently proposed to exist in magnetar magnetospheres \citep{Lyutikov02}.
Observationally seen as SGRs and anomalous X-ray pulsars (AXPs),
magnetars are young, strongly magnetized ($B \gtrsim 10^{14}$\,G),
and slowly spinning ($P \sim 1-10$\,s) neutron stars.  By extrapolating
the scales known for solar flares and magnetically active T-Tauri
stars, \citet{Lyutikov02} proposed that magnetars should exhibit
short ($<1$\,s), coherent, strong ($\sim 0.1 - 100 \times D_{10{\rm\
kpc}}^{-2}\,{\rm Jy})$, drifting ($\nu_{max} \propto t^{\pm 2}$)
decimetric radio-bursts.  The expected signal drift of $t \propto
\nu^{-1/2}$ is in disagreement with the strongly constrained
observation of $ t \propto \nu^{-2}$, but this may or may not be a
serious problem given the heuristic derivation of the burst properties
that \citet{Lyutikov02} employs.

However, the known magnetars are all in the Galactic plane whereas
the {\it Sparker} and the FRBs are found at high latitude regions
and so we do not consider the Galactic magnetar model to be reasonable.
Parenthetically, as can be gleaned from this discussion, it would
be useful to search  for chirped bursts with different chirp signals
($t\propto \nu^n$ with values other than $n=-2$) in archival pulsar
data, especially at low Galactic latitudes.

\section{Unifying Perytons \&\ FRBs}
\label{sec:Perytons}

In this section we attempt to unify Perytons and FRBs.  We are
motivated by the  fact that Perytons which are a $\nu^{-2}$ chirped
signal are somehow produced either in our atmosphere or by an
artificial source or sources.  Perytons must be nearby because they are
seen in almost all beams.  FRBs are also chirped signals but since
they appear almost always in single beams  they must be located in
the far-field.  Naturally, it is tempting to unify the two classes
of chirped signals by putting Perytons nearby and FRBs further away.
It is the exploration of this simple idea that constitutes the
primary focus of this section.

We submit that examining the detailed properties of radio telescope optics is helpful in our quest for unification. Since Perytons are generally considered to be ``nearby'' it is possible that the events are not sufficiently far away to assume that they are in the Fraunhofer regime, as would normally be the case for celestial events.  In addition to helping unify these phenomena, these details inform us that some care is needed in interpreting Pertyon rates. 

This section is organized as follows. In (\S\ref{sec:Primer}) we
summarize what we know about Perytons.  The necessary background of the Fresnel-Fraunhofer regimes in optical theory is given in \S\ref{sec:Fresnel}.
We then summarize searches for Perytons at other Observatories
(\S\ref{sec:Bleien}--\ref{sec:Other}). We end the section by
constructing a unified model for Perytons and FRBs
(\S\ref{sec:WorkingHypothesis}).

\subsection{A Primer on Perytons}
\label{sec:Primer}

\begin{figure}[htbp] 
   \centering
   \includegraphics[width=2.5in]{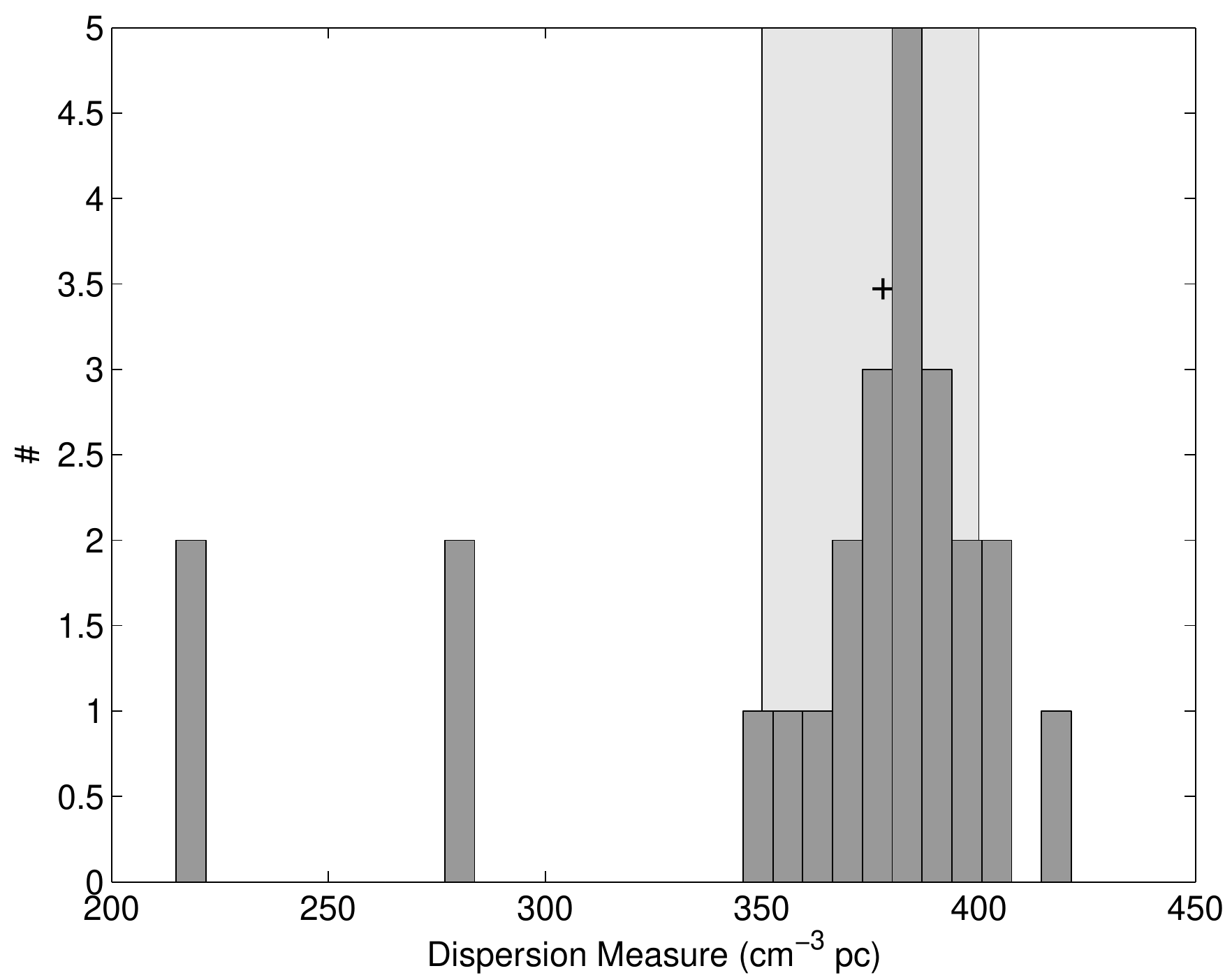}
   \caption{\small
   Histogram of the Perytons observed at Parkes.  The {\it Sparker}
   with a DM=$375\,{\rm cm^{-3}\,pc}$ is shown by a `+' sign. The
   four daytime Perytons found at the Bleien Observatory
   (\S\ref{sec:Bleien}) span the range 350--400$\,{\rm cm^{-3}\,pc}$
   (this range is shown by light shading).  
   }
\label{fig:PerytonHistogram} 
\end{figure}

To date Perytons have been reported from two observatories: Parkes
(\S\ref{sec:Parkes}) and the Blein Observatory (\S\ref{sec:Bleien}).
\cite{kbb+12} provide a succinct description: ``Perytons are signals
with swept-frequency characteristics that mimic the dispersion of
a pulsar, are detected in multiple receiver beams with approximately
the same signal-to-noise ratio ({\it sic}), and cannot be traced
to an astronomical source." It is worth noting that some of the
Perytons show a $\nu^{-2}$ arrival time delay to within experimental
errors (e.g., Peryton 12 and 13 listed in Table~1 of \citealt{bbe+11})
and that others show an approximate quadratic sweep.  The  DMs
inferred from the frequency sweeps lie in the range 200--400\,cm$^{-3}$\,pc
with a mode at about $380\,{\rm cm^{-3}\,pc}$
(Figure~\ref{fig:PerytonHistogram}).

Perytons show symmetric pulses with pulse widths which are tens
of milliseconds. The widths remain the same across the 1.28--1.52\,MHz
band of the Parkes pulsar spectrometer.  In contrast, the pulse
widths of FRBs are less than ten milliseconds with many being
unresolved at the millisecond scale.  The brightest FRB exhibits
an exponential decay which is also  frequency dependent.  The  {\it
Sparker} shows a frequency dependent width but not an exponential
tail.

Perytons show a strong propensity to occur during daytime and many
occur during clear days  \citep{bnm12}.  Furthermore, some Perytons
occur closely spaced in time: five Perytons within a two-minute
interval \citep{kbb+12} and two Perytons within a minute of each
other \citep{bnm12}. In contrast FRBs are not seen to recur despite
several hour-long stares at the same position \citep{lbm+07,tsb+13}.
We defer the discussion of the rates of Perytons to later subsections.

\subsection{Fresnel \&\ Fraunhofer Regimes}
\label{sec:Fresnel}

There are two considerations that matter when observing nearby
objects with large aperture telescopes. First, the beam response
of a large aperture (diameter, $\mathcal{D}$) telescope  depends
strongly on whether the source is ``near-field'' (Fresnel regime)
or ``far-field'' (Fraunhofer regime; Fourier optics). Next, the
angular resolution of a telescope is $\theta_D=\lambda/\mathcal{D}$
where $\lambda$ is the wavelength of the radio signal. We have  no
knowledge of the angular sizes of Perytons and it may well be that
Perytons will be resolved by sufficiently large telescopes (and
this may account for their presence in several beams).

\begin{figure}[htbp]
   \centering
   \includegraphics[width=2.5in]{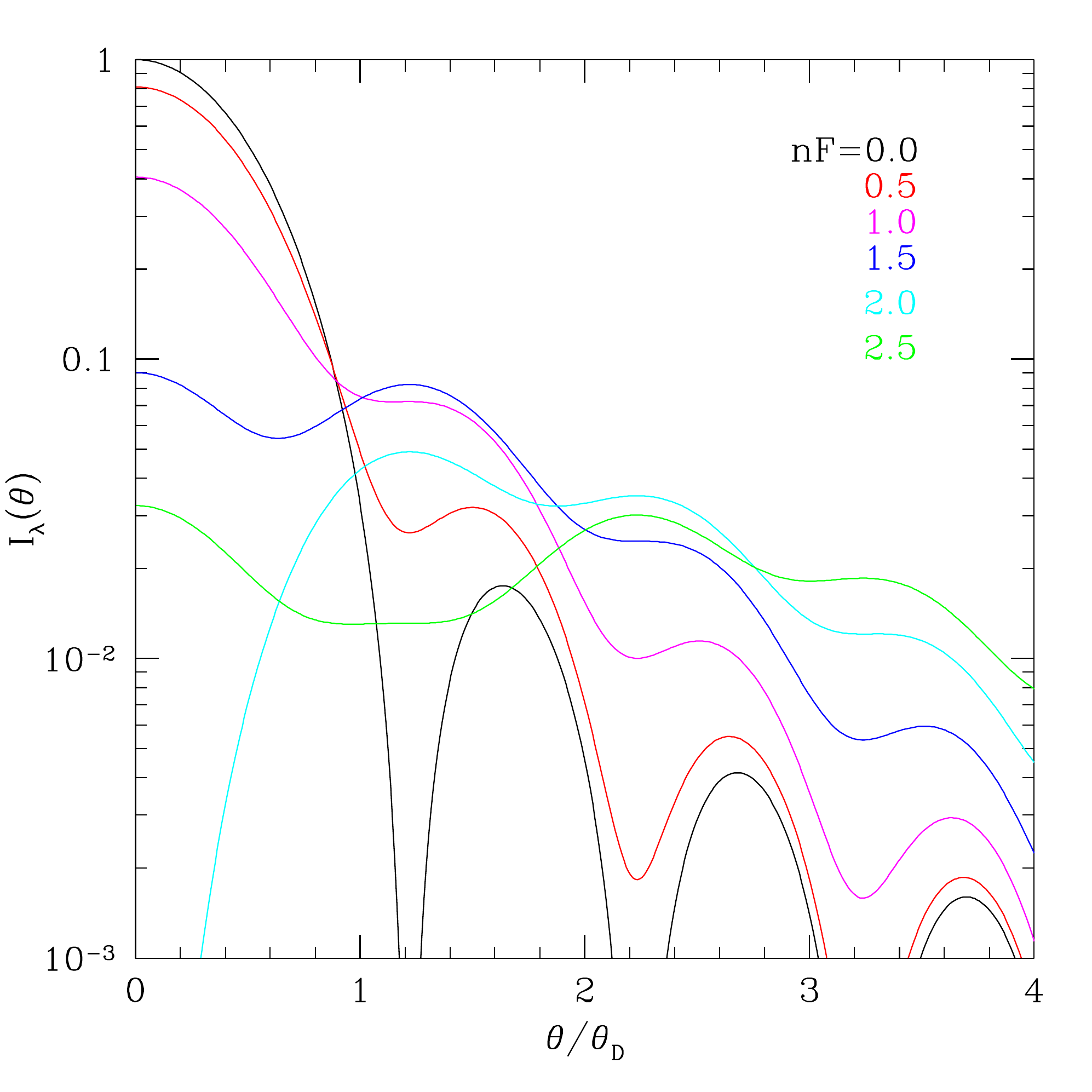} 
   \includegraphics[width=2.5in]{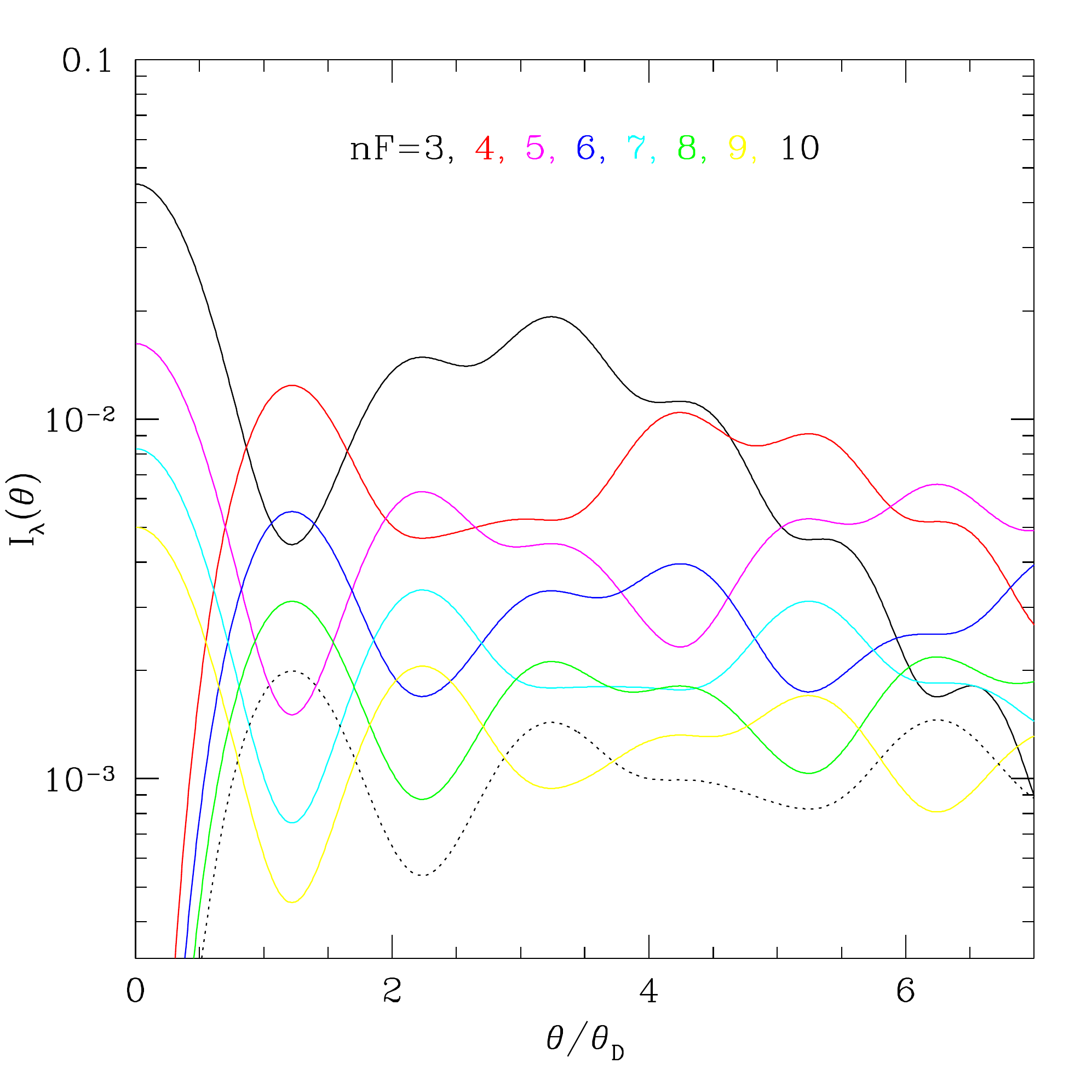}
   \caption{\small
	The response of a circular aperture (diameter, $\mathcal{D}$)
	to a point source located at a distance $D=a_F/n_F$ where
	$a_F=\mathcal{D}^2/\lambda$ is the Fresnel scale.  The
	horizontal axis is in units of $\theta_D=\lambda/\mathcal{D}$
	with $\lambda$ being the wavelength. The normalization is
	such that the beam response for a point source at infinity and
	on axis is unity.  }
   \label{fig:FresnelPSF}
\end{figure}

The Fresnel scale and Fresnel zone number are, respectively,
\begin{eqnarray} 
	a_F = \frac{\mathcal{D}^2}{\lambda} \ \ \ \ 
		{\rm and}\ \ \ 
	&& n_F = \frac{a_F}{D},
	\label{eq:aFnF}
\end{eqnarray}
where $D$ is the distance to the source.  The Fraunhofer approximation
is applicable when $n_F\rightarrow 0$.  The Fresnel formulation is
applicable when $n_F$ is in the vicinity of unity (with ray optics
applicable when $n_F\rightarrow \infty$).  As a matter of reference,
at $\lambda=21\,$cm, the Fresnel scales for a 6-m (ATA), 25-m (VLA
or VLBA antennas), 64-m (Parkes) and 305-m telescope (Arecibo) are
0.18\,km, 3\,km, 20\,km and 440\,m, respectively. This wide variation
in $a_F$ means that care must be taken when comparing Peryton
detections and statistics at the various facilities.

The response, at wavelength $\lambda$, of a telescope with a circular
aperture (diameter, $\mathcal{D}$), to a point source located at
distance $r$ is given by
\begin{equation}
	I_\lambda(\theta|n_F,\theta_D) = {\rm abs}\Big(\int_0^1 J_0[\pi\rho(\theta/\theta_D)]\exp(i n_F\pi \rho^2)2\rho d\rho\Big)^2.
	\label{eq:BesselJ}
\end{equation}
Here, $\theta$ is the angular offset of the receiving beam with
respect to the bore-sight.  This response is graphically summarized
in Figure~\ref{fig:FresnelPSF} for a range of $n_F$. As can be seen
from this figure, even with a  modest Fresnel number a point source
will appear as extended for a telescope which is focused for observing
sources at infinity.

In the Fraunhofer regime the only way a distant compact source can
be seen in multiple beams is by side-lobe ``pick up''. For an
unobscured circular aperture the beam response, in the Fraunhofer
regime, is give by
\begin{eqnarray}
	I_\lambda(\theta) &=&
	\Bigg[\frac{2J_1(\pi\theta/\theta_D)}{\pi\theta/\theta_D}\Bigg]^2,\cr
 	&\approx& \frac{2}{\pi^4} \Big(\frac{\theta}{\theta_D}\Big)^{-3}\ \
	\  {\rm for} \ \ \theta/\theta_D\gg 1.
	\label{eq:FarSideLobes}
\end{eqnarray}
As before, here $I(\theta)$ is normalized to unity for a point
source at infinity and located on axis.

However, structures which obscure the aperture cause additional
side-lobes (and in some cases result in side-lobes with responses
greater than expected from Equation~\ref{eq:BesselJ}).  Let $\eta_m$
be the beam response obtained by integrating from say $\theta=0$
to a few $\theta_D$ (``main beam response''). Then $1-\eta_m$ must
account for the integrated response of the these wayward side lobes.
The smallest response by these side-lobes is obtained by spreading
$1-\eta_m$ uniformly over a solid angle $\Omega_{\rm SL}$ which can
reasonably account for most of the side-lobes. With these two
simplifying assumptions the side-lobe response is
	\begin{equation}
	I_{\rm SL} = (1-\eta_m)\frac{\theta_D^2}{\Omega_{\rm SL}}.
	\label{eq:ISL}
	\end{equation}
For the Parkes telescope we find $I_{\rm SL}=2\times 10^{-6}\Omega_{\rm SL}^{-1}$
where we assume $\eta_m=0.8$ and $\Omega_{\rm SL}$ has the units of steradian.

\subsection{Perytons from Bleien Observatory}
\label{sec:Bleien}

An important very recent development is the detection of Peryton-like
events at the Bleien Observatory located 50\,km west of Zurich,
Switzerland \citep{sbm14}. These authors recorded the radio spectrum
of the sky  with a log-periodic antenna in the band 1.15--1.74\,GHz.
The spectrometer channel width and dump time was 1\,MHz and 10\,ms,
respectively. The beam of the antenna was 110$^\circ$ in the
North-South direction and 70$^\circ$ in the East-West direction.
Over 288 days (from 3 June 2009 to 18 March 2010) the authors found
four {\it day-time} pulsed events with pulse widths of about 20\,ms
and peak fluxes ranging from 250 to 840\,kJy, exhibiting a trajectory
in the frequency-time plane consistent with a $\nu^{-2}$ sweep.

The inferred DMs are in the range 350--400\,cm$^{-3}$\,pc even
though the search covered the range 50--2000\,cm$^{-3}$\,pc.  The
DM determinations are necessarily crude, being limited by coarse
time binning and low SNR (8 to 16). Apart from their apparent
brilliance, these events appear to share all the properties of
Perytons including the strong clustering of the inferred DMs around
300\,cm$^{-3}$\,pc.  It is not unreasonable to conclude that these
events are also Perytons.  With this independent detection at an
Observatory far away from Parkes we can reasonably conclude that
Perytons are truly a world-wide phenomenon.\footnote{ The Parkes
data certainly required Perytons to be of local origin (the Shire
of Parkes). The observations of \cite{sbm14} elevate Perytons to
world-wide or terrestrial status.}

The mean time between the bright Perytons detected at the Bleien
Observatory is 72\,days. Next, the beam of the log-period antenna
is 1.75 steradians. Thus the daily {\it all-sky} rate of the bright
Bleien Perytons is 0.1 per day or 36 per year.

\subsection{Search for Perytons at ATA}
\label{sec:ATA}

A search for FRBs was undertaken at the Allen Telescope Array
\citep[ATA,][]{sbf+12}.  This array consists of 42 dishes each of
6-m diameter and operates in the 1.4-GHz band.  We note the Fresnel
radius for the ATA antennas is 0.18\,km. This length scale is small
enough that we can assume that the intensity response of each antenna
to Perytons, $I_\lambda(\theta)$, is securely in the Fraunhofer
regime.

In the Fly's Eye experiment each antenna was pointed to a different
region of sky and search was undertaken for dispersed pulses
\citep{sbf+12}.  The resulting instantaneous field-of-view was an
impressive 150 square degrees and the experiment lasted 450 hours.
The authors  state ``This wide-field search yielded no detections,
allowing us to place a limiting rate of less than ${\rm
2\,sky^{-1}\,hr^{-1}}$ for 10\,ms duration pulses having mean
apparent flux densities greater than 44\,Jy." Apparently, despite
a gain of nearly $10^4$ in peak flux sensitivity the ATA experiment
could not detect Perytons.

Adopting a mean peak flux of 440\,kJy for the Bleien  sample we
deduce that the all-sky daily rate of Perytons as a function of
peak flux ($S$), $\mathcal{N}_P(S)\propto S^q$ would require that
$q\gtrsim -0.67$.  We appreciate that this inference is subject to
Poisson errors but nonetheless are intrigued by the fact that the
value of $q$ hints at a disk  or even a curved atmosphere geometry
for the distribution of Perytons (\S\ref{sec:SourceCounts}).

We now estimate whether a typical bright Bleien Peryton could have
been detected by the ATA dishes via side-lobes
(Equation~\ref{eq:FarSideLobes}).  Thus a 440\,kJy compact source
would be detectable to a single ATA antenna via off-axis response
provided that $\theta/\theta_D<6$. In this case, the effective
field-of-view of the  Fly's Eye can be as large as 5400 square
degrees.  The product of the solid angle (where all sky is set to
unity) and the exposure time of the Fly's Eye experiment is 2.53
day-sky. The mean Poisson expectation is 0.25. Thus the lack of
detection of a single bright Peryton ($S\gtrsim 440$\,kJy) via a
side-lobe does not violently violate the Bleien rate.

\begin{figure}[htbp]
   \centering
   \includegraphics[width=2.5in]{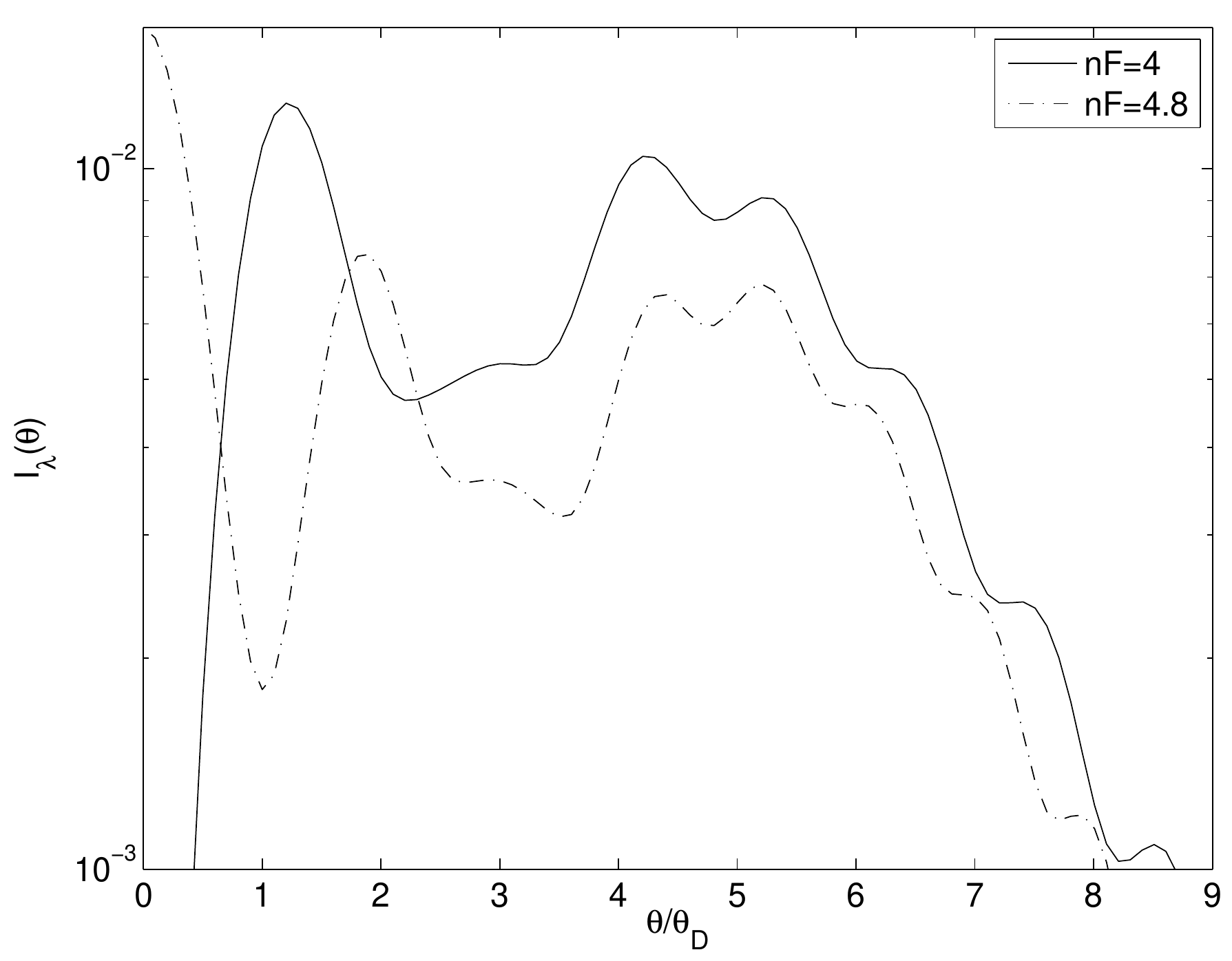} 
   \caption{\small
   The response of the Parkes 64-m telescope to a point source
   located at $D=4.4\,$km. The corresponding Fresnel zone number
   at the lower edge (1.28\,GHz) of the frequency band
   (1.28\,GHz--1.52\,GHz) is 4 and that at the higher edge is 4.8.}
\label{fig:FresnelBandEdge}
\end{figure}

\subsection{Perytons from Parkes}
\label{sec:Parkes}

\cite{bbe+11} and \cite{kbb+12} report Perytons found in the analysis
of the high-latitude data while \cite{bnm12} report events found
in Galactic Plane survey.  In both cases, the same (analog-filter
bank) backend  that was used to detect the {\it Sparker} was used
with the Multi-Beam receiver.  Perytons have been found with the
new digital filter bank\footnote{ from which FRBs were found and
reported by \cite{tsb+13}} (S. Burke Spolaor \&\ M. Bailes, pers.
comm).

The observed rate of Perytons appears to be dependent on which
survey the search (and analysis) was based upon.  \cite{bbe+11}
spent 45 days of observing and found 16 Perytons (or 6 if those
which occurred in a short period counts as only one Peryton). Thus
their observed rate is 0.36 (0.13) per day where the quantity in
brackets refers to ``independent'' Perytons.  The typical peak flux
density for this sample is 0.1\,Jy.  \cite{bnm12} analyzed the
Galactic Plane data and found four Perytons over 75 days. The typical
flux density is higher, 0.5\,Jy.  The daily observed rate is thus
0.05 (0.04) per day\footnote{\cite{bnm12} quote a rate that is smaller than
those quoted here because they treat each beam as an independent stream.
Perytons are found in all beams and thus the beams should not counted
as being independent.}
At low Galactic latitudes the system temperature ($T_{\rm sys}$) is higher
than at high latitudes. So one expects a higher limiting flux and thus
fewer Perytons but the large difference between the rates of the
two surveys (admittedly subject to severe Poisson errors) needs
careful investigation.

The above rates are {\it observed} rates. Translation of  these
rates to {\it all-sky} rates depends upon the location of Perytons
(near-field or far field).  \cite{bbe+11} assume that Perytons are
via pickup of bright source by side-lobe which are severely off-axis:
$\theta\gg 5^\circ$ from the principal pointing axis.  \cite{bbe+11}
go further and assume that the instantaneous field-of-view for the
Perytons is the visible sky ($\Omega=2\pi$ steradians).  This would,
via Equation~\ref{eq:ISL}, require a pickup level of about $10^{-7}$
and thus in this framework Perytons are Mega Jansky sources.  The
apparent coincidence of the Bleien rate and the Peryton rate of
\cite{bbe+11} could be seen as supporting the implicit assumption
of \cite{bbe+11}. However, if  we accept this conclusion then it
would mean that there are very few Perytons which are fainter than
those found at Bleien Observatory -- which, while convenient, we
find to be discomforting.

We take the following two views. (1) It is by no means clear that
Perytons are distant sources, or that one can securely assume that the
Perytons are in the far-field.  (2) We assume that there is a range
of luminosities for the Perytons and the apparent coincidence between
the Bleien and the Parkes rate is a victim of small number statistics.
Consistent with this view we have to consider the possibility that
some Perytons will be nearby and some far away.  With these views
we now re-interpret the Parkes Peryton data.

The Parkes Multi-Beam field-of-view is circumscribed by a circle
of radius\footnote{We give the radius in units of degrees but when
computing solid angles we switch to radians.} $\theta_{\rm
MB}=1.3^\circ$. A source located at $a_F/4$ will easily illuminate
all the thirteen beams.  Thus the Parkes Peryton all-sky rate can
be as high as $\dot\mathcal{N}_{\rm P}\times 4\pi/\Delta\Omega$ per
day where $\dot\mathcal{N}_{\rm P}$ is the daily observed Parkes
Peryton rate and  $\Delta\Omega$ is the average angular ``size''
of Perytons (as seen by the Parkes telescope).  Since all the Parkes
Perytons reported to date are found in all thirteen beams we can
safely conclude that  $\Delta\Omega>\pi\theta_{\rm MB}^2$. The
typical flux density of the Parkes Peryton (in each beam) is a few
tenths of Jansky.  The integrated flux density is larger by at least
$\pi\theta^2_{\rm MB}/\theta_D^2 \approx 150$. Thus the Parkes
Perytons have a peak flux of 15\,Jy or larger.

Since most of what we know about Perytons has come from Parkes it is
worth our time to study the Parkes beam response in some detail.
The beam response function across the Parkes bandwidth is shown in
Figure~\ref{fig:FresnelBandEdge}. We draw the reader's attention
to three points. First, in the Fresnel approximation, the source is
picked up at a level of $10^{-2}$  as opposed to the much smaller
pickup hypothesized by \cite{bbe+11}.  Next, for  most of the solid
angle (angular offset, $\theta \gtrsim 2\theta_D$), the spectrum
can be approximated by a power law with a small value of
$\alpha$.\footnote{Conversely, we note that strong spectral indices,
positive or negative and with large magnitudes can also be obtained.}
Third, as can be seen from Figure~\ref{fig:FresnelPSF}, it is
difficult (in the absence high SNR), to distinguish a Peryton with
$n_F=1/2$ from that located at infinity ($n_F=0$).

 \subsection{Searches for Perytons at Other Observatories}
\label{sec:Other}

Currently, a search for FRBs is being carried out at the Expanded
Very Large Array (EVLA)\footnote{Principal Investigator: Casey
Law.}.  The array has 27 antennas with $\mathcal{D}=25\,$m.  The
Fresnel scale for a single antenna, at a wavelength of 21\,cm, is
3\,km. Likely most Perytons will be in the far-field regime.  A
search for Perytons in the signal streams from each antenna would
be useful. Perytons as nearby objects, given the spatial width of
the B-array, will have substantial parallax.  For instance, the sky
angular position of a Peryton hypothetically located at 5\,km will
vary by $\pm 45^\circ$ as we go from one end of the array to the
other.  Thus, curiously enough, for the study of Perytons, the
instant field of view of the VLA is 27 times that of a single 25-m
telescope.  This total field-of-view exceeds that of the Parkes
Multi-Beam system. Furthermore, given a smaller $a_F$ Perytons are
likely to be in ``focus'' (relative to the situation at Parkes) and
thus the Perytons will be brighter. Going forward it appears to us
that it would be quite promising to undertake commensal or archival
analysis of L-band data.  A single detection of a Peryton will
immediately inform us of its parallax.

The same comments apply to the search for FRBs with the Very Long
Baseline Array (VLBA) system  -- the V-FASTR experiment \citep{wtd+12}.
An additional advantage of the V-FASTR experiment is that it can
simultaneously search for Perytons in ten different weather regions.

\begin{figure}[htbp]
   \centering
   \includegraphics[width=2.5in]{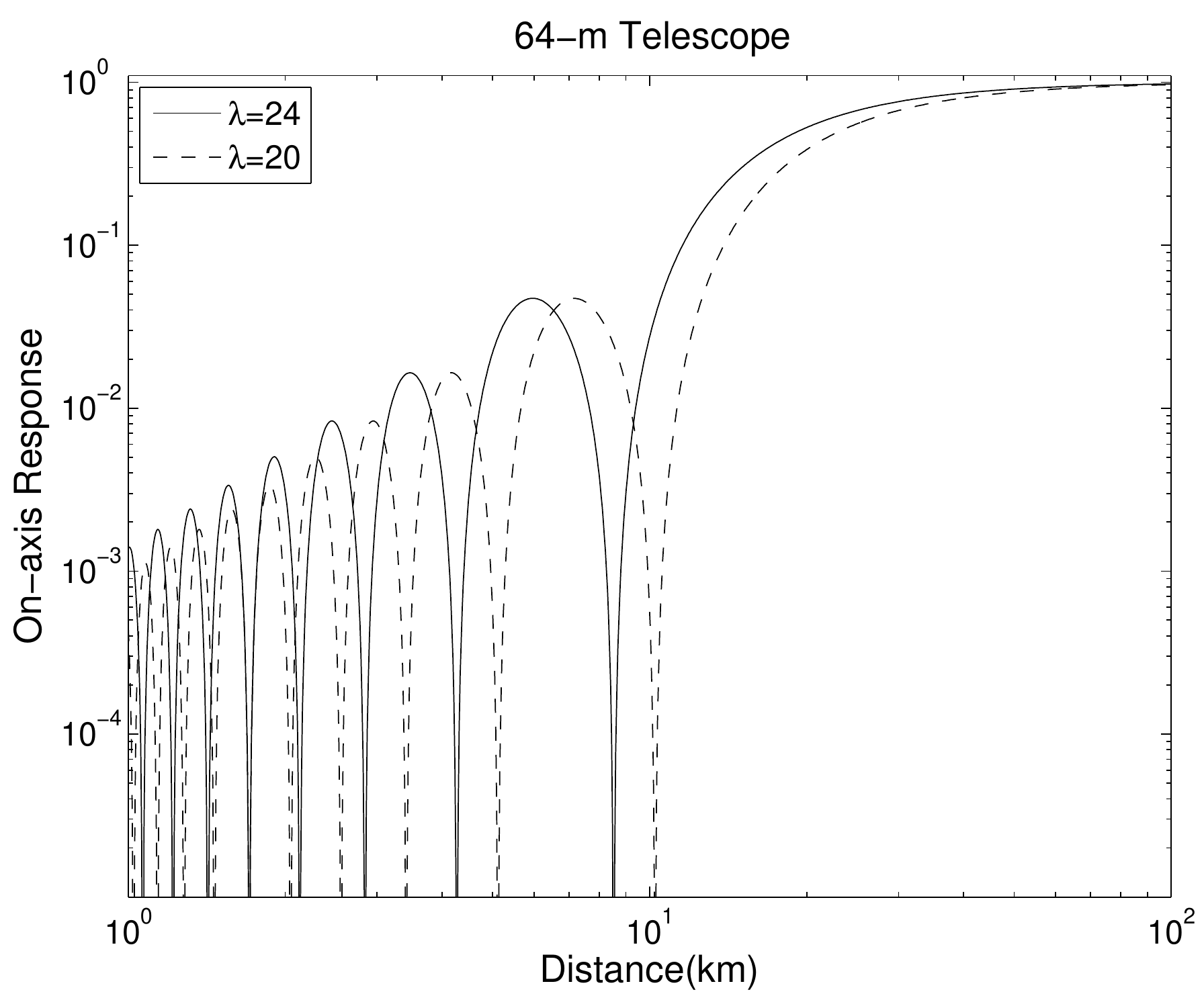}
   \includegraphics[width=2.5in]{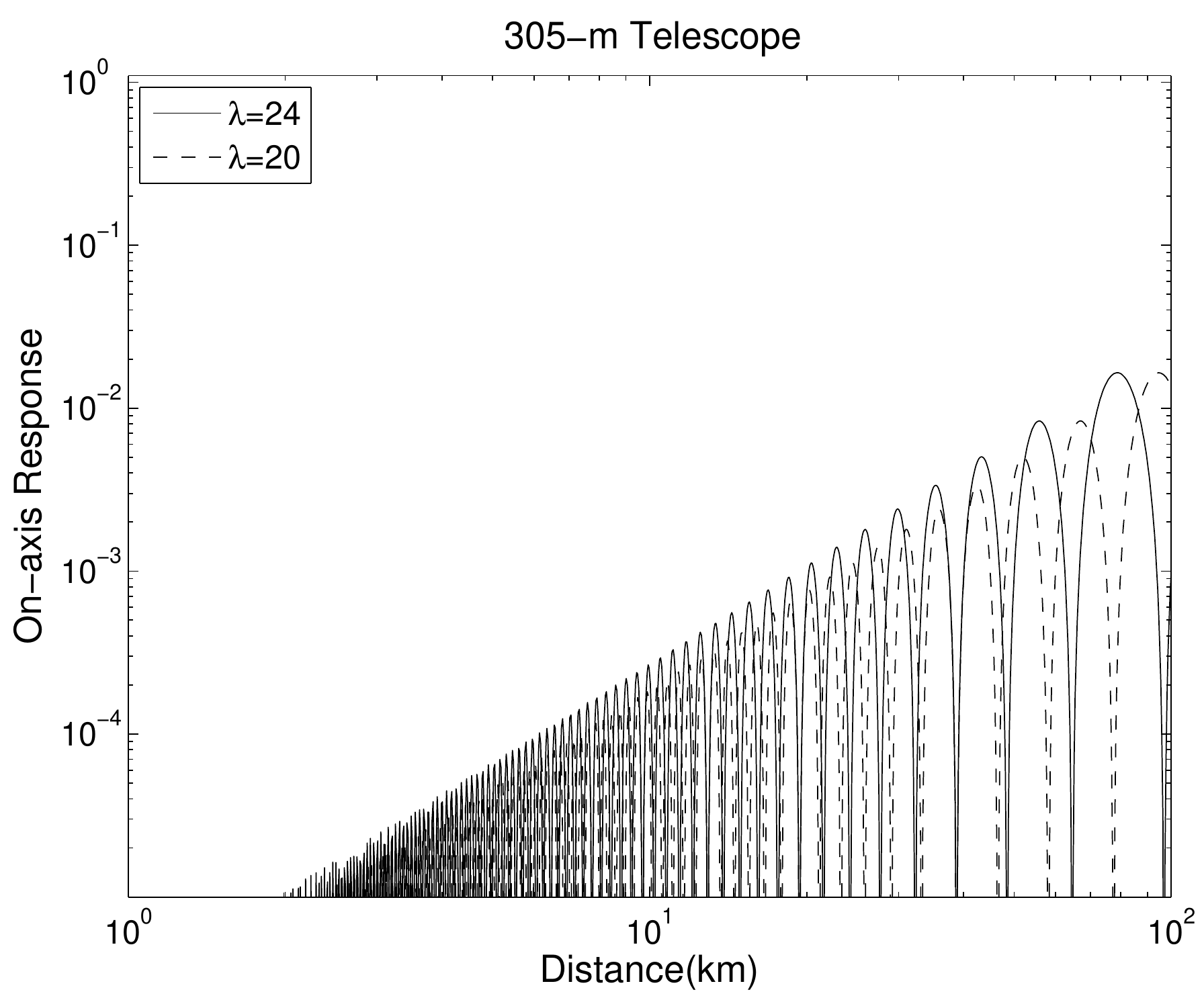}
   \caption{\small The on-axis response for a point source by the
   Parkes 64-m telescope (left) and the Arecibo 305-m telescope
   (right) operating in 1.4-GHz band for point sources located
   between 1\,km and 100\,km.  For {\it each} telescope, the response
   is normalized to be unity for a source at a very large distance
   ($n_F\rightarrow 0$).} 
\label{fig:FresnelOnAxis}
\end{figure}

The Arecibo 305-m radio telescope is also equipped with a Multi-Beam
pulsar receiver and signal processing
system\footnote{\texttt{http://www.naic.edu/alfa/pulsar/}}.  For
the Arecibo telescope $a_F=443\,$km at $\lambda=21\,$cm. Thus,
relative to Parkes, the Perytons will be considerably out of focus
(see Figure~\ref{fig:FresnelOnAxis}) and it may be well be that
Arecibo, despite its larger collecting area, will not detect any
Perytons.

 \subsection{A Working Hypothesis}
\label{sec:WorkingHypothesis}

We propose a {\it working hypothesis} aimed at unifying Perytons, the
{\it Sparker} and FRBs.  The underpinnings are the following:
\begin{enumerate}
\item Perytons are atmospheric phenomena which are detected
essentially on axis (and not via side lobes located a radian or two away from
the bore-sight).
\item  Perytons are seen in many beams for the Parkes multi-beam receiver. 
Ergo we deduce that they
be located in the near-field (``out of focus'').  Thus we infer that Perytons are
located at distances not beyond the first Fresnel zone for the Parkes
telescope at 21-cm wavelength.
\item The {\it Sparker} is a Peryton that probably occurred close to
the Fresnel radius of the telescope, $a_F$ (Equation~\ref{eq:aFnF}).
The higher distance ensures that the {\it Sparker} will appear more or
less in good focus.  Our primary motivation for claiming that the
{\it Sparker} is a Peryton is that the DM of the {\it Sparker} coincides with 
the  peak of the
DM distribution for Perytons  (see also \citealt{bnm12}).
\item The FRBs appear to be in good focus and therefore in this
hypothesis have to  occur beyond the Fresnel radius of the Parkes
telescope at $\lambda=21\,$cm.  As can be seen from
Figure~\ref{fig:FresnelPSF} the beam response for a source with
$n_F\lesssim 1$ is not different from that of cosmic sources
($n_F\rightarrow 0$).  
\end{enumerate}

In this framework, for Perytons, the effective field-of-view is the
larger of the circumscribed circle discussed above (5.3 square
degrees) and the solid angle covered by the Fresnel point-spread
function. The all-sky Peryton rate is then
$(4\pi/\Delta\Omega)\dot{\mathcal{N}}_{\rm P}$ where $\Delta\Omega$
is the larger of 5.3 square degrees and the apparent angular size
of Perytons (as seen by the Parkes telescope in the 20-cm band).
The bulk of the Perytons in this hypothesis would be intrinsically
weak signals, perhaps 100\,Jy to a few kJy.

We are acutely aware that our working hypothesis glosses over many
key issues. To start with, we have provided no strong reasons for
a natural, or atmospheric origin for Perytons as opposed to a man
made origin.  Next, we have not provided any physical model for the
Perytons, nor have we even suggested why Peryton-like phenomenon
(the {\it Sparker} and FRBs), occurring at supposedly larger
heights\footnote{We caution that what matters is the distance to
the Peryton as opposed to the height. A Peryton at an altitude of
say 3\,km can be beyond the first Fresnel zone if viewed at low
elevation angles.} in the atmosphere, should exhibit narrower pulses,
show a $\nu^{-2}$ sweep of arrival time, nor why the  {\it Sparker}
and one of the FRBs exhibit a frequency-dependent pulse width. In
our defense, we note that Perytons are accepted to be local events
and some of them show a $\nu^{-2}$ sweep of arrival time (within
experimental errors; see above).  So our suggestion has some basis
in reality.

We end this section with two  observations. In the proposed framework,
for the Arecibo telescope, the Perytons, the {\it Sparker}, and the
FRBs will be deeply into the Fresnel region. Assigning nominal
heights\footnote{Arecibo is a transit instrument and so the sources
are observed over head.} of 5\,km, 20\,km  and 40\,km for Perytons,
the {\it Sparker}, and FRBs, we find Fresnel zone numbers of 86,
21 and 11.  As can be seen from Figure~\ref{fig:FresnelOnAxis}, the
pick up of Perytons by the giant Arecibo reflector, relative to
the Parkes telescope, is diminished severely.  Next, we note that
a source even at a height of 100\,km has $n_F\approx 4$ and the
Fresnel beam would be quite out of focus (see
Figure~\ref{fig:FresnelBandEdge}).  Our proposed working hypothesis
would have great difficulty (perhaps even to a fatal level) in explaining a
single beam detection of an FRB by the Arecibo multi beam system.
We do note that no Peryton and for that matter no robust
FRB was reported from the archival analysis of the Arecibo data
described in \cite{dcm+09}.  Since the submission of the paper we
became aware of a detection of an FRB candidate at Arecibo
\citep{Spitler:14:2934}. We will assume that this Arecibo event is
not a local artificial signal. In that case, the event must originate
above the atmosphere (see discussion in \S\ref{sec:Fresnel} and also
above).  We point out the broad band spectrum of the event is extremely
unusual having a spectral index, $\alpha$, ranging from 7 to 11!
Spitler and coworkers explain this spectral index by positing that
the event was seen in the side-lobe. This is a plausible explanation.
However, we note that the event is located close to the Galactic
equator. A small intervening ionized nebula (e.g. compact HII region)
could also produce such a strong positive spectral index. Referring to
Equation~\ref{eq:beta}, we find a free-free absorption of $\tau_0\approx 4$
would be sufficient to convert an intrinsic spectral index of $-1$
to the observed spectral index. In this case, the Arecibo event
would be an RRAT with an intervening compact nebula.

We conclude this section by noting that the Fresnel scale for the VLBA
is $3\times 10^3\,$AU and that for the VLA is approximately the
distance to the moon.  In that sense, a {\it single} detection of
an FRB by the VLA will immediately establish an extra-lunar origin
and that by the VLBA an extra-solar-system origin.

\section{Conclusions}
\label{sec:Conclusions}

From analysis of archival pulsar data obtained at the Parkes
Observatory astronomers have reported radio pulses with millisecond
duration and with a frequency dependent arrival time which if
interpreted as due to propagation would require dispersion measures
considerably exceeding that expected from the Galactic interstellar
medium \citep{lbm+07,tsb+13}. The short durations of these events
require a high brightness temperature, even if their origin is
Galactic, let alone extragalactic. The all-sky rate of FRBs is an
astounding $10^4\,$ per day.  In this paper we have explored a wide
range of scenarios capable of explaining the properties and suggested
progenitors of the {\it Sparker} \citep[which has the lowest inferred
DM;][]{lbm+07} and Fast Radio Bursts (FRBs; \citealt{tsb+13}).
Complicating this discussion is the presence of ``Perytons'' which
share the properties of FRBs but are conclusively identified as
arising locally (terrestrial origin). The inferred DMs of the
Perytons are strongly clustered in the range 300--400\,cm$^{-3}$\,pc.

We started our investigation of these sources by accepting that the
large inferred DM for the {\it Sparker} is indeed a result of a
signal propagating through a cold plasma.  We arrived at the following
conclusions.  
\begin{enumerate}
	\item Based on available archival imaging, the nebula that
	produces the large DM for the {\it Sparker} can be no closer
	than 300\,kpc. The minimum distance for the four FRBs (with
	their larger inferred DMs) would be higher. This conclusion
	led us to investigate extra-galactic models for the sources.

	\item We consider a host of plausible extra-galactic
	progenitors including supernovae, blitzars, short hard
	bursts, white dwarf mergers and soft gamma-ray repeaters.
	The models either are physically inconsistent (lack a
	suitable clean and relativistic environment to produce high
	brightness temperature bursts or suffer from free-free
	absorption in the general vicinity of the progenitor) or
	are unable to account for the high all-sky FRB rate
	($10^4\,$day$^{-1}$).  

	\item Of all the possible progenitors, giant flares from
	young magnetars present the most attractive physical model.
	This model has the advantage of naturally explaining why
	some FRBs show frequency-dependent pulse widths. The model
	can also account for the rates provided that an efficiency
	of $10^{-5}$ can be achieved in converting the energy release
	in giant flares into radio emission.  
\end{enumerate}

We believe that we have explored all reasonable {\it stellar} models
for FRBs. Thus, should it turn out that FRBs are not of stellar origin
then non-stellar models (e.g., quasars - E.~S.~Phinney, pers. comm.;
cosmic superconducting strings - \citealt{Vachaspati08}) have to
be considered.

Consistent with our agnostic exploration of the FRB phenomenon we
drop the requirement that the {\it Sparker}'s large DM was produced
by propagation through a cold plasma. In this framework the source
produce a ``chirped'' signal (frequency dependent arrival time).
Chirped signals are used by the military (radar), communications
(spread spectrum) and also arise from natural phenomenon (e.g.,
bursts from the Sun, atmospheric events). We propose an empirical
model unifying Perytons with FRBs with the Perytons being in the
near-field of the Parkes telescope (where the Fresnel approximation
holds) and FRBs being in the far-field (where the traditional Fourier
optics assumed by radio astronomers holds).

The inferred DM for the {\it Sparker} is similar to the mode of the
Peryton distribution (Figure~\ref{fig:PerytonHistogram}).  Next,
it is not obvious to us (from the signal level in different beams)
that the {\it Sparker} has to be a source at a very large distance
(\S\ref{sec:BetterLocalization}).  Economy of hypotheses lead us
to suggest that the {\it Sparker}  itself is a Peryton that occurred
at a height of about 20\,km (the Fresnel scale for the Parkes 64-m
telescope at a wavelength of 21\,cm). In order to explain FRBs as
Perytons we require that the chirp rate of Perytons must scale
proportionally with their distance (height). We offer no explanation
for this requirement.

Perytons form a formidable foreground for FRBs. As such further
progress will require astronomers to understand the distribution
of and distances to Perytons.  Perytons show clearly that Nature
can produce chirped signals in the 21-cm band and so a thorough
understanding of the Perytons will only help astronomers distinguish
local sources from cosmic sources. Since Perytons are local sources
with as yet unknown distances some care is needed prior to comparing
the rates of Perytons from different telescopes (with differing
Fresnel scales).

In summary, there is no compelling evidence to support an
extra-terrestrial origin for FRBs. A plausible argument can be made
to relate giant flares from SGRs to FRBs.  In this picture the
typical redshift of an FRB is $z\approx 0.5$.  An interferometric
localization of FRBs will immediately rule out a local origin. The
same data will either show a host galaxy (which would then revive
stellar models or quasar models) or no host galaxy (which will favor
truly exotic origins). A modest investment in several clusters of
simple dipoles tuned to the 1--2\,GHz band and separated  moderately
(tens to hundreds of kilometers) would be a worthwhile investment (if
only to explore strong decimetric pulses  not only from Galactic
giant flares but from the gamut of Galactic
sources).

Despite the current murky situation it is it tempting to think of
bountiful diagnostics that can be provided by millisecond bursts
of extra-galactic origin.  In \cite{Zheng14} we review a couple
of these diagnostics. In particular we draw the reader's attention
to a unique way by which astronomers can search for solar-mass
intergalactic MACHOs through FRBs.

We conclude by noting that in the title of the paper: ``Giant Sparks
at Cosmological Distances?", the adjective giant refers to the
nominal length scale of the emitting region (300\,km;
\S\ref{sec:ExtraGalacticOrigin}), and the word spark has the same
meaning as in pulsar phenomenology.  We point out that the traditional
outcome of papers that pose a question in their title is generally
in the negative. Nonetheless, one could take some comfort from the
history of gamma-ray bursts (GRBs). This was an exotic phenomenon
even for astronomers.  The history of GRBs started with searches
for possible terrestrial (artificial) signals.  Since their discovery
in 1967, the diversity of observed phenomena has grown tremendously.
Bursts of gamma-rays are now seen from atmospheric events \citep{fbm+94},
from the Sun \citep[Third Orbiting Solar Observatory,][]{krau72},
from compact stellar sources in our Galaxy \citep{mgi+79,cdp+80,kck+08},
and from cosmological distances \citep{mdk+97} and that too from
at least two distinct populations \citep{kmf+93}.  So, at early
times, what one could have  considered to be a single phenomenon
literally spans terrestrial to cosmological scales. It may well be
that astronomers are on a similar adventure in the radio band.

\acknowledgments

SRK would like to thank the hospitality of the Institute
for Advanced Study (IAS). The sylvan surroundings and verdant
intellectual ambiance of IAS resulted in a fecund mini-sabbatical
stay (Fall 2007).  We are grateful to M.~Bailes, S.~Burke-Spolaor and 
D.~Lorimer for sharing with us the details of the Parkes Multi-Beam
Survey,  We  thank M.~Putman and S.~Stanimirovic for help with the
\ion{H}{1} data on the SMC and S.~B.~Cenko, D.~.B.~Fox and J.~Kanner  discussions
about X-ray transients.  We gratefully acknowledge C.~Hirata for careful
reading,  feedback and instructions of basic physics; and P. Kumar and J-P
Macquart for discussions about relativistic flows.  We thank J.~Cordes for
sharing his ideas about detection strategies.  We acknowledge useful
discussions with D.~Bhattacharya, G.~Bower, Y-H~Chu, J.~Condon, D.~A.~Frail, 
P.~M.~Goldreich, A.~Gruzinov, G.~Hallinan, C.~Heiles, E.~S.~Phinney, 
S.~Thorsett, D.~Q.~Wang and  E.~Witten. We especially thank Y.~Cao, S.~Tendulkar and 
M.~H.~van Kerkwijk for a careful reading of the paper.
We acknowledge robust discussions with authors of several recent
papers attempting to explain the origin of FRBs:  
H.~Falcke,
J.~I.~Katz,
A.~Loeb, 
P.~M\'esz\'aros, 
T.~Totani and 
B.~Zhang.  Finally, we thank the anonymous referee whose thorough
reading and thoughtful comments helped to improve the paper.

EOO is incumbent of the Arye Dissentshik career development chair
and is grateful to support by grants from the Willner Family
Leadership Institute Ilan Gluzman (Secaucus NJ), Israeli Ministry
of Science, Israel Science Foundation, Minerva and the I-CORE Program
of the Planning and Budgeting Committee and The Israel Science
Foundation.

This work is supported in part by grants from NSF and NASA.  We
gratefully acknowledge the use of the following archives: Southern
H$\alpha$ Sky Survey, UK Schmidt Surveys, the Digital Sky Survey,
{\it GALEX} and the ATNF Pulsar catalog.  As usual, the authors are
indebted to the librarians who maintain ADS and {\it Simbad}.

\appendix

\section{A Better Localization of the Sparker}
\label{sec:BetterLocalization}

\begin{figure}
\centerline{\includegraphics[width=8.5cm]{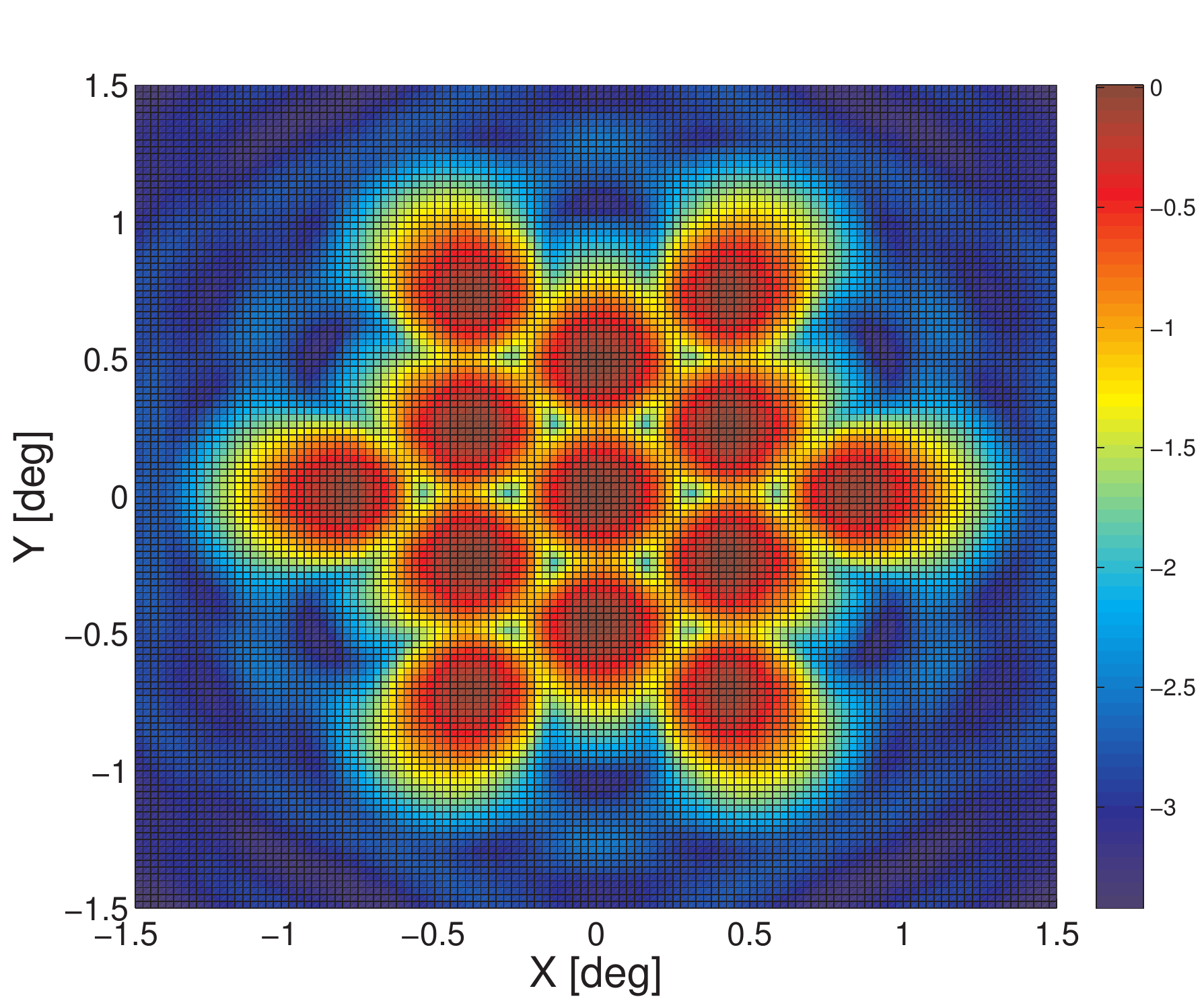}}
\caption{The Parkes Multi-Beam relative response pattern, based on
electromagnetic modeling, as a function of position in degrees
relative to the center of the field-of-view (L.~Staveley-Smith,
pers. comm; updated from \citealt{swb+96}).  The response function
is valid for point sources located well beyond the first Fresnel
zone.  } 
\label{fig:ParkesMB} 
\end{figure}

With a single detection of a pulse by a single beam, the localization
is necessarily poor -- no better than the area of sky illuminated
by the main beam.  However, the {\it Sparker} was detected in 3 out
of the 13 beams, with SNRs of $>100$, $\sim21$ and $\sim14$; a
summary can be found in Table~\ref{tab:Beams}. This pattern of
detections, in principle, should allow us to improve the position
of the {\it Sparker}. To this end we need the location of the beams
and the response of the beams.  We tried several assumptions which
we briefly summarize.  First, as a zero order approximation,
we assume that the beam shapes are Gaussian  with
the widths and gains specified In Table~\ref{tab:Beams},
and that the ratio in intensities in the different beams
are provided by the square of the SNR. We also assumed that the relative intensities are known to precision of about 5\% and that the real S/N ratio
of the saturated beam is smaller than about 1000.
Given these assumptions we find that the Sparker localization
is within the error region specified in Table~\ref{tab:ErrRegion}.
However, it is well known that the beam shapes of radio instruments
are non-Gaussian.  Therefore, next we attempt to use an electromagnetic model of the beam response supplied to us by L.~Staveley-Smith (updated
from \citealt{swb+96}, see Figure~\ref{fig:ParkesMB}).  The response function is valid for point
sources located well beyond the first Fresnel zone.

\begin{table}[hb]
\begin{center}
\caption[]{SNR of the {\it Sparker} in the Beams}
\begin{tabular}{cccccc}
\hline\hline
Beam & RA (J2000) & Dec (J2000) & SNR & FWHM & Gain\\
           &            &     & arcmin &   \\                                
\hline
1 & $01:21:18.0$ & $-74:46:01$ & $  <5   $ &  14.0 & 0.74 \\
2 & $01:17:09.8$ & $-74:22:04$ & $  <5   $ &  14.1 & 0.69 \\
3 & $01:24:19.5$ & $-74:19:33$ & $  <5   $ &  14.1 & 0.69 \\
4 & $01:28:37.7$ & $-74:43:00$ & $  <5   $ &  14.1 & 0.69 \\
5 & $01:25:39.1$ & $-75:09:40$ & $  <5   $ &  14.1 & 0.69 \\
6 & ${\bf01:18:06.0}$ & ${\bf-75:12:19}$ & $  >100 $ &  14.1 & 0.69 \\
7 & $01:13:55.8$ & $-74:48:09$ & $  14   $ &  14.1 & 0.69 \\
8 & $01:09:53.4$ & $-74:23:32$ & $  <5   $ &  14.5 & 0.58 \\
9 & $01:20:13.1$ & $-73:55:27$ & $  <5   $ &  14.5 & 0.58 \\
10 & $01:31:30.8$ & $-74:15:57$ & $  <5   $ &  14.5 & 0.58 \\
11 & $01:33:14.5$ & $-75:06:15$ & $  <5   $ &  14.5 & 0.58 \\
12 & $01:22:30.3$ & $-75:36:34$ & $  <5   $ &  14.5 & 0.58 \\
13 & $01:10:25.8$ & $-75:14:14$ & $  21   $ &  14.5 & 0.58 \\
\hline
\end{tabular}
\label{tab:Beams}
\end{center}
Notes: The entry in bold-face is the beam in which the {\it Sparker}
signal is saturated (SNR$>$100).  FWHM stands for Full Width at
Half Maximum of the beam. The adopted FWHM values are 14,
14.1 and 14.5\,arc\,minute, for the central beam, inner-ring beams
and outer-ring beams, respectively (see \citealt{mlc+01}).  The
gain is the mean aperture efficiency of each beam ({\it ibid}).
The positions of the beams were provided to us  by M.~Bailes and
D.~Lorimer.
\end{table}

\begin{table}[hb]
\begin{center}
\caption[]{{\it Sparker} Error region}
\begin{tabular}{cc}
\hline\hline
RA (J2000) & Dec (J2000)  \\                                
\hline
18.85619 & -75.12665 \\
19.34551 & -75.18275 \\
19.37912 & -75.19694 \\
19.34551 & -75.20249 \\
18.83199 & -75.14258 \\
18.83064 & -75.13011 \\
\hline
\end{tabular}
\label{tab:ErrRegion}
\end{center}
\end{table}

We adopted the SNRs given above for  beams 6, 7, and 13, and $<5$
in the rest of the beams (Table~\ref{tab:Beams}).  Since the SNR
values are subject to Poisson errors, we allow for $3\sigma$
uncertainties in the SNR values that we used.  However, we were not
able to find any position within the Parkes multi beam field-of-view
which can reproduce the observed detections.  This failure could
be  due to ({\it i}) the electromagnetic model is not adequate to
model responses at large angles ($2\,\theta_D$ to $3\,\theta_D$)
or ({\it ii}) the {\it Sparker} is not located at a great distance
(in which case  our use of the Multi-Beam pattern is incorrect).

\cite{bbe+11} used an empirical beam response (by scanning the
multi-beam receiver across a bright pulsar) and found a best-fit
position: RA=$19.44^\circ\pm 0.08^\circ$ and Dec=$-75.17^\circ\pm
0.08^\circ$ (J2000).  This position almost coincides with beam\,6
(see Figure~\ref{fig:SparkerLocation} and also Table~\ref{tab:Beams}).
The {\it Sparker} is detected in beam\,7 (due North West) and
beam\,13 (due West). However, given the claimed location we would
have expected the {\it Sparker} to be detected by the beams due
South East, due East and due North East with SNRs similar to those
seen in beams 7 and 13 or at least with SNR $>$ 5.  The lack
of detection in these three beams is troubling.

In order to deduce the most conservative localization of the {\it
Sparker} we adopted an approach based primarily on symmetry. We
assumed that the beam pattern has circular symmetry. Since the
{\it Sparker} was detected in three beams, but not in all the other beams,
we conclude that the {\it Sparker} should be in the region between the
three beams. Beam 7, 6  and 12 are on a straight line, therefore
the lower part of the localization region is perpendicular to the
line connecting these beams.  We assumed that the region is symmetric,
mostly because beam 6 had the strongest detection.  These considerations
led us to a polygon (aka ``kite''). The vertexes of this polygon
are listed in Table~\ref{tab:Polygon}.  The North-East side of the
polygon is defined by the centers of beams 6 and 7 (see
Table~\ref{tab:Beams}), while the North-West side is defined by the
centers of beams 7 and 13.  The South-East side is perpendicular
to the line joining beams 6 and 7, and the South-West side is the
intersection of the line joining beams 13 and 12 and the South-East
side.

\begin{table}
\begin{center}
\caption[]{\small The vertices of the polygon which encloses all possible positions ofthe {\it Sparker}.}
\begin{tabular}{cc}
\hline
\hline
RA (J2000) & Dec (J2000) \cr
\hline
  $19.5250$ & $-75.2053$ \\
  $18.4825$ & $-74.8025$ \\
  $17.6075$ & $-75.2372$ \\
  $18.5900$ & $-75.3647$ \\
\hline
\end{tabular}\
\label{tab:Polygon}
\end{center}
\end{table}

\section{Mono-energetic particle synchrotron spectrum}
\label{sec:MonoEnergetic}

The simplest model for producing an arbitrarily steep spectrum radio
emission is to have mono-energetic electrons gyrating in a magnetic
field. Starting at lower frequencies the spectrum rises as $x^{1/3}$,
peaking at $x=0.29$ and declining as
\begin{equation}
        S(x)= A \sqrt{x} \exp(-x), \qquad x\gg 1.
        \label{eq:SimplifiedMonoEnergetic}
\end{equation}
Here, $A$ is a normalization factor, $x=\nu/\nu_c$ and
\begin{equation}
        \nu_c = \frac{3}{4\pi}\gamma^3\omega_B\sin(\alpha)
\end{equation}
is the so-called ``gyro-synchrotron'' frequency.  Here, $\omega_B
= eB/(\gamma m_ec)$ is the gyro-frequency of an electron with Lorenz
factor $\gamma$ and gyrating in a magnetic field of strength $B$
and moving in the mean at an angle $\alpha$ with respect to the
field lines (\citealt{rl79}, Chapter 6).

For $x\gg 1$ the power law slope is given by
\begin{equation}
        \alpha \equiv \frac{d{\rm ln}S_\nu}{d{\rm ln}\nu} =
        \frac{1}{2}-\frac{\nu}{\nu_c}.
\end{equation}
At high frequencies, an arbitrarily large spectral index can be
obtained by invoking a smaller value of $\nu_c$.

\section{Stars \&\ Supernovae: DM \&\ EM}
\label{sec:StarsSupernovae}

Consider a star with a mass loss rate of $\dot M$ and radius $R_*$.
In a steady state this leads to a wind with a density radial distribution,
$\rho(r)$, given by
\begin{equation}
	\dot M = 4\pi r^2 {\rm v_w}\rho(r).
\end{equation}
Here, ${\rm v_w}$ is the radial velocity of the wind many stellar
radii away from the star.  The dispersion measure, DM, emission measure,
EM, and plasma frequency, $\nu_p$, are then given by
\begin{eqnarray}
	{\rm DM} &=& \int_{R_*}^\infty \frac{\rho(r)}{\mu m_H}dr=
	 \frac{\dot M}{4\pi {\rm v_w}\mu m_H}R_*^{-1},\\
	{\rm EM} &=& \int_{R_*}^\infty \Big(\frac{\rho(r)}{\mu m_H}\Big)^2dr
	= \Big(\frac{\dot M}{4\pi {\rm v_w}\mu m_H}\Big)^2 \frac{R_*^{-3}}{3},\\
	\nu_p &=& \frac{1}{2\pi}\sqrt{\frac{4\pi n_e e^2}{m_e}}\,{\rm Hz},
\end{eqnarray}
where $\mu$ is mean molecular weight of electrons.
	
The stellar wind velocity is clearly greater than the escape velocity.
For stars on the lower main sequence, the escape velocity is constant
since $R\propto M$. We set  ${\rm v_w}=10^3\,$km\,s$^{-1}$ and
for simplicity let $\mu=1$. Then we find
	\begin{eqnarray}
		{\rm DM} &=& 17B
		\Big(\frac{R_*}{R_\odot}\Big)^{-1}\,{\rm cm^{-3}\,pc},\cr
		{\rm EM} &=& 4\times 10^{9}B^2
		\Big(\frac{R_*}{R_\odot}\Big)^{-3}\,{\rm cm^{-6}\,pc},\cr
		\nu_p &=& 223
		B^{1/2}\Big(\frac{R_*}{R_\odot}\Big)^{-1}\,{\rm MHz},
		\label{eq:StarDMEM}
	\end{eqnarray}
where $B=\dot M_{-10}/({\rm v_w}/10^3\,{\rm km\,s}^{-1})$ and $\dot
M_{-10} = \dot M/10^{-10}\,M_\odot\,{\rm yr}^{-1}$.   These equations
show why stellar models cannot produce  sufficient dispersion measure
without producing a very large emission measure leading to free-free
absorption in the decimetric band.

In the model of \cite{lsm13}, the radio pulse is produced at some
radius within an extended corona and the DM results from the pulse
propagating to the surface.  Such an extended corona cannot be
stably bound to the star and it is reasonable to assume a wind
solution as above. However, we will not assume a steady state.  Let
the radio pulse be emitted at radius $R_*$ and the edge of the
corona be at $L$.  In this case and the DM and EM are
\begin{eqnarray}
	{\rm DM} &=& n_*R_*\Big[1-(R_*/L)\Big],\cr
	{\rm EM} &=&  \frac{n_*^2R_*}{3}\Big[1-(R_*/L)^3\Big],\cr
                &=& \frac{\rm DM^2}{3R_{\rm
		pc}}\frac{[1-(R_*/L)^3]}{[1-(R_*/L)]^2},
\end{eqnarray}
where $n_* = n_e(R_*)$ and $R_{\rm pc}=R_*/(1\,{\rm pc})$.  Even
if $L$ is greater than $R_*$, by as little as a factor of 1.3, we
have ${\rm EM\approx DM^2/R_{\rm pc}}$.

One of the models suggested for the Parkes events is the merger of
two white dwarfs that eventually forms a magnetar \citep{lev06}. Our current
understanding of the  merger is as follows: the lower mass white
dwarf is tidally disrupted and accretes onto the other (``primary'')
white dwarf. During the mass buildup of the primary white dwarf, a
fraction of the accretion energy drives a very strong stellar wind.
The relevant outflow velocity is the escape velocity of the primary
star and so ${\rm v_w}\approx 5\times 10^8\,{\rm cm\,s^{-1}}$. Using
the convention from supernovae, we have $A_*\equiv \dot M/(4\pi
{\rm v_w})/5\times 10^{11}\,{\rm gm\,cm^{-1}}$ \citep{cf06}.  $A_*=1$
for  ${\rm v_w}=5\times 10^3\,{\rm km\,s}^{-1}$ and $\dot M=5\times
10^{-5}\,M_\odot\,{\rm yr}^{-1}$.  Rescaling from
Equation~\ref{eq:StarDMEM} and using Equation~\ref{eq:tauff} we
obtain the free-free optical depth at $\nu_0=1.4$\,GHz to be
	\begin{equation}
		\tau_{\rm ff}(\nu_0) = 1.9 A_*^2r_{15}^{-3}.
		\label{eq:SNtauff}
	\end{equation}
Here, the radial distance $r=10^{15}r_{15}\,$cm.  The run of plasma
frequency with density is
	\begin{equation}
		\nu_p = 5A_*^{1/2}r_{15}^{-1}\,{\rm MHz}.
	\end{equation}
These two equations inform us that only radio emission emitted after
the blast wave has crossed the radius at which the free-free optical
depth is sufficiently small will reach the observer.  For instance,
even if the stellar wind lasts for a day the circumstellar medium
will be optically thick to decimetric radiation (provided $A_*$ is
comparable to unity).

Next, we consider the case of a merger product transmuting to the
next level of compactness: merged white dwarfs to magnetar, or
merged neutron stars to a rapidly spinning black hole. We will
assume a mass $\Delta M$ is ejected at sub-relativistic velocities
${\rm v}$ in a spherical geometry.  Numerical simulations suggest
$10^{-4}\lesssim \Delta M\lesssim 10^{-2}\,M_\odot$ \citep{hkk+13}.

Let us assume that the debris is a shell of radius, $R={\rm v}t$
and has a  width, $\Delta R=fR$ with $f$ being assumed to be a
constant (with time). Then
\begin{eqnarray}
	n_e &=& 9.5\times 10^{9}f_{-1}^{-1}\Delta M_{-2}{\rm v_{10}}^{-3}t_5^{-3}\,{\rm cm^{-3}},
	\ \  \nu_p = 878 \big(f_{-1}^{-1}\Delta M_{-2}{\rm
	v}_{10}^{-3}t_5^{-3}\big)^{1/2}\,{\rm MHz},\\
	{\rm DM} &=& 3.1\times 10^{5}\Delta M_{-2}{\rm v}_{10}^{-2}t_5^{-2}\,{\rm cm^{-3}\,pc},\ 
	{\rm EM} = 3\times 10^{15}f_{-1}^{-1}\Delta M_{-2}{\rm
	v}_{10}^{-5}t_5^{-5}\,{\rm cm^{-6}\,pc},
\end{eqnarray}
where $\Delta M=10^{-2}\Delta M_{-2}\,M_\odot$, ${\rm v}=10^{10}{\rm
v_{10}}\,{\rm cm\,s^{-1}}$, $f=0.1f_{-1}$ and $t=10^5t_5\,$s.   Thus,
for these nominal parameters one would have to wait many months
before any radio emission from the central source can successfully
propagate to the outside world.

\section{Source Counts}
\label{sec:SourceCounts}

Here we review the source count for several geometries (in particular
curved atmosphere).  This section may be useful in inferring the
geometry of Perytons (from observations).

\bigskip
\noindent{\bf Spherical Geometry.}
Consider the following case: a homogeneous population of sources,
density $s$ per unit volume, with identical luminosity, $L$ in
Euclidean geometry. Then the volume within distance $r$ is  $V(<r)
= \frac{4}{3}\pi r^3$.  % following assumption not required according
to Zheng %Assume that the luminosity function is clustered around
$L$.  The flux density at Earth is $S=L/(4\pi r^2)$. Thus number
of sources with flux density less than $S$ is
	\begin{equation}
		N(<S) = s\frac{4\pi}{3} r^3 \propto S^{-p}
	\end{equation}
with $p=3/2$.

\begin{figure}[htbp] 
   \centering
   \includegraphics[width=3in]{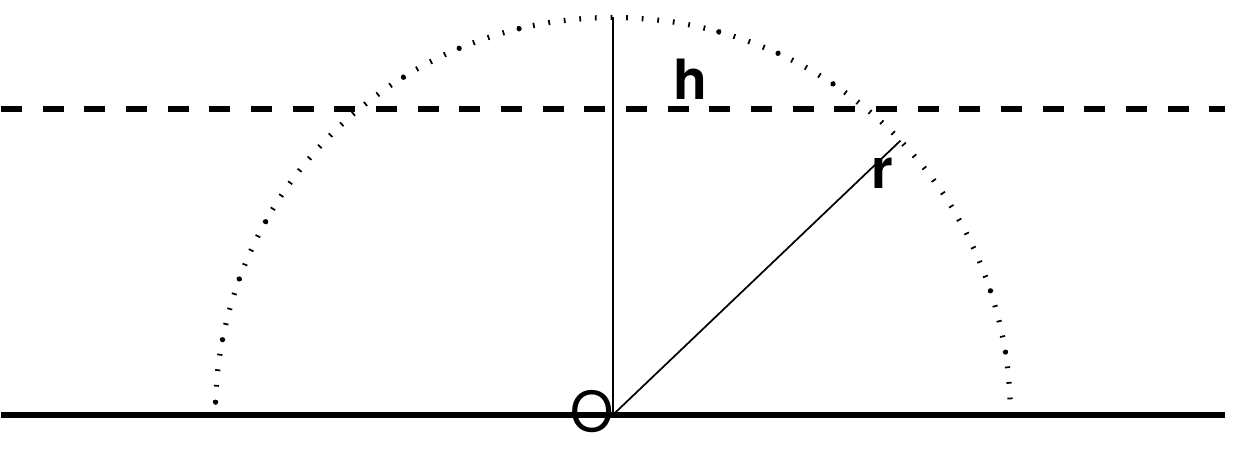}
   \caption{\small
   Geometry of plane-parallel atmosphere. The scale height is $h$.
   The solid line indicates the horizontal plane on which the
   telescope (marked at `O') is located. The dotted line shows the
   hemisphere of radius $r$.  
   }
   \label{fig:PlaneParallel} 
\end{figure}

\bigskip
\noindent{\bf Plane Parallel Geometry.}
Now consider, a slab of height $h$ and extending indefinitely along
its length as shown in Figure~\ref{fig:PlaneParallel}.  For $r<h$
we the same scaling as in the spherical case. For $r>h$ the volume
of the atmosphere is the difference between the volume of the
hemisphere of radius $r$ and the volume of the polar cap whose
height is $r-h$. This volume is
	\begin{equation}
		V(<r) = \pi h^3\Big[\frac{r^2}{h^2}-\frac{1}{3}\Big].
		\label{eq:PlaneParallel}
	\end{equation}
Thus in this case $V(<r)$ asymptotically approaches $r^2$ (as $r\gg h$) and 
$N(S)\propto S^{-p}$ with $p\rightarrow 1$.

\begin{figure}[htbp] 
   \centering
   \includegraphics[width=3in]{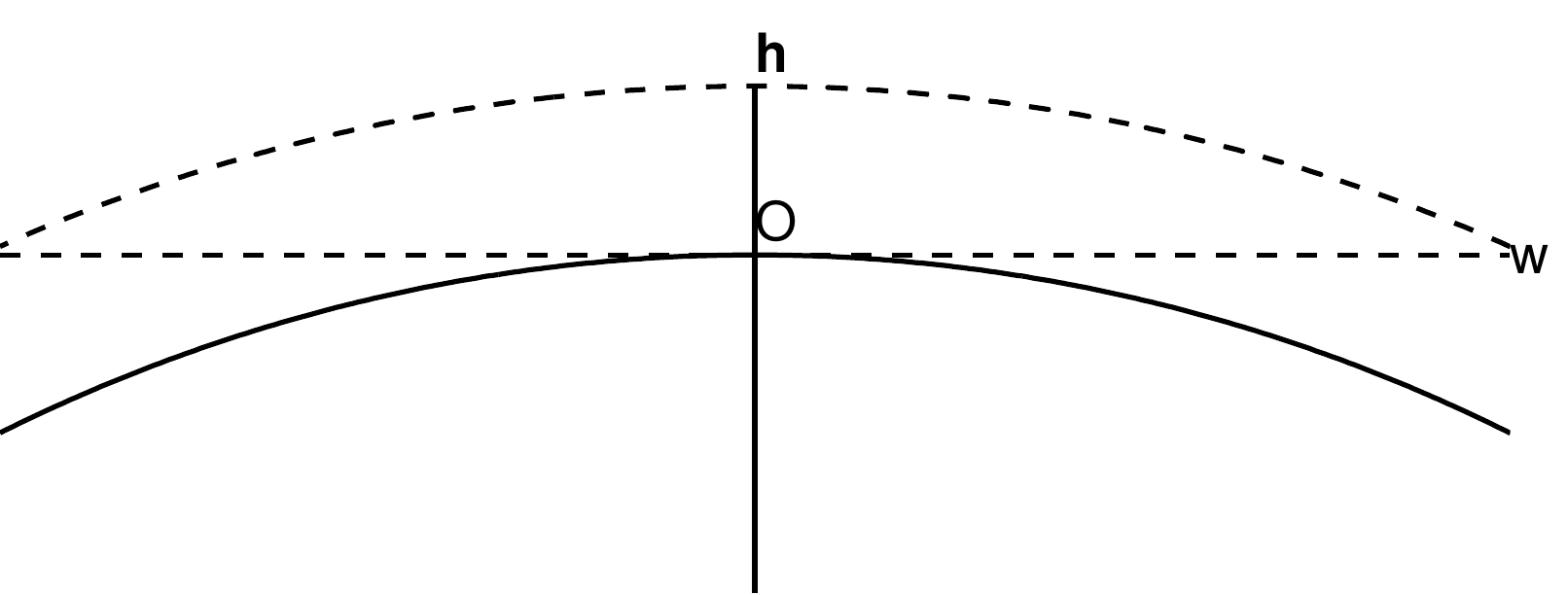} 
   \caption{\small 
   The scale height of the atmosphere is $h$ and the radius of Earth
   is $R$ (not shown).  Only sources are above the local horizontal
   plane are visible to the radio telescope (located at `O'). The
   maximum horizontal distance is $w=\sqrt{2hR+h^2}$.  
   }
   \label{fig:CurvedAtmosphere} 
\end{figure}

\bigskip
\noindent{\bf Curved Atmosphere Geometry.}
Now consider an atmosphere enclosing a sphere (as in the case of
our atmosphere).  In this case as can be seen from
Figure~\ref{fig:CurvedAtmosphere} $V(r)$ has a maximum value. For
$r\lesssim h$ a sphere of radius $r$ will within the atmosphere and
thus $p\approx 3/2$. Next, the maximum value for $r$ is $w=
\sqrt{2Rh+h^2}$. Clearly, we run out of volume, $V(<r)$, when $r>w$.
Thus $p=0$, asymptotically.  Thus a flat power law index (especially
$p\lesssim 1/2$) would indicate a population within a curved
atmosphere.

An elegant   derivation of the differential and the integral was
obtained by E.~S.~Phinney and is given below
	\begin{eqnarray}
		\frac{dV}{dr} &=&  2\pi r^2\Big[\frac{(R+h)^2-R^2-r^2}{2Rr}\Big]=
		\frac{\pi r}{R}\Big[2Rh+h^2-r^2]\cr
		V(<r) &=& \frac{2\pi}{3}h^3 + 
		\frac{\pi(r^2-h^2)}{2R}\Big[2Rh+h^2\Big]
		+ \frac{\pi}{4R}\big(h^4-r^4\big)
	\label{eq:Phinney}
	\end{eqnarray}
and valid for $h<r<w$. In the limit of $h\ll R$  the formula for
plane parallel distribution is recovered (Equation~\ref{eq:PlaneParallel}).
The differential formula is well suited for computing the population
distribution with a specified vertical dependence for the density
of the sources.

\section{Glossary}

To assist the reading in navigating the paper, we provide the following
tables of frequently used symbols, their meanings, their units, and where
they are first discussed in the paper.

\begin{deluxetable}{rlrr}
\tabletypesize{\scriptsize}
\tablecaption{Frequently Used Symbols, Part 1}
\tablehead{
\colhead{Symbol} & \colhead{Meaning} & \colhead{Units/Value} &
\colhead{Section}
}
\startdata
.\\
.\\
.\\
.\\
.\\
.\\
DM & Dispersion Measure & cm$^{-3}$\,pc &
	\S\ref{sec:Introduction}\\
$t$ & Arrival time of radio signal & sec &
	\S\ref{sec:Introduction}\\
$\nu$ & Frequency of radio signal & GHz &
	\S\ref{sec:Introduction}\\
$n$ & Exponent of relation between $t$ and $\nu$, $t(\nu)\propto \nu^n$ &  &
	\S\ref{sec:Introduction}\\
$\dot{\mathcal{N}}$  & All-sky daily event rate & day$^{-1}$ &
	\S\ref{sec:Introduction}\\
$\Delta\tau$ & measured pulse width at frequency $\nu$ & ms &
	\S\ref{sec:Introduction}\\
$L$ & Thickness (size) of nebula causing most of DM & pc &
	\S\ref{sec:RationaleLayout}\\
$d$ & Distance to the nebula & pc &
	\S\ref{sec:RationaleLayout} \\
$D$ & Distance to the source ($>d$) & pc &
	\S\ref{sec:RationaleLayout}\\
$\nu_0$ & center frequency of observing band & 1.4\,GHz &
	\S\ref{sec:TheSpark}\\
$S_0$ & spectral flux density at frequency $\nu_0$ & Jy &
	\S\ref{sec:Energetics}\\ 
$\Delta\tau_0$ & measured pulse width at $\nu_0$ & ms &
	\S\ref{sec:Energetics}\\
$\Delta t$ & intrinsic pulse width at $\nu_0$ & 1\,ms &
	\S\ref{sec:Energetics}\\
$S(\nu)$ & spectral flux density at frequency, $\nu$ & Jy &
	\S\ref{sec:Energetics}\\
$\mathcal{F}(\nu)$ & Fluence, $S(\nu)\Delta\tau(\nu)$ & Jy\,ms &
	\S\ref{sec:Energetics}\\
$\alpha$ & spectral index of the spectrum of the fluence, $F(\nu) \propto \nu^\alpha$  & -1 &
	\S\ref{sec:Energetics}\\
$\tau_0$ & free-free optical depth at frequency, $\nu=\nu_0$& &
	\S\ref{sec:Energetics}\\
$\nu_l$ & the lowest frequency of source emission & GHz &
	\S\ref{sec:Energetics}\\
$\nu_u$ & the highest frequency of source emission & GHz &
	\S\ref{sec:Energetics}\\
$\mathcal{E}_R$ & isotropic total (integrating from $\nu_l$ to $\nu_u$) energy release& erg &
	\S\ref{sec:Energetics}\\
$D_{\rm kpc}$ & Distance to the source in units of kpc & kpc &
	\S\ref{sec:Energetics}\\
$R$ & Radius of the source & pc &
	\S\ref{sec:BrightnessTemperature}\\
$T_B(\nu)$ & Brightness temperature at frequency $\nu$ & K &
	\S\ref{sec:BrightnessTemperature}\\
$\Gamma$ & Bulk Lorentz factor of the expanding source & &
	\S\ref{sec:BrightnessTemperature}\\
EM & Emission Measure of the nebula & cm$^{-6}\,$pc &
	\S\ref{sec:DistanceSize}\\
$n_e$ & mean electron density in nebula & cm$^{-3}$ &
	\S\ref{sec:DistanceSize}\\
$L_{\rm pc}$ & The thickness of nebula in pc units & pc &
	\S\ref{sec:DistanceSize}\\ 
$\tau_{\rm ff}(\nu)$ & Free-free optical depth of the nebula at frequency $\nu$& &
	\S\ref{sec:EM}\\
$T_e$ & temperature of nebula & K &
	\S\ref{sec:EM}\\
$\alpha^\prime$ & the log-derivative of $\mathcal{F}(\nu)$ & &
	\S\ref{sec:EM}\\
$\nu_c$ & the characteristic frequency of an exponential spectrum & GHz &
	\S\ref{sec:EM}\\
$L_{\rm ff}$ & the size of the nebula for which $\tau(\nu_0)=5$ & pc &
	\S\ref{sec:EM}\\
$\mathcal{F}$ & Bolometric Fluence & erg\,cm$^{-2}$ &
	\S\ref{sec:EM}\\
$\theta_{\rm DM}$ & Maximum angular size of nebula & degree &
	\S\ref{sec:GalacticDM}\\
Superscript $^S$ & Object located in SMC & &
	\S\ref{sec:SMCDM}\\
Superscript $^G$ & Object located in Milky Ways & &
	\S\ref{sec:SMCDM}\\
$F_{H\alpha}$ & line-integrated H$\alpha$ emission & erg\,cm$^{-2}$\,s$^{-1}$ &
	\S\ref{sec:Halpha}\\
$d_{\rm min}$ & minimum distance to the nebula & pc &
	\S\ref{sec:Halpha}\\
$\dot{N}_I$ & rate of ionization of the nebula & s$^{-1}$ &
	\S\ref{sec:UVContinuum}\\
$\dot{N}_R$ & rate of recombination within the nebula & s$^{-1}$ &
	\S\ref{sec:UVContinuum}\\
$\alpha_B$ & Case-B recombination rate & cm$^{3}$\,s$^{-1}$ &
	\S\ref{sec:UVContinuum}\\
$h\nu_1$ & energy of a photon at the Lyman edge & eV &
	\S\ref{sec:UVContinuum}\\
$\Delta_{UV}$ & {\it GALEX} color: FUV - NUV & AB mag &
	\S\ref{sec:UVContinuum}\\
$\phi_V$ & volume filling factor of the nebula & &
	\S\ref{sec:PorousNebula}\\
$n_0$ & particle density of ambient medium & cm$^{-3}$ &
	\S\ref{sec:RadiativeShocks}\\
$v_s$ & velocity of shock into the ambient medium & cm\,s$^{-1}$ &
	\S\ref{sec:RadiativeShocks}\\
$N_e$ & total number of electrons in (flash-ionized) nebula &  &
	\S\ref{sec:Flash}\\
$\tau_{\rm ion}$ & timescale for ionization of a neutral atom at the edge of the
nebula & yr &
	\S\ref{sec:Flash}\\
$\Delta t_X$ & duration of the soft X-ray flash & ms &
	\S\ref{sec:Flash}\\
$E_{\rm ion}$ & energy release of the soft X-ray flash & erg &
	\S\ref{sec:Flash}\\
$\tau_{\rm R}$ & recombination timescale within the nebula & s &
	\S\ref{sec:Flash}\\
$R_*$ & radius of the stellar corona & pc &
	\S\ref{sec:StellarCorona}\\
$R_{pc}$ & radius of the stellar corona in units of pc & pc & 
	\S\ref{sec:StellarCorona}\\
$\alpha(\nu)$ & free-free absorption coefficient per unit length & cm$^{-1}$ & 
	\S\ref{sec:StellarCorona}\\
$\tau(\nu)$ & free-free optical depth at frequency $\nu$ & & 
	\S\ref{sec:StellarCorona}\\
$\dot M$ & mass loss from corona & $M_\odot$\,yr$^{-1}$ & 
	\S\ref{sec:StellarCorona}\\
$\epsilon_{\rm ff}$ & free-free luminosity per unit volume & erg\,cm$^{-3}$\,s$^{-1}$ & 
	\S\ref{sec:StellarCorona}\\
*$\mathcal{L}_{\rm ff}$ & free-free luminosity & erg\,s$^{-1}$ & 
	\S\ref{sec:StellarCorona}\\
$f_{\rm ff}$ & bolometric flux density from corona & erg\,cm$^{-2}$\,s$^{-1}$ & 
	\S\ref{sec:StellarCorona}\\
$F_{\rm ff}$ & bolometric fluence from corona & erg\,cm$^{-2}$ & 
	\S\ref{sec:StellarCorona}\\
$\tau_X$ & Duration of hard X-ray emission & s & 
	\S\ref{sec:StellarCorona}\\
.\\
.\\
.\\
.\\
.\\
.\\
.\\
.\\
.\\
.\\
.\\
.\\
\enddata
\end{deluxetable}

\begin{deluxetable}{rlrr}
\tabletypesize{\scriptsize}
\tablecaption{Frequently Used Symbols, Part 2}
\tablehead{
\colhead{Symbol} & \colhead{Meaning} & \colhead{Units/Value} &
\colhead{Section}
}
\startdata
.\\
.\\
.\\
.\\
.\\
.\\
.\\
.\\
.\\
$\mathcal{E}_S$ & Isotropic energy release from FRBs in the radio band & erg & 
	\S\ref{sec:ExtraGalacticOrigin}\\
$\Phi_{\rm FRB}$ & volumetric annual rate of Fast Radio Bursts (FRBs) & Gpc$^{-3}$\,yr$^{-1}$
& 
	\S\ref{sec:Progenitors}\\
$A$ & mass loss parameter, $A=\dot M/(4\pi v_w)$ & g\,cm$^{-1}$ & 
	\S\ref{sec:CoreCollapseSupernovae}\\
$A_*$ & mass loss parameter, $A$,  in units of $5\times 10^{11}\,{\rm g\,cm}^{-1}$ & &
	\S\ref{sec:CoreCollapseSupernovae}\\
$P_1$ & Period of neutron star just prior to collapse into a black hole & ms & 
	\S\ref{sec:Blitzar}\\
$P_0$ & Period of newly born massive neutron star &  ms & 
	\S\ref{sec:Blitzar}\\
$\tau$ & time taken to go from $P_0$ to $P_1$ & yr & 
	\S\ref{sec:Blitzar}\\
$B$ & Dipole field strength of pulsar & G & 
	\S\ref{sec:Blitzar}\\
$\dot E$ & spin down luminosity of pulsar & erg\,s$^{-1}$ & 
	\S\ref{sec:Blitzar}\\
$R_S$ & radius of blast wave (supernova) & cm & 
	\S\ref{sec:Blitzar}\\
$I$ & moment of inertia of neuron star & cm$^2$\,g & 
	\S\ref{sec:Blitzar}\\
$\theta$ & half of the opening angle of conical jet for GRBs & rad & 
	\S\ref{sec:ShortHardBursts}\\
$f_b$ & beaming factor of conical GRBs (=$1-\cos(\theta)$) & & 
	\S\ref{sec:ShortHardBursts}\\
$\mathcal{E}_\gamma$ & isotropic gamma-ray energy release of a giant flare & erg & 
	\S\ref{sec:GiantFlares}\\
$\mathcal{E}_* $& $\mathcal{E}_\gamma$ of giant flares (characteristic value) & erg &  
	\S\ref{sec:GiantFlares}\\
$\Phi_{GF}$ & volumetric rate of giant flares & ${\rm Gpc^{-3}\,yr^{-1}}$ &  
	\S\ref{sec:GiantFlares}\\
$\tau_{GF}$ & mean time between Galactic giant flares & yr &  
	\S\ref{sec:GiantFlares}\\
$\eta_R$ & ratio of energy emitted in radio to that in gamma-rays for Giant Flares & &  
	\S\ref{sec:EfficiencyRadioEmission}\\
$m$ & exponent of relation between pulse width ($\Delta t$) frequency ($\nu$), $\Delta t \propto \nu^m$ & & 
	\S\ref{sec:ISS}\\
$q$ & spatial frequency of turbulence power spectrum & &  
	\S\ref{sec:ISS}\\
$\beta_K$ & power law index of turbulence power spectrum & & 
	\S\ref{sec:ISS}\\
$l_1$ & the length scale at which turbulence energy is dissipated & cm & 
	\S\ref{sec:ISS}\\
$l_0$ & the length scale at which energy is injected for turbulence & cm & 
	\S\ref{sec:ISS}\\
$r_0$ & the spatial coherence scale of the scattering screen & cm & 
	\S\ref{sec:ISS}\\
$\lambda$ & the wavelength of the propagating radio signal ($c/\nu$) & cm & 
	\S\ref{sec:ISS}\\
$d_s$ & distance from the observer to the scattering screen & kpc & 
	\S\ref{sec:ISS}\\
$\theta_s$ & angle by which a ray is typically bent by the scattering screen & rad & 
	\S\ref{sec:ISS}\\
$\Delta\tau$ & the temporal spread induced by the scattering screen & ms & 
	\S\ref{sec:ISS}\\
SM & scattering measure & kpc\,m$^{-20/3}$ & 
	\S\ref{sec:ISS}\\
$\omega_p$ & plasma frequency & GHz & 
	\S\ref{sec:SolarFlares}\\
$\mathcal{D}$ & diameter of receiving antenna & m & 
	\S\ref{sec:Fresnel}\\
$\theta_D$ & full width at half maximum of receiving antenna at wavelength $\lambda$ & m & 
	\S\ref{sec:Fresnel}\\
$a_F$ & Fresnel scale ($\mathcal{D}^2/\lambda$) & km & 
	\S\ref{sec:Fresnel}\\
$n_F$ & Fresnel number ($a_F/D$) & & 
	\S\ref{sec:Fresnel}\\
$\theta$ & angle between the principal axis of the telescope and source & rad & 
	\S\ref{sec:Fresnel}\\
$\mathcal{N}_{\rm P}(S)$ & all-sky daily rate of Perytons with flux density $S$ & d$^{-1}$ & 
	\S\ref{sec:ATA}\\
$\Delta\Omega$ & survey area: maximum of Peryton size in degrees$^2$ and 5.3 degrees$^2$ & degrees$^2$ & 
	\S\ref{sec:WorkingHypothesis}\\
$\dot M$ & spherical stellar mass loss rate & $M_\odot\,{\rm yr}^{-1}$ &
	\S\ref{sec:StarsSupernovae}\\
$v_{\rm w}$ & radial velocity of the stellar wind & km\,s$^{-1}$ &
	\S\ref{sec:StarsSupernovae}\\
$\nu_p$ & plasma frequency ($\omega_p=2\pi\nu_p$)& MHz &
	\S\ref{sec:StarsSupernovae}\\
$L$ & the corona extends between $R_*$ and $L$ & cm &
	\S\ref{sec:StarsSupernovae}\\
.\\
.\\
.\\
.\\
.\\
.\\
.\\
.\\
.\\
.\\
.\\
.\\
\enddata
\end{deluxetable}

\clearpage

\bibliographystyle{apj1b}
\bibliography{bibsparker}

\end{document}